\documentclass[a4paper,11pt]{article}
\pdfoutput=1

\usepackage{jheppub}

\usepackage[T1]{fontenc}
\usepackage{graphicx}
\usepackage{epstopdf}
\usepackage{bm}
\usepackage{epsfig}
\usepackage{graphics}
\usepackage{xspace}
\usepackage{hyperref}
\usepackage{slashed}
\usepackage{xcolor}
\usepackage{longtable}
\usepackage[small]{subfigure}
\usepackage[normalem]{ulem}

\newcommand{\MSb}{\overline{\mathrm{MS}}}

\newcommand{\dd}{\mathrm{d}}

\title{\boldmath Secondary massive quarks with the Mellin-Barnes expansion}

\preprint{\begin{flushright} IFT-UAM/CSIC-24-12\end{flushright}\vspace*{-2cm}}

\author[a,b]{Alejandro Bris}
\author[c]{and Vicent Mateu}

\affiliation[a]{Departamento de F\'isica Te\'orica, Universidad Aut\'onoma de Madrid,\\Cantoblanco, 28049, Madrid, Spain}
\affiliation[b]{Instituto de F\'isica Te\'orica UAM-CSIC, E-28049 Madrid, Spain}
\affiliation[c]{Departamento de F\'isica Fundamental e IUFFyM,\\Universidad de Salamanca, E-37008 Salamanca, Spain}

\emailAdd{alejandro.bris@uam.es}
\emailAdd{vmateu@usal.es}

\abstract{Processes involving only massless or massive quarks
at tree-level get corrections from massive (lighter, heavier, or equal-mass) secondary quarks starting at two-loop order, generated by a virtual gluon splitting into a massive quark anti-quark pair.
One convenient approach to compute such two-loop corrections
is starting with the one-loop diagram considering the virtual gluon massive. Carrying out a dispersive integral with a suitable kernel over the gluon mass yields the desired two-loop result. On the other hand, the Mellin-Barnes representation can be used to compute the expansion of Feynman integrals in powers of a small parameter. In this article we show how to combine these two ideas to obtain the corresponding expansions for large and small secondary quark masses to arbitrarily high orders in a straightforward manner. Furthermore, the convergence radius of both expansions can be shown to overlap, being each series rapidly convergent. The advantage of our method is that the Mellin representation is obtained directly for the full matrix element from the same one-loop computation one needs in large-$\beta_0$ computations, therefore many existing results can be recycled. With minimal modifications, the strategy can be applied to compute the expansion of the one-loop correction coming from a massive gauge boson. We apply this method to a plethora of examples, in particular those relevant for factorized cross sections involving massless and massive jets, recovering known results and obtaining new ones. Another bonus of our approach is that, postponing the Mellin inversion, one can obtain the small- and large-mas expansions for the RG-evolved jet functions. In many cases, the series can be summed up yielding closed expressions.}

\begin{document}
\maketitle
\flushbottom

\section{Introduction}\label{sec:intro}
The precision of experimental measurements achieved by current particle-physics colliders, together with the expected updated accuracy of upcoming facilities, makes equally precise theoretical predictions mandatory. Lattice QCD simulations uncertainties are becoming increasingly small~\cite{FlavourLatticeAveragingGroupFLAG:2021npn} and one expects this tendency will continue in the near future. Therefore, perturbative computations must keep up with the pace set by both experimental and lattice results. In practice, since the strong coupling is larger than the electroweak one, and given that the gluon is massless, better precision translates in computing higher-order QCD corrections. While multi-loop computations for processes involving a single scale are showing tremendous progress
, most physical situations naturally involve several scales, making computations beyond one-loop very involved, in particular when it comes to find analytic results. In the situation of having only two scales, the result (up to overall factors) can only depend on their (dimensionless) ratio, and one can rely on expansions if there is a clear hierarchy between the scales. The Mellin-Barnes (MB for short) representation (see Ref.~\cite{Smirnov:2004ym} for a complete review of the subject) can be used to obtain such expansion for an arbitrarily high order, as was first noticed in Ref.~\cite{Friot:2005cu}. The idea is, for a given master integral, integrate all loop momenta using Feynman parameters $\{x_i\}$, apply the MB representation as many times as necessary, and after carrying out the remaining integrals over $\{x_i\}$, apply the inverse mapping theorem to obtain the desired expansion.

Quark masses are important parameters of the strong sector of the Standard Model and play a key role in flavor physics and even in searches beyond the Standard Model. Knowing them with high precision is then of utmost importance, and for such endeavor it is practical having ready-to-use theoretical expressions that depend on them. Even if the process of interest depends on a single scale at lowest order (e.g.\ because it only involves massless quarks and gluons), at $\mathcal{O}(\alpha_s^2)$ one can produce massive quarks through the splitting of a virtual gluon into a quark-antiquark virtual or real massive pair. Even though at very high energies quarks can be considered approximately massless, at lower energies (or if aiming at high precision) the effects of their mass cannot be ignored. Furthermore, such secondary contributions may contribute in different ways for a given process depending on how the size of the mass compares to the various energy scales involved, giving rise to a sequence of effective field theories (EFTs for short) in which the massive quark may or may not be dynamic. This is the core of the so-called ACOT scheme~\cite{Aivazis:1993kh,Aivazis:1993pi}, also known as the variable-flavor number scheme \cite{Gritschacher:2013pha,Pietrulewicz:2014qza} when final-state jets are produced.

For virtual massive bubbles, or for the real radiation of secondary quarks in quantities that only depend on the total momentum of the quark pair, the bulk of these corrections can be computed with a dispersive integral over the ``fake'' mass of a (virtual or real) gluon, see Refs.~\cite{Kniehl:1988id,Hoang:1994it,Hoang:1995ex}. The secondary virtual bubble contribution is simply an insertion of the lowest-order massive vacuum polarization function $\Pi(\ell^2)$ in the gluon propagator carrying loop momentum $\ell$, and the dispersive integral accounts for the contribution of this function in the on-shell (OS for short) scheme, $\Pi^{\rm OS}(\ell^2)=\Pi(\ell^2)-\Pi(0)$, which is ultraviolet (UV) finite. On top of the dispersive contribution, one needs to add a term proportional to the one-loop result that accounts for the $\Pi(0)$ contribution and strong coupling renormalization, which can be combined in an ${\overline {\rm MS}}$ renormalized $\Pi(0)$, dubbed $\Pi_0^{\overline {\rm MS}}$, free from UV divergences. While this procedure is straightforward for quantities which do not carry anomalous dimensions, and a clear separation of ultraviolet (UV for short) convergent (dispersive integral) and divergent (proportional to $\Pi_0^{\overline {\rm MS}}$) pieces is achieved, obtaining analytic expressions in terms of the secondary mass is in general complicated if at all possible. Closed expressions often depend on polylogarithms, hypergeometric, or even less familiar functions which are not friendly to code in high-level computer programs such as \texttt{C++} or \texttt{Fortran}. In the worse case, the integral can always be carried out numerically, although in limiting cases it might get unstable. When the method is applied to EFTs one encounters that the term involving $\Pi^{\rm OS}(\ell^2)$ contains UV divergences, and hence the dispersive integral cannot be computed numerically ``out of the box'': some strategy to subtract the divergent UV behavior must be used. No matter if UV divergences pollute or not the dispersive integral, there is no clear way of obtaining an expansion around large or small values of the secondary mass. In this article we aim at filling this gap.

In the case of EFTs for jets, namely Soft-Collinear Effective Theory (SCET for short)~\cite{Bauer:2000ew,Bauer:2001ct,Bauer:2001yt} or boosted heavy-quark effective theory (bHQET for short)~\cite{Fleming:2007qr,Fleming:2007xt}, some of the matrix elements as the jet or soft functions, must be convolved with an evolution kernel to sum up large logarithms. In the case of bHQET, if dealing with the unstable top quark, an additional convolution with a Breit-Wigner becomes necessary. While these two convolutions can be carried out analytically when all particles are massless, see e.g.\ Refs.~\cite{Becher:2008cf,Abbate:2010xh,Hoang:2014wka,Bachu:2020nqn}, even if the jet function with mass effects is known in a closed form, it is in general not possible to obtain analytical expressions for the RG-evolved function. Numerical implementations of the convolution are unpractical since, depending on the hierarchy among the scales involved in the factorization theorem, analytical continuation though subtractions may be necessary. In Refs.~\cite{Fleming:2007xt,Bris:2020uyb} the one loop jet function for a massive primary quark was computed in various event-shape schemes, and its RG-evolved counterparts could be expressed in terms of ${}_3F_2$ and ${}_4F_3$ hypergeometric functions, which are difficult to code in high-level programming languages. Likewise, it seems impossible to find the analytic Fourier or Laplace transform of these primary or secondary massive jet functions, expressions that become useful in some circumstances. In Ref.~\cite{Bris:2020uyb}, expansions for small and large masses to arbitrary order were found for the RG-evolved SCET jet functions, and the MB representation played an important role in deriving those for the small mass limit. On the other hand, even though secondary mass corrections to the jet function for a massless primary quark were computed analytically in Ref.~\cite{Pietrulewicz:2014qza}, it was not possible to obtain a closed form for its RG-evolved version. Likewise, the correction to the massless jet function due to a massive vector boson\footnote{Even though most of the time ``massive vector boson'' will actually refer to a ``massive gluon'', for clarity we use the latter expression to denote an infinitesimal gluon mass used as a regulator, whereas the former will account for a finite (not necessarily small) gluon mass.} was computed analytically in Ref.~\cite{Gritschacher:2013pha}, but once again no closed form could be found for its RG-evolved counterpart. Therefore, another purpose of this article is devising a general method to obtain expansions for the corrections to the Fourier-space or RG-evolved jet functions due to secondary massive quarks or massive vector bosons.

For processes with infrared (IR) or collinear singularities,\footnote{For brevity, we often use IR to refer to both infrared and collinear singularities.} which usually show up for the first time at one loop, the result will depend on the regulator used to tame such divergences (usual choices are dimensional regularization or off-shellness). Since a gluon mass can also be used as a regulator, these singularities are obviously absent when considering a massive vector boson. Likewise, the OS vacuum polarization function insertion also regulates IR divergences, but one still has to choose a regulator for the term proportional to the one-loop amplitude. On the other hand, when including massive vector bosons or secondary quarks in SCET (and, as will be shown in a forthcoming publication, also bHQET), soft mass-mode bin subtractions~\cite{Chiu:2009yx} need to be accounted for in order to cancel rapidity divergences appearing in individual diagrams. This requires regulating intermediate steps using e.g.\ the $\Delta$-regulator, which involves additional energy scales complicating the computations.

As we shall show, the MB representation can be readily applied to the case of virtual massive secondary bubbles or vector bosons. Our method will not follow the usual sequence of steps, that is: a)~expressing the result in terms of master integrals (MI for short), b)~writing down each MI in terms of Feynman-parameters, c)~applying the MB transform at every MI to pull the mass out of the integrations, d)~using the converse mapping theorem. On the contrary, our strategy uses the MB identity at a very early stage of the computation: after expressing the massive vacuum polarization function in terms of an integral over a Feynman parameter $x$ ---\,keeping the exact dependence on $d=4-2\varepsilon$\,--- but before any other loop or Feynman integration is carried out. After applying the MB representation, the integration over $x$ can be performed trivially giving rise to gamma functions, and only a single loop integral remains. Such loop computation involves a modified (massless) gluon propagator which is exactly the same one employed in large-$\beta_0$ computations (that is, the denominator of the gluon propagator is raised to a non-integer power $1-h$), such that many existing results can be recycled, see e.g.\ Ref.~\cite{Gracia:2021nut} where renormalon calculus has been adapted to SCET and bHQET. If need be, the exact dependence on $\varepsilon$ can be retained at each order of the large or small mass expansions. Furthermore, one can apply the converse mapping theory after Fourier transforming or RG evolving the jet functions, such that easy-to-use expansions are obtained. An additional nice feature of this methodology is the fact that one does not really need to use any regulator since no rapidity divergence appears in any loop integral. Indeed, the Mellin variable $h$ in the modified gluon propagator effectively acts as an analytic regulator~\cite{Becher:2011dz} which does not involve any additional energy scale. Moreover, soft mass-mode bin subtractions identically vanish. The downside of our method is that it cannot be applied to quantities which need explicit regularization of IR singularities at one-loop, such as e.g. the quark form factor, but it will turn out particularly good at computing matching coefficients between dijet current operators in two EFTs, since those are IR-safe.\footnote{The method can be adapted to compute such IR-divergent quantities in a straightforward manner: one simply needs to add an IR regulator, for instance, an off-shellness. After the computation, the regulator can be set to zero to recover the IR-finite result. In some cases, this limit can be accomplished within the Mellin-Barnes paradigm itself.} Nevertheless, using consistency conditions we will be able to distinguish the QCD and EFT pieces of the matching computations in all cases under study. Finally, extracting the UV poles, taking $m\to 0$ or $m\to \infty$ limits, or figuring out the matching condition between two consecutive EFTs is completely trivial in our method: these simply correspond to the residues of some poles close to the origin in the complex $h$ plane. There is also a nice connection to the large-order behavior or, conversely, with non-perturbative physics. Since a pole located at $h=-1/2$ implies a term proportional to the gluon mass, it signals linear sensitivity to soft momenta: an $\mathcal{O}(\Lambda_{\rm QCD})$ renormalon.

This article is organized as follows: In Sec.~\ref{sec:bubble} we review the computation of the one-loop massive quark vacuum polarization function and write it in a form amenable to compute the Mellin-Barnes transform. We present the small- and large-mass expansion for this quantity using the inverse mapping theorem, along with the exact result in $d=4$ dimensions. In Sec.~\ref{sec:Mg} we apply the MB transform to one-loop computations with massive vector bosons and set the stage for those computations that shall be carried out in the rest of this article. In Sec.~\ref{sec:2loop} we exploit our \mbox{$d$-dimensional} expression for the massive vacuum polarization function to write down the MB transform for the two-loop contribution coming from secondary massive quarks. We discuss renormalization and explain how to match to a quantum field theory in which the secondary quark has been integrated out. In Sec.~\ref{sec:MSbar} we apply our formalism to the relation between the pole and $\overline {\rm MS}$ masses, a quantity without cusp anomalous dimension, and recover known results. In Sec.~\ref{sec:SCET} the formalism is applied to computations in SCET: the hard matching coefficient in Sec.~\ref{sec:hard}, and the jet function in Sec.~\ref{sec:JetSCET}. From the former computation, taking suitable limits we isolate the QCD and SCET form factors. In this section we derive a series of constraints that renormalization factors must satisfy in order to render UV-finite anomalous dimensions, and make a thorough review of SCET factorization and evolution, along with the scenarios introduced in Ref.~\cite{Pietrulewicz:2014qza}. Computations in bHQET are presented in Sec.~\ref{sec:bHQET}: the matching between SCET and bHQET (Sec.~\ref{sec:Hm}) and the bHQET jet function (Sec.~\ref{sec:Bjet}). From the former computation we obtain separately the SCET and bHQET form factors, which serves to compute the contribution of a primary quark massive bubble to the matching coefficient. In this section we also review the factorization of the cross section in bHQET and streamline the scenarios that appear when secondary masses are present. In Secs.~\ref{sec:MSbar} through \ref{sec:bHQET} the method is applied to the case of a massive vector boson and a secondary massive quark, including the computation of $Z$ factors, anomalous dimensions, and matching conditions for all matrix elements showing up when the secondary quark is integrated out. Our conclusions are summarized in Sec.~\ref{sec:conclusions}.

\section{Massive Quark Vacuum Polarization Function}\label{sec:bubble}
\begin{figure}[t]\centering
\includegraphics[width=0.5\textwidth]{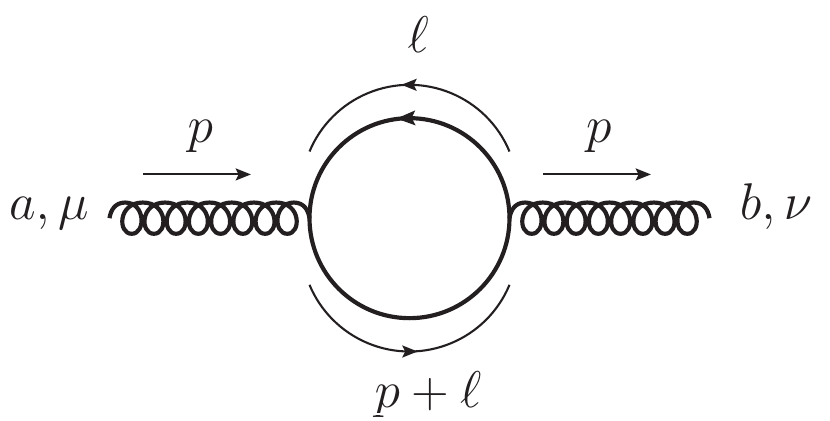}
\caption{Self-energy of an off-shell gluon due to a massive quark bubble. Here $a$ and $b$ denote the color indices on both sides of the gluon propagator, $\mu$ and $\nu$ are Lorentz indices, and $p$ and $\ell$ are the gluon and loop momenta, respectively.
\label{fig:bubble}}
\end{figure}
In this section we compute the contribution of a massive quark bubble to an off-shell gluon's self-energy. Of course, this result is textbook material and known since long time ago, but it is nevertheless instructive to review the computation as it will be necessary to keep the full dependence on $d=4-2\varepsilon$ and $p^2$. Our aim is to bring it to a form amenable to MB transform. Along the way, we provide a derivation of the dispersive integral over a fake gluon mass which does not rely on analytic properties of the vacuum polarization function. In what follows, the secondary heavy quark mass is denoted by $m$.

The diagram we need to compute is shown in Fig.~\ref{fig:bubble}. Denoting the initial/final color indices with the Roman characters $a$ and $b$, the initial/final Lorentz indices with the Greek letters $\mu$, $\nu$, and the off-shell gluon's momenta with $p$, and considering $n_f=1$ heavy flavors, one can show that the vacuum polarization function is diagonal in color space and, due to gauge invariance, transverse in Lorentz space:
\begin{align}
\Pi_{a b}^{\mu \nu} (p, m,\mu,\varepsilon) = \, &\delta_{ab} \Pi^{\mu \nu} (p, m,\mu,\varepsilon)\,,\\
\Pi^{\mu \nu} (p, m,\mu,\varepsilon) =\, & \Pi (p^2, m,\mu,\varepsilon)(p^2 g^{\mu \nu} - p^{\mu} p^{\nu})\,.\nonumber
\end{align}
The vacuum polarization function is analytic everywhere in the complex $p^2$ plane except for a cut running along the positive real axis, starting at $p^2=4m^2$. In particular, this implies that $\Pi_0(m,\mu,\varepsilon)\equiv \Pi(0, m,\mu,\varepsilon)$ is a finite distance away from the branch cut, hence IR finite. Furthermore, $\Pi_0(m,\mu,\varepsilon)$ contains all UV divergences of $\Pi(p^2, m,\mu)$ if dimensional regularization is used. After inserting the vacuum polarization function, assuming $\ell$ is the loop momentum of the gluon internal line, the usual gluon propagator in the Feynman gauge will be replaced by
\begin{equation}
\frac{-i g_{\mu\nu}}{\ell^2}\to -i \frac{g_{\mu\alpha}}{\ell^2}\Pi^{\alpha\beta}(\ell,m,\mu,\varepsilon) \frac{ g_{\beta\nu}}{\ell^2}=
\frac{-i \Pi(\ell^2,m,\mu,\varepsilon)}{\ell^2}\biggl(g^{\mu\nu} - \frac{\ell^\mu\ell^\nu}{\ell^2}\biggr)\,.
\end{equation}
The term proportional to $\ell^\mu\ell^\nu$ will vanish after adding all Feynman diagrams due to gauge invariance, hence it will be ignored in the following.

After a straightforward tensor decomposition of loop integrals, the transverse (or $J=1$) polarization function $\Pi (p^2, m, \mu,\varepsilon)$ can be brought to the following form:
\begin{align}
\Pi (p^2, m,\mu,\varepsilon) &\,= - \frac{4 T_F g_s^2}{3 - 2 \varepsilon} \biggl[ \biggl( 1 - \varepsilon +\frac{2 m^2}{p^2} \biggr) B (p^2, m,\mu,\varepsilon)
- \frac{2 (1 - \varepsilon)}{p^2} A(m,\mu,\varepsilon) \biggr],\\
B(p^2, m,\mu,\varepsilon)&\,= -i\tilde{\mu}^{2 \varepsilon} \!\!\int\! \frac{{\rm d} \ell}{(2 \pi)^d}\frac{1}{(\ell^2 - m^2) [(\ell + p)^2 - m^2]}\,,\nonumber\\
A(m,\mu,\varepsilon)&\,= -i\tilde{\mu}^{2 \varepsilon} \!\!\int\! \frac{{\rm d} \ell}{(2 \pi)^d}\frac{1}{(\ell^2 - m^2)}=
\frac{m^2}{(4 \pi)^2} \biggl( \frac{4 \pi\tilde{\mu}^2}{m^2} \biggr)^{\!\!\varepsilon}\frac{\Gamma (\varepsilon)}{1 - \varepsilon}\,,\nonumber
\end{align}
where $\tilde{\mu}^2=\mu^2 e^{\gamma_E}/(4\pi)$ and $A(m,\mu,\varepsilon)$ and $B(p^2, m,\mu,\varepsilon)$ are usually referred to as the tadpole and bubble scalar loop integrals. The QCD color factors for $N_c=3$ take the values $T_F=1/2$ and $C_F=4/3$. For our purposes, $g_s$ can be regarded as the renormalized quark-gluon coupling. Even though individual coefficients diverge for $p^2=0$, we shall see that $\Pi (0, m^2,\varepsilon)$ is actually IR finite. For our purposes, it is convenient to write $B(p^2, m,\mu,\varepsilon)$ as follows:
\begin{equation}
B (p^2, m,\mu,\varepsilon) = - \frac{4m^2}{(4 \pi)^2} \biggl( \frac{\pi\tilde{\mu}^2}{m^2} \biggr)^{\!\!\varepsilon} \frac{\Gamma (1 -\varepsilon)}{\Gamma (2 - 2 \varepsilon)}\!\!
\int_0^1 \!\text{d} x \frac{x^{- 2 +\varepsilon} (1 - x)^{\frac{1}{2} - \varepsilon}}{p^2 - \frac{4 m^2}{x}}\,.
\end{equation}
To arrive at such expression one uses a Feynman parameter $x$ and integrates over loop momentum. After changing integration variables to $x = (1 - y) / 2$, the integral's symmetry under $y \to - y$ is used to map back to the unit segment. An additional change of variables $y^2=1-z$ brings the integral into the canonical form of a hypergeometric function. Using the symmetry of ${}_2F_1$ under the exchange of its two first arguments and using again the integral representation of the hypergeometric function yields the displayed result. Using partial fraction

\begin{equation}
\frac{2 m^2}{p^2} \frac{1}{p^2 - \frac{4 m^2}{x}} = \frac{x}{2} \Biggl(\frac{1}{p^2 - \frac{4 m^2}{x}} - \frac{1}{p^2} \Biggr), \label{eq:partFrac}
\end{equation}
one can write down
\begin{equation}
B (p^2, m,\mu,\varepsilon) = (1 - \varepsilon)\frac{A(m,\mu,\varepsilon)}{m^2} - \frac{1}{(4 \pi)^2} \biggl( \frac{\pi\tilde{\mu}^2}{m^2} \biggr)^{\!\!\varepsilon} \frac{ \Gamma (1 -\varepsilon)}{\Gamma (2 - 2 \varepsilon)}
\!\!\int_0^1 \!\text{d} x \frac{x^{- 1 +\varepsilon} (1 - x)^{\frac{1}{2} - \varepsilon}}{1- \frac{4 m^2}{xp^2 }} \,.
\end{equation}
The last relation can be used to write the vacuum polarization as a single integral which is manifestly convergent for $p^2=0$, and to check that the $1/\varepsilon$ divergence arising as $\varepsilon\to 0$ does not depend on $m$. In order to ease the upcoming discussion, we present results for $\Pi^{\rm OS}(p^2, m,\mu,\varepsilon)=\Pi(p^2, m,\mu,\varepsilon)-\Pi_0(m,\mu,\varepsilon)$ and $\Pi_0(m,\mu,\varepsilon)$:\footnote{For $p^2>4m^2$ one needs to add a small imaginary part $p^2\to p^2+i 0^+$ to pick the upper side of the branch cut, as dictated by the $i\epsilon$ prescription of Feynman propagators.}
\begin{align}\label{eq:PiOS}
\Pi^{\rm OS} (p^2, m,\mu,\varepsilon) &\,= \frac{2T_F \alpha_s(\mu)}{\pi} \biggl( \frac{\pi\tilde{\mu}^2}{m^2} \biggr)^{\!\!\varepsilon} \frac{\Gamma (2 - \varepsilon)}{\Gamma (4 - 2 \varepsilon)} \int_0^1 \text{d} x \biggl( 1 - \varepsilon + \frac{x}{2} \biggr) \frac{x^{- 1 + \varepsilon} (1 - x)^{\frac{1}{2} - \varepsilon}}{1 - \frac{4 m^2}{xp^2}} \,,\nonumber\\
\Pi_0 (m,\mu,\varepsilon)&\,= - \frac{T_F\alpha_s(\mu) \Gamma (\varepsilon)}{3 \pi} \biggl(\frac{\mu^2e^{\gamma_E}}{m^2} \biggr)^{\!\!\varepsilon}\,,\nonumber\\
\Pi^{\rm \overline{MS}}_0(m,\mu,\varepsilon) &\,= - \frac{T_F}{3} \frac{ \alpha_s(\mu) }{\pi}\biggl[\biggl(\frac{\mu^2e^{\gamma_E}}{m^2} \biggr)^{\!\!\varepsilon}\Gamma (\varepsilon)-\frac{1}{\varepsilon}\biggr]\,,
\end{align}
where $\Pi^{\rm \overline{MS}}_0$ is the ${\rm \overline{MS}}$-renormalized version of $\Pi_0$, finite as $\varepsilon\to 0$. Changing variables to $x=4m^2/\lambda$ on the first line, being $\lambda$ the squared of a fake gluon mass, reproduces the ``classical'' dispersive integral. Using the MB identity
\begin{equation}\label{eq:MB}
\frac{1}{(1 + X)^{\nu}} = \!\!\int_{c - i \infty}^{c + i \infty} \frac{{\rm d}h}{2 \pi i} (X)^{- h} \frac{\Gamma (h) \Gamma (\nu - h)}{\Gamma (\nu)}\,,
\end{equation}
where $0\leq c\leq \nu$ is the fundamental strip, on the first line of Eq.~\eqref{eq:PiOS} with $\nu=1$ and $X=-4m^2/(xp^2)$, after integration over $x$ one arrives at the following representation for the subtracted vacuum polarization function:
\begin{align}\label{eq:master}
\Pi^{\rm OS}(p^2, m,\mu,\varepsilon) &\,= \frac{ \alpha_s(\mu) }{\pi}T_F \biggl(\frac{\mu^2e^{\gamma_E}}{m^2} \biggr)^{\!\!\varepsilon} \!\int_{c - i \infty}^{c + i \infty}\frac{{\rm d}h}{2 \pi i} \biggl(\! - \frac{m^2}{p^2} \biggr)^{\!\!- h} G (h,\varepsilon)\,,\\
G (h,\varepsilon) &\,= \frac{\Gamma (h) \Gamma (1 - h) \Gamma (2 + h) \Gamma (h + \varepsilon)}{(3 +2 h) \Gamma (2 h + 2)}\,.\nonumber
\end{align}
Since, as already discussed in Sec.~\ref{sec:intro}, the method can only be applied to IR-finite quantities, $\varepsilon>0$ in order to regulate UV divergences, hence the fundamental strip is $0<c<1$. From Eq.~\eqref{eq:master} one can read the form of the ``effective'' gluon propagator which will yield the MB representation from a one-loop computation: the insertion of $\Pi^{\rm OS}(p^2, m,\mu)/(-p^2)$ in the MB representation modifies the denominator of the gluon propagator from $(-p^2)$ to $(-p^2)^{1-h}$ and adds an overall factor.

Our expansions will become useful to compute the matching between two consecutive EFTs, one in which the secondary massive quarks are dynamic, another one in which they are not. In order to compute the matching condition we need to relate the strong coupling in the two EFTs. Such matching condition for $\alpha_s(\mu)$ is obtained from $\Pi^{\rm \overline{MS}}_0(m,\mu,\varepsilon)$ and reads
\begin{equation}
\alpha_s^{(n_f + 1)} (\mu) = \alpha_s^{(n_f)} (\mu) \biggl[ 1 + \frac{\alpha_s^{(n_f)} (\mu) T_F }{3 \pi} \log \biggl( \frac{\mu^2}{m^2} \biggr) \biggr]\! +\mathcal{O} (\alpha_s^3)
\equiv \alpha_s^{(n_f)} (\mu) \bigl( 1 + \delta^{(n_f)}_\alpha\bigr)\, .
\end{equation}
Indeed one has that
\begin{equation}
\Pi^{\rm \overline{MS}}_0(m,\mu, \varepsilon) + \delta_\alpha =
- \frac{T_F\varepsilon}{6} \frac{ \alpha_s(\mu) }{\pi} \biggl[\log ^2\biggl(\frac{\mu ^2}{m^2}\biggr)+\frac{\pi ^2}{6}\biggr]+ \mathcal{O}(\varepsilon^2)\,.
\end{equation}

Before discussing any further the computation of the two-loop massive bubble diagrams, we apply the converse mapping theorem to $\Pi^{\rm OS}(p^2, m,\mu,0)$, as it will serve as an illustration. Since the on-shell vacuum polarization function is UV finite, we can set $\varepsilon\to 0$ in Eq.~\eqref{eq:master}.\footnote{If keeping a non-zero $\varepsilon$, the UV-finiteness is transparent when closing towards ${\rm Re}(h)\to+\infty$. When closing towards ${\rm Re}(h)\to-\infty$ the poles at $h=0$ and $h=-\varepsilon$ generate $1/\varepsilon$ singularities which however cancel when both terms are added up.} The inverse mapping theorem is nothing less than closing the contour towards the positive or negative real axes and using the residue theorem. To see which side one must pick, Jordan's lemma must be applied, what implies expanding $G (h,0)$ for large $h$:
\begin{equation}
\lim_{h\to\infty} \Biggl|G (h,0)\biggl(\! - \frac{m^2}{p^2} \biggr)^{\!\!- h}
\Biggr| = \frac{\pi^{3 / 2} \csc (\pi h)}{4 h^{3 / 2}} \biggl( \frac{4 m^2}{| p^2 |}\biggr)^{\!\!- h}.
\end{equation}
It is clear that for $|p^2| < 4m^2$ ($|p^2| > 4m^2$) the contour must be closed towards the positive (negative) real axis, resulting in an expansion for big (small) masses. This is exactly what one could have guessed from the analytic behavior of $\Pi(p^2,m,\mu,0)$: the distance from the origin of the $p^2$ complex plane to the branch point, which sets the convergence radius, is exactly $|p^2|=4m^2$. For $|p^2| = 4m^2$ the contour can be closed on either side since the damping factor $h^{3 / 2}$ ensures Jordan's lemma will be satisfied. When closing towards ${\rm Re}(h)>0$, simple poles will be found at positive integer values of $h=n$, hence no non-analytic terms will be present. This is expected, since for large $m$ one is always below the branch cut of the vacuum polarization function. Each pole generates a term in the power expansion for large $m$, namely $(-p^2/m^2)^n$ with $n\geq 1$ such that $\Pi^{\rm OS}(0, m,\mu,\varepsilon)=0$. On the other hand, when closing towards ${\rm Re}(h)<0$ one finds double poles at all negative integer values of $h=-n$, except for $h=-1$ where the pole is simple. Logarithms are expected since, for small $m$ and positive $p^2$, an imaginary part should appear. Each pole generates a term in the power expansion for small $m$, namely $(-m^2/p^2)^n$. Double poles generate a power of $\log(-p^2/m^2)$, whereas simple poles do not. The expansions read
\begin{figure*}[t!]
\subfigure[]
{\includegraphics[width=0.511\textwidth]{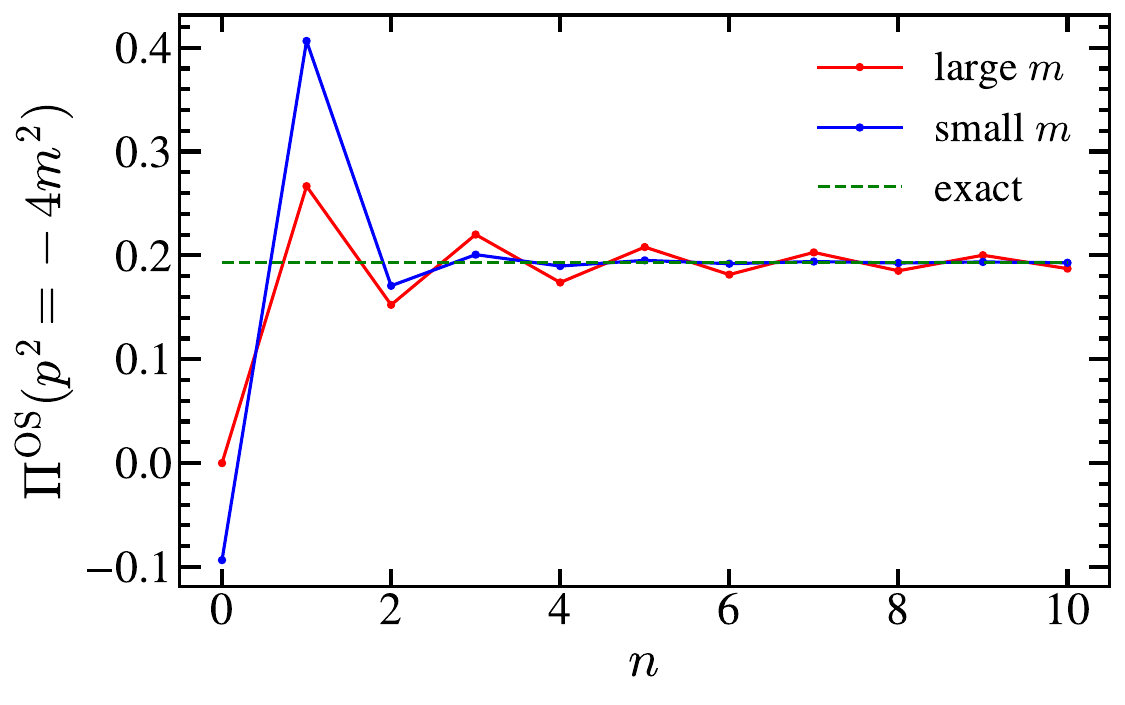}
\label{fig:Pi01}}
\subfigure[]{\includegraphics[width=0.489\textwidth]{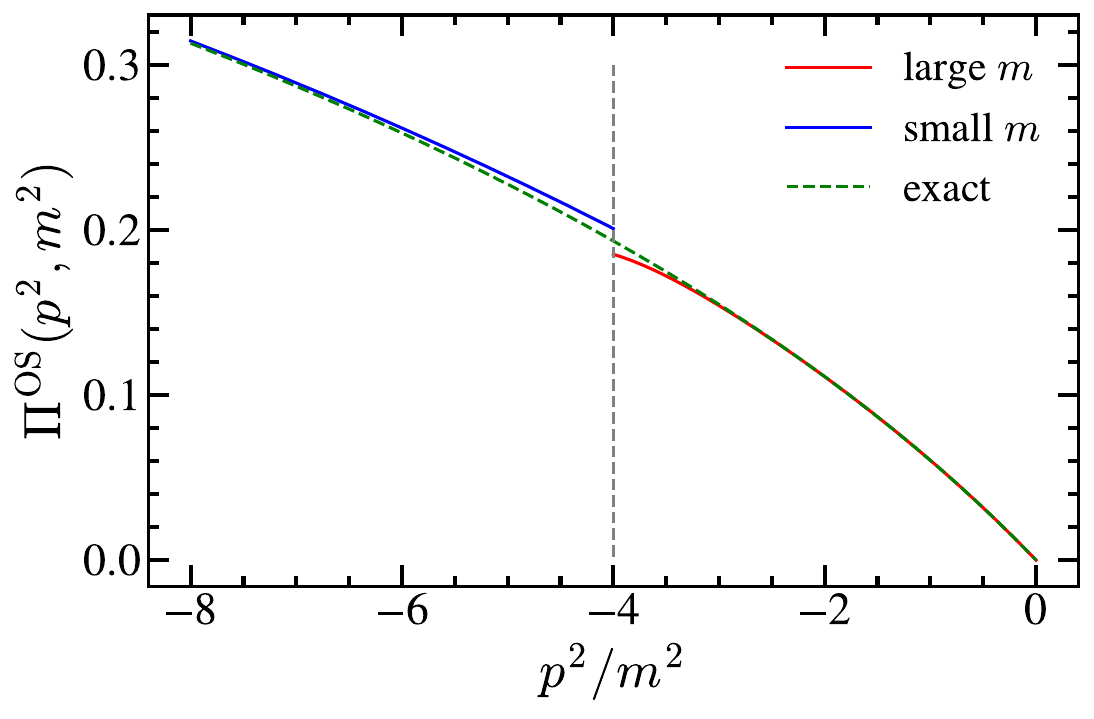}
\label{fig:Pi0}}
\caption{On-shell subtracted vacuum polarization function for a massive quark measured in units of $\alpha_s(\mu)T_F/\pi$ in its exact form (dashed green), small- (blue) and large-mass (red) expansions for $\varepsilon=0$. Left panel: $\Pi^{\rm OS}$ at the boundary between the mass expansions $p^2 = -4m^2$, as a function of the expansion order $n$ of each series. Right panel: Dependence of $\Pi^{\rm OS}(p^2,m,\mu,0)$ with the ratio $p^2/m^2$ including $4$ and $8$ non-zero terms in the small- and large-mass expansions, respectively.}
\label{fig:Pi}
\end{figure*}
\begin{align}
\Pi^{\rm OS}\biggl(\frac{p^2}{m^2},\mu,0\biggr) =\,&-\!\frac{ \alpha_s(\mu) }{\pi} T_F\sum_{n = 1}^{\infty} \frac{(n - 1) ! (n + 1)!}{(2 n + 3) (2 n + 1) !} \biggl( \frac{p^2}{m^2} \biggr)^{\!\!n}\\
=\, & \!\frac{ T_F \alpha_s}{\pi}
\sum_{n = 0} \biggl( \frac{m^2}{p^2} \biggl)^{\!\!n} \frac{(2 n) !}{(2 n - 3)(2n-1) (n!)^2} \biggl[2(n - 1) (H_{2 n} -H_{n}) \nonumber\\
& - (n - 1)\log \biggl(\! - \frac{p^2}{m^2} \biggr) -\frac{4 n^2-8 n+5}{(2 n-3) (2 n-1)} \biggr] ,\nonumber
\end{align}
where $H_i=\sum_{n=1}^i n^{-1}$ is the harmonic number. The first series can be summed up analytically and we obtain the well-known result
\begin{align}
\Pi^{\rm OS}\biggl(\frac{p^2}{m^2},\mu,0\biggr) =\,&\frac{ \alpha_s(\mu) }{\pi}\frac{T_Fm^2}{9p^2}\Biggl[6 \sqrt{1 - \frac{4 m^2}{p^2}}\, \biggl(2+\frac{p^2}{ m^2}\biggr)\!
\log \biggl(\sqrt{1 - \frac{p^2}{4 m^2}}+\sqrt{- \frac{p^2}{4 m^2}}\biggr) \nonumber \\
&-12 - \frac{5p^2}{m^2}\Biggr].
\end{align}
The vacuum polarization function diverges at $p^2\to+\infty$, which is nothing less than the massless limit, not well defined due to the OS subtraction. We compare the exact form to the two series expansions in Fig.~\ref{fig:Pi} where it can be seen that both series can be used at $p^2=-4m^2$. We observe an oscillatory behavior of the large-mass expansion.

\section{One-loop Computations with a Massive Vector Boson}\label{sec:Mg}
Before we discuss in detail the two-loop contribution from a secondary massive quark bubble, we pause to describe how the MB representation can be applied to generate large and small mass expansions to one-loop computations involving a massive vector boson. For simplicity, we consider a gluon with a non-zero mass $m_g$, but the method can be generalized to other massive mediators. In this case, one simply uses the MB identity Eq.~\eqref{eq:MB} directly to the massive gluon propagator
\begin{equation}
\frac{1}{- p^2 + m_g^2} = \frac{1}{- p^2} \int_{c - i \infty}^{c + i \infty} \frac{{\rm d} h}{2 \pi i} \biggl( \!- \frac{m_g^2}{p^2} \biggr)^{\!\!-h} \Gamma (h) \Gamma (1 - h)\,,
\end{equation}
where the fundamental branch is $0 < c < 1$. This result again implies that for obtaining the Mellin representation one modifies the gluon propagator shifting the power of its denominator in exactly the same way, $(-p^2)$ to $(-p^2)^{1-h}$, and multiplies by the factor $\Gamma(h)\Gamma(1-h)=\pi \csc (\pi h)$ which changes sign under $h\to-h$. Let us assume that our matrix element is dimensionless: then, the one-loop computation with a massless gluon whose propagator has been ``shifted'', and where $d=4-2\varepsilon$ has been kept unexpanded, can be written as
\begin{equation}\label{eq:genericM1}
M_1(h, \mathcal{Q},\mu,\varepsilon) = \frac{g_s^2 C_F}{4 \pi^2} \mathcal{Q}^{2 h} \biggl( \frac{\mu^2\varepsilon^{\gamma_E}}{\mathcal{Q}^2} \biggr)^{\!\!\varepsilon}\, m_1 (h,\varepsilon)\,,
\end{equation}
where $\mathcal{Q}$, with mass-dimension 1, is the only scale in the matrix element being computed ---\,necessary to render the one-loop result non-zero\,--- and $g_s$ is the bare strong coupling constant. The function $m_1 (h,\varepsilon)$ is dimensionless and does not depend on $\mathcal{Q}$, while the prefactor $\mathcal{Q}^{2h}$ accounts for the overall dimension caused by the shifted gluon propagator. All in all, the one-loop result with a massive vector boson takes the following form:
\begin{align}\label{eq:meromorphic}
M_1 (m_g, \mathcal{Q},\mu, \varepsilon) & \,= \frac{\alpha_s(\mu)}{\pi} C_F F_1(m_g, \mathcal{Q},\mu, \varepsilon)\,,\\
F_1 (m_g, \mathcal{Q},\mu, \varepsilon) & \,= \biggl( \frac{\mu^2 \varepsilon^{\gamma_E}}{\mathcal{Q}^2} \biggr)^{\!\!\varepsilon}
\!\! \int_{c - i \infty}^{c + i \infty} \! \frac{\dd h}{2 \pi i}
\mathcal{M}_1\! \biggl( h, \frac{m_g}{\mathcal{Q}},\varepsilon \biggr),\nonumber \\
\mathcal{M}_1\! \biggl( h, \frac{m_g}{\mathcal{Q}},\varepsilon \biggr) & \,= \biggl( \frac{\mathcal{Q}^2}{m_g^2} \biggr)^{\!\!h} \, \Gamma (h) \Gamma (1 - h) m_1 (h,\varepsilon) \,, \nonumber
\end{align}
where, at this order, $\alpha_s(\mu)$ is already the renormalized strong coupling.

Let us discuss some generic features. The function $m_1 (h,\varepsilon)$ can modify the fundamental strip, and whenever the matrix element needs renormalization (that is, when the one-loop computation with an unmodified gluon propagator generates $1/\varepsilon^n$ poles), it gets narrowed down to $0<c<\varepsilon$. This is easy to understand: since the Mellin parameter $h$ acts as a UV regulator (it is well-known that large-$\beta_0$ calculations can be carried out setting $\varepsilon=0$), UV poles manifest themselves as singularities of the type $1/(h-\varepsilon)^n$ with $n=2$ for quantities carrying a cusp anomalous dimension, $n=1$ otherwise.\footnote{It is not hard to deduce the poles' form. Let us assume a generic scalar one-loop bubble containing a regular and a modified gluon propagator. The $d$-dimensional integration measure after Wick rotation is $\dd^d\ell =\ell^{3-2\varepsilon} \dd \Omega\dd \ell$, while the product of propagator denominators behaves as $\ell^{4-2h}$. When combined, one has $\ell^{-1+2(h-\varepsilon)}\dd \Omega\dd \ell$, which upon integration diverges like $1/(h-\varepsilon)$.} The massless result $F_1 (0,\mathcal{Q},\mu,\varepsilon)$ is trivial to obtain: it corresponds to the $h=0$ pole's residue.\footnote{One trivially gets $F_1 (0,\mathcal{Q},\mu,\varepsilon)=\Bigl( \frac{\mu^2 \varepsilon^{\gamma_E}}{\mathcal{Q}^2} \Bigr)^{\!\!\varepsilon} m_1(0,\varepsilon)$ from the converse mapping theorem or directly from Eq.~\eqref{eq:genericM1}.} Since the massless limit is manifest, there will be no logarithms of $m_g/\mathcal{Q}$ in this limit. The poles at $h=0$ and $h=\varepsilon$ contain the same ($m_g$-independent) UV poles, but have different finite terms. Since the correction to the massless result $\Delta_0 F_1(m_g/\mathcal{Q}) \equiv F_1 (m_g,\mathcal{Q},\mu,\varepsilon)-F_1 (0,\mathcal{Q},\mu,\varepsilon)$ is UV finite ---\,and $\mu$-independent as well\,--- we can ``move'' the fundamental strip to $-1/2 < {\rm Re}(h)<0$ and set $\varepsilon\to 0$ to obtain a closed form:\footnote{From renormalon calculus considerations one can convince oneself there are not singularities for \mbox{$h\in(-1/2,0)$}. In most cases, there are no singularities between $h=-1$ and $h=-1/2$ either, but to be general we stick to the smaller range.}
\begin{equation}\label{eq:MassCorrMg}
\Delta_0 F_1\biggl(\frac{m_g}{\mathcal{Q}}\biggr) =\!\! \int_{e - i \infty}^{e + i \infty} \! \frac{\dd h}{2 \pi i}
\biggl(\frac{\mathcal{Q}^2}{m_g^2} \biggr)^{\!\!h} \, \Gamma (h) \Gamma (1 - h) m_1 (h,0)\,.
\end{equation}
The $m_g\to\infty$ limit corresponds to minus the residue of the $h=\varepsilon$ pole, and since the decoupling limit is not manifest in the ${\rm \overline{MS}}$ scheme, it will contain powers of $\log(m_g/\mathcal{Q})$. The correction to this limit $\Delta_\infty F_1(m_g/\mathcal{Q}) \equiv F_1 (m_g,\mathcal{Q},\mu, \varepsilon) - F_1 (m_g\to \infty,\mathcal{Q},\mu, \varepsilon)$ is also UV finite and $\mu$-independent, and can be cast in the same way as Eq.~\eqref{eq:MassCorrMg} (that is, with $\varepsilon=0$) moving the fundamental strip to $0<{\rm Re}(h)<1$. Finally, the difference of the $m_g\to\infty$ and $m_g\to0$ limits $\Delta_0^\infty F_1(m_g/\mathcal{Q})\equiv F_1(m_g\to \infty,\mathcal{Q},\mu, \varepsilon) - F_1 (0,\mathcal{Q},\mu, \varepsilon)$ is once more UV-finite and $\mu$-independent, and given by the contribution of the pole sitting at $h=0$ obtained if $\varepsilon$ is set to zero prior to computing the residue. This increases the pole's multiplicity, generating the expected logs of $m_g/\mathcal{Q}$.

We will consider the matching between two EFTs: the high-energy one, containing a massive and a massless gluon, and the low-energy one, with a massless gluon only. At one-loop, the coupling in the two theories coincides, and since there are massless gluons in both, such contributions cancel in the matching. Since the two theories should yield the same answer in the $m_g\to\infty$ limit, the relevant quantity for the matching is
\begin{equation}\label{eq:generalMatching}
[F_1 (m_g\to\infty, \mathcal{Q},\mu,\varepsilon)]_{\rm fin} = [F_1 (0, \mathcal{Q},\mu,\varepsilon)]_{\rm fin} + \Delta_0^\infty F_1\biggl(\frac{m_g}{\mathcal{Q}}\biggr)\,,
\end{equation}
where, since the matching is performed using renormalized matrix elements, the subscript ``fin'' has been added to signify the $1/\varepsilon^n$ poles have been stripped away.

\section{Two-loop massive Bubble Computations}\label{sec:2loop}
In this section we derive the general expression for the renormalized two-loop matrix element due to the insertion of a massive bubble. On top of the dispersive integral, which we have written as an inverse Mellin transform, one has to account for the contribution due to $\Pi_0(m,\mu,\varepsilon)$ and the strong coupling renormalization, which can be combined as a term proportional to $\Pi_0^{\rm \overline{MS}}(m,\mu,\varepsilon)$. When inserting the vacuum polarization into the gluon internal line, the contribution from $\Pi_0$ corresponds to the replacement $1/\ell^2\to \Pi_0(m,\mu,\varepsilon)/\ell^2$ in the gluon propagator. Since $\Pi_0(m,\mu,\varepsilon)$ does not depend on the loop momentum $\ell$, this contribution is proportional to the one-loop result computed with a massless gluon propagator. The two-loop result can be then written as
\begin{align}\label{eq:M2}
M_2(m, \mathcal{Q},\mu, \varepsilon) &\,= \biggl[ \frac{\alpha_s(\mu)}{\pi} \biggr]^2 C_F T_F F_2 (m, \mathcal{Q},\mu,\varepsilon)\,,\\
F_2 (m, \mathcal{Q},\mu,\varepsilon) &\,= H_2 (m, \mathcal{Q},\mu, \varepsilon) - \frac{1}{3}F_1 (0,\mathcal{Q},\mu,\varepsilon) \biggl[\! \biggl( \frac{\mu^2e^{\gamma_E}}{m^2} \biggr)^{\!\!\varepsilon} \Gamma (\varepsilon) -\frac{1}{\varepsilon} \biggr],\nonumber\\
H_2(m, \mathcal{Q},\mu,\varepsilon) &\,= \biggl( \frac{\mu^2 e^{\gamma_E}}{\mathcal{Q}^2} \biggr)^{\!\!\varepsilon}
\biggl(\frac{\mu^2 e^{\gamma_E}}{m^2} \biggr)^{\!\!\varepsilon} \!\!\int_{c - i \infty}^{c + i \infty} \!\frac{\dd h}{2 \pi i}
\mathcal{M}_2\! \biggl( h, \frac{m_g}{\mathcal{Q}},\varepsilon \biggr), \nonumber\\
\mathcal{M}_2\! \biggl( h, \frac{m_g}{\mathcal{Q}},\varepsilon \biggr) &\,= \biggl( \frac{\mathcal{Q}^2}{m^2} \biggr)^{\!\!h} \,G (h, \varepsilon)m_1 (h,\varepsilon)\,.\nonumber
\end{align}
where $G (h, \varepsilon)$ has been given in the second line of Eq.~\eqref{eq:master}. The coupling constant $\alpha_s(\mu)$ is already renormalized and runs with $n_f=n_\ell + 1$ active flavors, where $n_\ell$ is the number of massless quarks. The function $\mathcal{M}_2$ [\,same as $\mathcal{M}_1$ in Eq.~\eqref{eq:meromorphic}\,] is meromorphic in the complex $h$ plane. We will refer to the term $H_2$ as the ``dispersive contribution''.

Let us again discuss some generic features. As argued in Sec.~\ref{sec:Mg}, $m_1 (h,\varepsilon)$ narrows the fundamental strip to $0<h<\varepsilon$ since single or double singularities appear at $h=\varepsilon$. We denote the second term in the second line of Eq.~\eqref{eq:M2} as ``the $\Pi_0$ contribution''. The UV singularities are contained in the contribution from the pole at $h=\varepsilon$ for the large $m$ expansion, and in the sum of residues of the poles located at $h=0$ and $h=-\varepsilon$ for the small $m$ expansion, to which one has to add the divergent terms coming from the $\Pi_0$ insertion. Once again, the divergences for the two expansions are $m$ independent and coincide, but the finite remainders differ. The massless result $F_2(0,\mathcal{Q},\mu,\varepsilon)$ can be obtained as the sum of the residues of poles at $h=0$ and $h=-\varepsilon$ plus the $\Pi_0$ contribution. Since the massless limit is manifest, no logs of $m/\mathcal{Q}$ arise. The correction to the massless limit $\Delta_0 F_2(m/\mathcal{Q})\equiv F_2(m,\mathcal{Q},\mu,\varepsilon) - F_2(0, \mathcal{Q},\mu,\varepsilon)$ is UV-finite, $\mu$-independent, and can be obtained moving the fundamental strip to $-1/2<{\rm Re}(h)<0$:
\begin{equation}\label{eq:massless}
\Delta_0 F_2\biggl(\frac{m}{\mathcal{Q}}\biggr)=\int_{e - i \infty}^{e + i \infty} \!\frac{\dd h}{2 \pi i}
\biggl( \frac{\mathcal{Q}^2}{m^2} \biggr)^{\!\!h} \,\frac{h(h+1)\Gamma^3(h) \Gamma (1 - h)}{(3 +2 h) \Gamma (2 h + 2)}m_1(h,0)\,.
\end{equation}
The $m\to\infty$ limit is simply the $\Pi_0$ contribution minus the residue at $h=\varepsilon$. Since the decoupling limit is not manifest in the ${\rm \overline{MS}}$ scheme, it will contain powers of $\log(m/\mathcal{Q})$. The correction to the decoupling limit $\Delta_\infty F_2(m/\mathcal{Q})\equiv F_2(m,\mathcal{Q},\mu,\varepsilon) - F_2(m\to\infty, \mathcal{Q},\mu,\varepsilon)$ is also UV-finite and $\mu$-independent, and can be written as in Eq.~\eqref{eq:massless} moving the fundamental strip to $0<{\rm Re}(h)<1$. Subtracting the $m\to \infty$ and $m\to0$ limits yields a UV-free and $\mu$-independent quantity $\Delta_0^\infty F_2(m/\mathcal{Q}) \equiv F_2(m\to \infty,\mathcal{Q},\mu,\varepsilon) - F_2(0,\mathcal{Q},\mu,\varepsilon)$ which again can be obtained as the contribution form the pole at $h=0$ setting $\varepsilon$ to zero before computing the residue, what increases the multiplicity of the pole. This quantity is related to the matching between two consecutive EFTs: one where the massive secondary quark is an active degree of freedom, another one in which it is not.

Let us succinctly describe how such matching condition is computed. In the theory where the massive quark is no longer active there are $n_\ell$ active flavors. To carry out the matching it is convenient to express the renormalized matrix elements as a series in powers of $\alpha_s^{(n_\ell)}$. After the conversion, the renormalized two-loop term takes the form
\begin{equation}\label{eq:matching}
F_{2,{\rm ren}}^{(n_\ell)}(m,\mathcal{Q},\mu) = \!\biggl\{\!H_2 (m, \mathcal{Q},\mu,\varepsilon) - \frac{F_1 (0,\mathcal{Q},\mu,\varepsilon)}{3} \biggl[\! \biggl( \frac{\mu^2e^{\gamma_E}}{m^2} \biggr)^{\!\!\varepsilon} \Gamma (\varepsilon) -\frac{1}{\varepsilon} -\log\biggl(\frac{\mu^2}{m^2}\biggr)\!\biggr]\!\biggr\}_{\!\rm fin}.
\end{equation}
Interestingly, if $F_1 (0,\mathcal{Q},\mu,\varepsilon)$ is UV-finite, the second term vanishes. Furthermore, in such cases the decoupling limit is manifest $H_2 (m\to\infty, \mathcal{Q})=0$, and the matching condition is trivial: the effects from massive quark bubbles are fully captured in the $\alpha_s$ decoupling relation. If one assumes $F_1 (0,\mathcal{Q},\mu,\varepsilon)$ has the following divergent structure:
\begin{equation}\label{eq:F1}
F_1(0,\mathcal{Q},\mu,\varepsilon) =\frac{m_2}{\varepsilon^2} + \frac{m_1}{\varepsilon} + m_0 + \mathcal{O}(\varepsilon) \,,
\end{equation}
where $m_0$ can potentially depend on an IR regulator, the relevant quantity for the matching coefficient
\begin{align}\label{eq:limit}
F_{2,{\rm ren}}^{(n_\ell)}(m\to\infty,\mathcal{Q},\mu) =\,& -\!\biggl\{\!\biggl( \frac{\mu^2 e^{\gamma_E}}{\mathcal{Q}^2} \biggr)^{\!\!\varepsilon}\!
\biggl(\frac{\mu^2 e^{\gamma_E}}{m^2} \biggr)^{\!\!\varepsilon} {\rm Res}_{h=\varepsilon}\biggl[ \mathcal{M}_2\! \biggl( h, \frac{m_g}{\mathcal{Q}},\varepsilon \biggr)\biggr]\biggr\}_{\!\rm fin}\!\!\!\!\!
-\frac{m_2}{18} \log ^3\biggl(\frac{\mu ^2}{m^2}\biggr) \nonumber\\
&-\!\frac{m_1}{6} \log^2\biggl(\frac{\mu ^2}{m^2}\biggr)-\frac{\pi ^2 m_2}{36} \log\biggl(\frac{\mu ^2}{m^2}\biggr)+\frac{m_2 \zeta_3}{9}-\frac{\pi ^2 m_1}{36}\,,
\end{align}
which can be rewritten as $[F_2(0,\mathcal{Q},\mu,\varepsilon)]_{\rm fin} + \Delta_0^\infty F_2(m/\mathcal{Q}) + 2 [F_1(0,\mathcal{Q},\mu,\varepsilon)]_{\rm fin} \log(\mu/m)/3$, does not depend on $m_0$.

\section[Relation between the pole and ${\mathbf {\rm \overline{MS}}}$ masses]{Relation between the pole and ${\mathbf {\rm \overline{MS}}}$ masses}\label{sec:MSbar}
Our first application is to a quantity which does not have a cusp part in its anomalous dimension: the perturbative relation between the pole and $ {\rm \overline{MS}}$ masses. Even though the results derived in this section are known, it is nevertheless worth re-deriving them within our formalism as it will illustrate the method on a simple example. To avoid confusion, we denote the primary quark mass (here and in the rest of the article) as $M$. We start by quoting the result for $\bigl[\,\overline M(\mu)-M_{\rm pole}\bigr]/M_{\rm pole}$ at one-loop using a modified gluon propagator, and identify $\mathcal{Q}=M_{\rm pole}$, where UV-divergences must be removed through a $Z$ factor in the ${\rm \overline{MS}}$ scheme
\begin{equation}\label{eq:m1Pole}
m_1 (h,\varepsilon) = - \frac{3 - 2 \varepsilon}{2} \frac{(1 + h -\varepsilon)\Gamma (\varepsilon - h)\Gamma (1 + 2 h - 2\varepsilon)}{\Gamma (3 + h - 2 \varepsilon) } \,,
\end{equation}
result that was recently computed in the form given above in Ref.~\cite{Gracia:2021nut}. Before presenting the 1- and two-loop results, we define the $\rm {\overline{MS}}$ mass anomalous dimension:
\begin{equation}
\mu \frac{\textrm{\text{d}} \overline{M}(\mu)}{\textrm{\text{d}}\mu} = 2 \overline{M}(\mu) \sum_{n = 1} \gamma^{\rm \overline{MS}}_{n - 1} \biggl[\frac{\alpha_s (\mu)}{4 \pi} \biggr]^n.
\end{equation}

\begin{figure*}[t!]
\subfigure[]
{\includegraphics[width=0.495\textwidth]{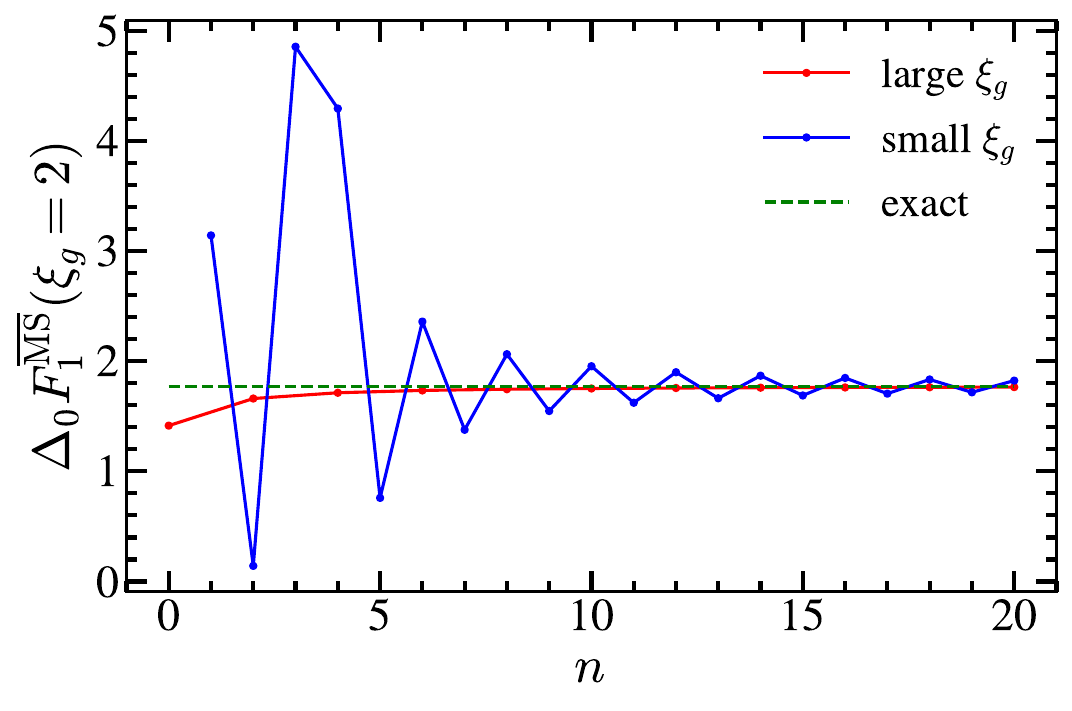}
\label{fig:MassMg1}}
\subfigure[]{\includegraphics[width=0.505\textwidth]{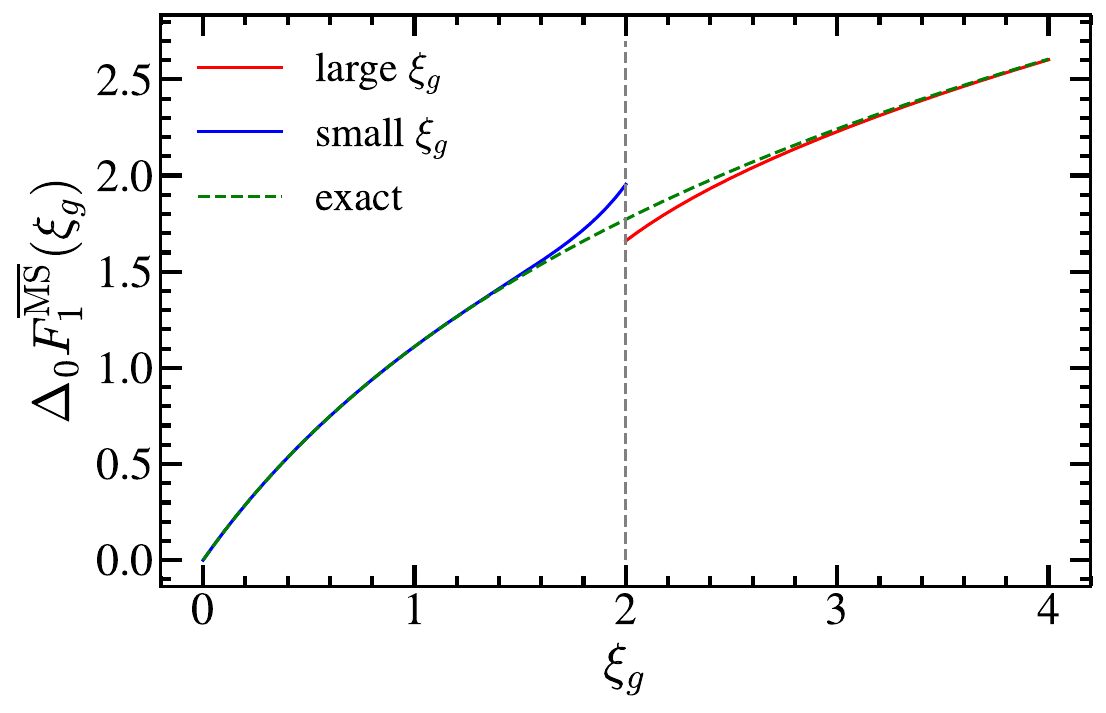}
\label{fig:MassMg}}
\caption{Gluon mass correction to the relation between the pole and $\rm {\overline{MS}}$ quark masses in its exact form (dashed green), small- (blue) and large-mass (red) expansions. Left panel: $\Delta_0 F_1^{\rm {\overline{MS}}}$ at the boundary between the mass expansions $\xi_g = 2$, as a function of the expansion order $n$ of each expansion. Right panel: Dependence of $\Delta_0 F_1^{\rm {\overline{MS}}}(\xi_g)$ with the parameter $\xi_g$ including $10$ and $3$ non-zero terms in the small- and large-mass expansions, respectively.}
\label{fig:MgMass}
\end{figure*}
\subsection{Massive gluon}\label{sec:MSglue}
Multiplying the result in Eq.~\eqref{eq:m1Pole} by the factor $\Gamma(h)\Gamma(1-h)$ we obtain the corresponding Mellin transform $\mathcal{M}^{\rm {\overline{MS}}}_1\bigl( h, \xi_g, \varepsilon \bigr)$. To figure out the convergence radius of both expansions, we look at the large-$h$ behavior of $\mathcal{M}^{\rm {\overline{MS}}}_1$ with $\varepsilon=0$. Defining $\xi_g=m_g/M_{\rm pole}$ we have
\begin{equation}
\bigl|\mathcal{M}^{\rm {\overline{MS}}}_1\bigl( h, \xi_g, 0 \bigr)\bigr|\xrightarrow[| h | \gg 1]{} \frac{3 \pi^{3 / 2} \csc^2 (\pi h)}{2 h^{3 / 2}} \biggl( \frac{\xi_g}{2}\biggr)^{\!-2 h}\,,
\end{equation}
from where we read that the small (large) gluon mass expansion works for $\xi_g$ smaller (larger) than $2$, while at $\xi_g=2$ both expansions are convergent, as can be seen in Fig.~\ref{fig:MassMg1}. After setting $\varepsilon = 0$, if closing towards ${\rm Re}(h)>0$ there are double poles at all positive integer values of $h$. When closing towards ${\rm Re}(h)<0$, one finds simple poles at all negative integers and half-integer values of $h$, except for $h = - 2$ where the pole is double. The residue at $h=-1/2$ yields a term linear in the gluon mass, which signals the $\mathcal{O}(\Lambda_{\rm QCD})$ renormalon ambiguity and fixes the residue of the $u=1/2$ pole in the Borel plane. The UV divergences of both expansions coincide:
\begin{equation}
F^{\rm {\overline{MS}}}_{1, \rm div} = - \frac{3}{4 \varepsilon} = \dfrac{\delta Z_{1,1}^{\rm \overline{MS}}}{\varepsilon} .
\end{equation}
This result correctly yields the well-known one-loop ${\rm \overline{MS}}$ mass anomalous dimension, but is removed after renormalization:
\begin{equation}
\gamma_1^{\rm \overline{MS}} =
4 C_F\delta Z_{1,1}^{\rm \overline{MS}}= - 3C_F=-4\,.
\end{equation}
The massless limit is given by the pole at $h= 0$, and after renormalization takes the form
\begin{equation}
F^{\rm {\overline{MS}}}_{1,\rm fin}(0,M_{\rm pole},\mu,\varepsilon) = - \frac{3}{2} \log \biggl( \frac{\mu}{M_{\rm pole}} \biggr) - 1\,.
\end{equation}
At this point we can compute both series expansions for the corrections to the massless gluon limit, obtaining
\begin{align}
\Delta_0 F_1^{\rm {\overline{MS}}}(\xi_g) =\, &3 \sum_{n = 0} \frac{\xi_g^{-2 n} (2 n)!}{(n + 2) (n!)^2}
\biggl[\log(\xi_g)+ H_{n} -H_{2 n} +\frac{1}{2 (n+2)}\biggr] \\
=\, & - \!\frac{\xi_g^4}{24} [ 5 - 6 \log (\xi_g) ] + \frac{3}{2} \sum_{n = 1}^{n \neq 4} \frac{( -\xi_g)^{n}\Gamma^2 \bigl( \frac{n}{2} \bigr)}{(n - 4) \Gamma (n)} \,.\nonumber
\end{align}
The bottom line can be summed up, and we find\footnote{For $\xi_g> 2$ one simply replaces
\begin{equation}
\sqrt{4 - \xi_g^{2}} \arccos \biggl(\frac{\xi_g}{2}\biggr) \to \frac{1}{2}\sqrt{\xi_g^2-4} \log \biggl[\frac{\xi_g}{2} \biggl(\xi_g -\sqrt{\xi_g^2-4}\,\biggr)-1\biggr]\,,
\end{equation}
to have every term manifestly real.}
\begin{equation}
\Delta_0 F_1^{\rm {\overline{MS}}} (\xi_g)= \frac{\xi_g}{4} \biggl[ \bigl( 2 + \xi_g^{2} \bigr) \sqrt{4 - \xi_g^{2}} \arccos \biggl(\frac{\xi_g}{2}\biggr) + \xi_g^{3} \log(\xi_g) - \xi_g \biggr] .
\end{equation}
This agrees with a direct computation whose details will be given elsewhere. To the best of our knowledge, this result has not been presented anywhere before. The large-mass expansion has only same-sign even powers. The series for small gluon masses has odd and even powers of $\xi_g$ and is oscillatory, hence it converges slowly, as can be seen in both panels of Fig.~\ref{fig:MgMass}.

Let us provide the matching coefficient between the ${\rm \overline{MS}}$ masses defined in the full theory, with massless gluons and a single massive vector boson $\overline{M}^{(n_g)}$, and in an EFT containing only massless gluons $\overline{M}^{(n_\ell)}$. The strategy to obtain the matching coefficient is through the condition of having a universal pole mass in the limit where both theories should be valid:
\begin{align}
M_{\rm pole}&\,=\overline{M}^{(n_g)} (\mu) \biggl\{ 1 - \frac{\alpha_s(\mu)}{\pi} C_F [2 F_1^{\rm {\overline{MS}}}(0,M_{\rm pole}) + \Delta_0 F_1^{\rm {\overline{MS}}} (\xi_g\to\infty)] \biggr\}_{\rm fin}
\\& \,= \overline{M}^{(n_{\ell})} (\mu) \biggl[ 1 - \frac{\alpha_s(\mu)}{\pi} C_F F_1^{\rm {\overline{MS}}}(0,M_{\rm pole}) \biggr]_{\rm fin} .\nonumber
\end{align}
Noting that $\Delta_0 F_1^{\rm {\overline{MS}}}(\xi_g\to\infty)=\Delta_0^\infty F_1^{\rm {\overline{MS}}}(\xi_g) = 3/8 + 3/2 \log(\xi_g)$, we easily obtain the matching condition, which moreover is independent of $M_{\rm pole}$:
\begin{align}
\frac{\overline{M}^{(n_g)} (\mu)}{\overline{M}^{(n_{\ell})} (\mu)} =\, & 1 + \frac{\alpha_s}{\pi} C_F [F_{1,\rm fin}^{\rm {\overline{MS}}}(0,M_{\rm pole},\mu,\varepsilon)
+\Delta_0^\infty F_1^{\rm {\overline{MS}}}(\xi_g) ] \equiv 1 + \frac{\alpha_s}{\pi} C_F \delta m^{(n_g \rightarrow n_{\ell})},\nonumber\\
\delta M^{(n_g \rightarrow n_{\ell})} =\, & -\! \frac{3}{2} \log \biggl(\frac{\mu}{m_g} \biggr) - \frac{5}{8} \,.
\end{align}

\begin{figure*}[t!]
\subfigure[]
{\includegraphics[width=0.495\textwidth]{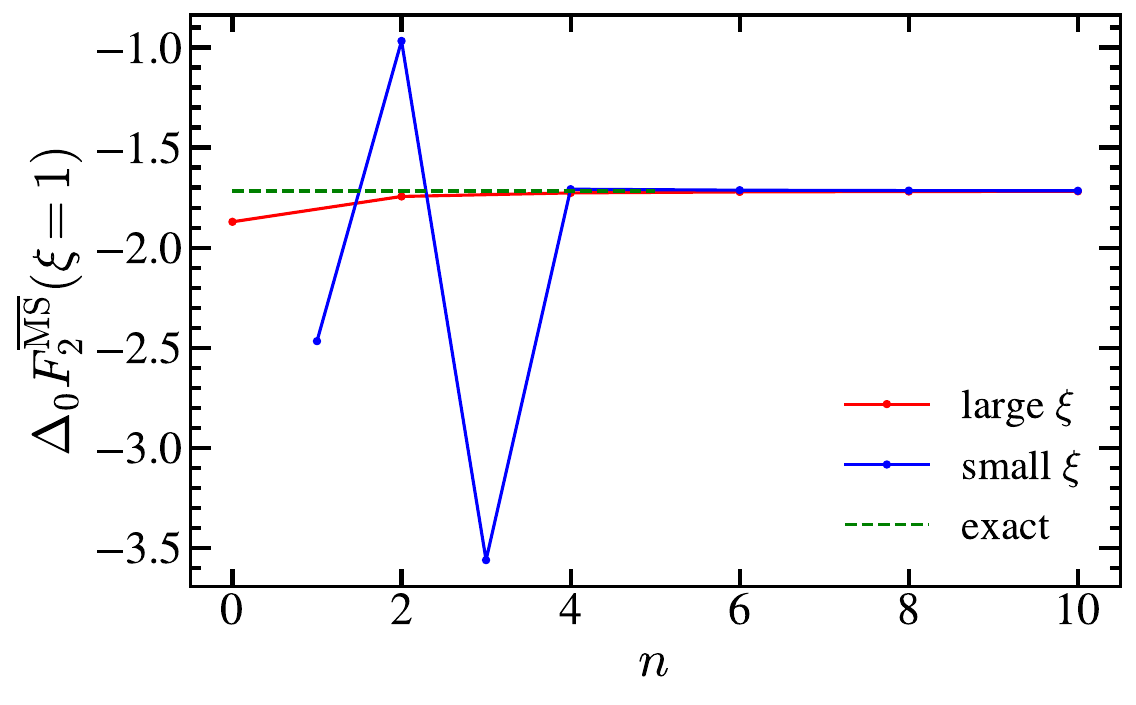}
\label{fig:MassSec1}}
\subfigure[]{\includegraphics[width=0.505\textwidth]{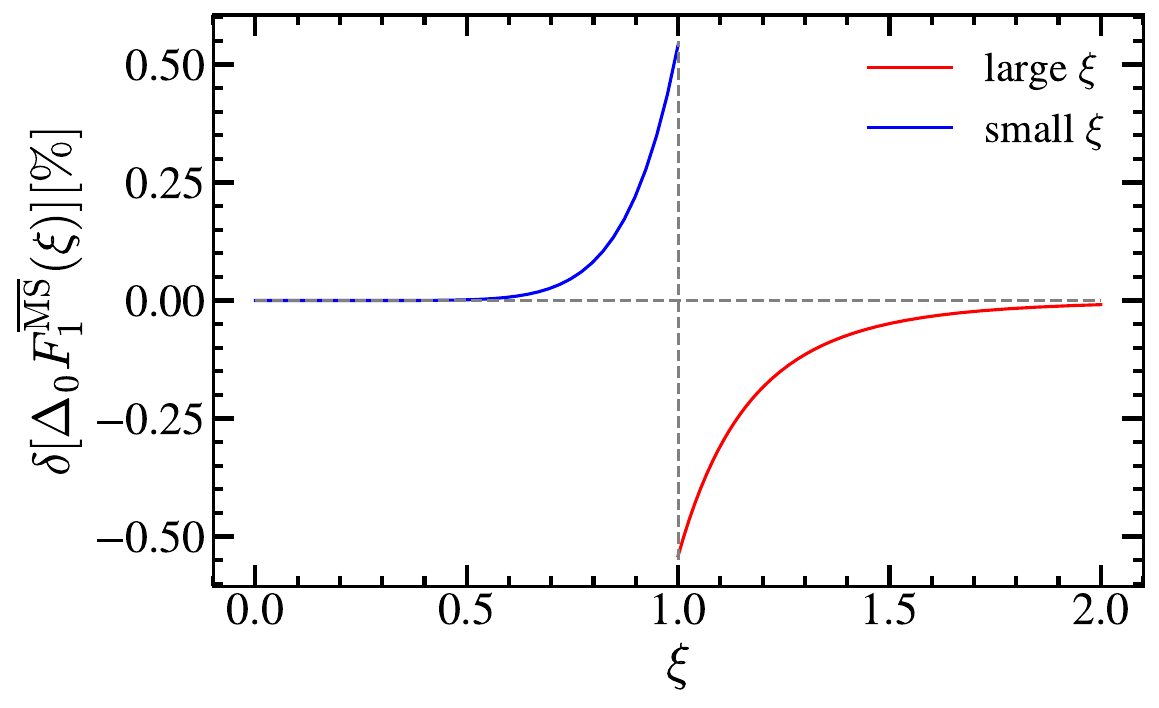}
\label{fig:MassSec}}
\caption{Secondary mass correction to the relation between the pole and $\rm {\overline{MS}}$ quark masses in its exact form (dashed green), small- (blue) and large-mass (red) expansions. Left panel: $\Delta_0 F_2^{\rm {\overline{MS}}}$ at the boundary between the mass expansions $\xi = 1$, as a function of the expansion order $n$. Right panel: percent deviation of the expansions compared to the exact result as a function of the parameter $\xi$, using $4$ and $3$ non-zero terms in the small- and large-mass expansions, respectively.}
\label{fig:SecMass}
\end{figure*}
\subsection{Secondary massive quark}\label{sec:MSbSec}
Multiplying Eq.~\eqref{eq:m1Pole} by $G(h,\varepsilon)$ we obtain $\mathcal{M}^{\rm \overline{MS}}_2( h, \xi,\varepsilon)$, and setting $\varepsilon=0$ we can figure out the convergence radius. Defining $\xi=m/M_{\rm pole}$ one has
\begin{equation}
\mathcal{M}^{\rm \overline{MS}}_2 ( h, \xi, 0 ) = \frac{3 \xi^{-2 h}(h + 1)\Gamma^2(h)\Gamma^2(1-h)}{2 (h + 2) h (2 h + 3) (2 h + 1)}\xrightarrow[| h | \gg 1]{} \frac{3 \pi^2 \!\csc^2 (\pi h)}{8 h^3} \xi^{-2 h}.
\end{equation}
It is clear that the small- and large-$m$ expansions converge for $m$ smaller and larger than $M_{\rm pole}$, respectively, and also for $\xi=1$.\footnote{We do not assume any hierarchy between $m$ and $M$ since our results are general and one could consider e.g.\ the secondary virtual top correction to the bottom mass in QCD with six active flavors.} The divergences of both expansions are $m$-independent and coincide:
\begin{equation}
F^{\rm \overline{MS}}_{2,{\rm div}} = - \frac{1}{8 \varepsilon^2} + \frac{5}{48\varepsilon} = \sum_{i=1}^2 \dfrac{\delta Z_{2,i}^{\rm \overline{MS}}}{\varepsilon^n}\,.
\end{equation}
This result correctly reproduces the $n_f$ color piece of the two-loop ${\rm \overline{MS}}$ mass anomalous dimension:
\begin{equation}
[\gamma_2^{\rm \overline{MS}}]_{n_f}
= \frac{4C_F T_F}{\varepsilon} \Bigl( 8\delta Z_{2,2}^{\rm \overline{MS}} + \beta_0^{(n_f)} \delta Z_{1,1}^{\rm \overline{MS}} \Bigr)
+ 32C_F T_F \delta Z_{2,1}^{\rm \overline{MS}}
= \frac{10}{3}\,C_FT_F = \dfrac{20}{9} \,,
\end{equation}
where $\beta_0^{(n_f)} = -4/3$ is the $n_f$ piece of the QCD beta function. The term in round brackets needs to cancel to yield a UV-finite anomalous dimension, and indeed the relation \mbox{$8\delta Z_{2,2}^{\rm \overline{MS}} + \beta_0^{(n_f)} \delta Z_{1,1}^{\rm \overline{MS}} =0$} is verified. From the sum of residues corresponding to the poles located at $h=0$ and $h=-\varepsilon$ and the $\Pi_0$ contribution ---\,after removing the UV divergences\,--- we reproduce the known two-loop massless result of Ref.~\cite{Gray:1990yh}, which accounts for the full $\mu$ dependence.
\begin{equation}
F^{\rm \overline{MS}}_{2, \rm fin}(0, M_{\rm pole}, \mu,\varepsilon) = \frac{1}{2} \log^2 \biggl( \frac{\mu}{M_{\rm pole}} \biggr)
+ \frac{13}{12} \log \biggl( \frac{\mu}{M_{\rm pole}} \biggr) + \frac{\pi^2}{12} + \frac{71}{96}\,.
\end{equation}
After setting $\varepsilon=0$ double poles are found at all positive and negative integer values of $h$, except for $h=-1$, and $h=-2$ for which the pole multiplicity is $1$ and $3$, respectively. Furthermore, there are simple poles at $h=-1/2$ and $h=-3/2$. The difference of the massless and $m\to\infty$ limits is UV finite and $\mu$-independent, but contains logarithms that blow up in any of those two limits. It can be computed as minus the contribution of the pole at $h=0$ obtained setting $\varepsilon=0$ before computing the residue
\begin{equation}
\Delta_0 ^\infty F^{\rm \overline{MS}}_2(\xi) = - \frac{1}{2}\! L_\xi^2 - \frac{13}{12}\! L_\xi- \frac{\pi^2}{12} - \frac{151}{144} \,,
\end{equation}
with $L_\xi = \log(\xi)$. Since the factor containing all gamma functions in $\mathcal{M}_2 ( h, \xi, 0 )$ is symmetric under $h\to -h$, and given that gamma functions are the only structures with an infinite number of poles, we expect that this symmetry will be manifest in the infinite sums of the `left' and `right' expansions. In fact this is what we find for the corrections to the massless limit:
\begin{align}
\Delta_0 F^{\rm \overline{MS}}_2(\xi) = \,& \Delta_0 ^\infty F^{\rm \overline{MS}}_2 + \frac{3}{2} \sum_{n = 1} G^{\rm \overline{MS}}_n(\xi) \\
= \,& -\! \frac{\pi^2}{4} \xi - \frac{\pi^2}{4} \xi^3 + \xi^4 \biggl(\frac{1}{2}L_\xi^2-\frac{13 }{12}L_\xi+\frac{\pi ^2}{12}+\frac{151}{144}\biggr)
- \frac{3}{2}\sum_{n = 1}^{n\neq 2}G^{\rm \overline{MS}}_{-n}(\xi) \,,\nonumber\\
G^{\rm \overline{MS}}_n(\xi) = \,& \frac{\xi^{-2 n} }{n (n + 2) (2 n +1) (2 n + 3)} \!\biggl[\frac{2 (3 + 10 n + 5 n^2)}{n (2 + n) (1 + 2 n) (3 + 2 n)}
+ 3 + 2 (n + 1) L_\xi \biggr].\nonumber
\end{align}
For $\xi=1$, that is, when it corresponds to the contribution of the heavy quark to its own self-energy, we can sum up either series obtaining the known result $\Delta_0 F_2(1)=(3-\pi^2)/4$. These expressions can be summed up to all orders, fully reproducing the known result of Ref.~\cite{Gray:1990yh}:
\begin{align}
\Delta_0 F^{\rm \overline{MS}}_2(\xi) = \,& \frac{1}{2} (1 - \xi) (1 - \xi^3) \biggl[ \text{Li}_2 (1 - \xi) + \frac{1}{2}L_\xi^2 + \frac{\pi^2}{6} \biggr]\!
-\frac{\pi^2}{12} + \frac{\xi^2}{2} \biggl( L_\xi + \frac{3}{2} \biggr) \\ &
- \!\frac{1}{2} (1 + \xi) (1 + \xi^3) \biggl[ \text{Li}_2 (- \xi) - \frac{1}{2}\! L_\xi^2 + \log (1 + \xi) L_\xi + \frac{\pi^2}{6} \biggr]
- \frac{1}{2}\! L_\xi^2 \,.\nonumber
\end{align}
The small- and large-mass expansions are shown in Fig.~\ref{fig:SecMass}. We observe that the small-mass expansion is badly convergent for the first orders. This pathological behavior can be blamed on the presence of odd-power terms which are related to $u=1/2$ and $u=3/2$ renormalon ambiguities, and to the fact that these do not include non-analytic behavior in $\xi$. After the fourth power of $\xi$ is included (which coincides with the first appearance of $L_\xi$), the accuracy improves drastically. The plots also reveal that the small- and large-mass expansions approach the exact result from above and below, respectively. Finally, the convergence of the small-mass expansion is much better behaved than it is for the gluon mass case. This can be understood since the gluon mass expansion contains an infinite number of odd-power corrections.

Let us compute the $\mathcal{O}(\alpha_s^2)$ matching condition between the ${\rm \overline{MS}}$ masses with $n_f + 1$ and $n_f$ flavors, $\overline M^{(n_f+1)}$ and $\overline M^{(n_f)}$, respectively, following the same strategy as in the previous section. Using Eq.~\eqref{eq:limit}, or equivalently the relation underneath, we find
\begin{align}
\frac{\overline{M}^{(n_f + 1)} (\mu)}{\overline{M}^{(n_f)} (\mu)} \equiv\, & 1 + \left[ \frac{\alpha_s (\mu)}{\pi} \right]^2 T_F C_F \delta M^{(n_f+1 \rightarrow n_{f})} \,,\\
\delta M^{(n_f+1 \rightarrow n_{f})} =\,& F_2(M_{\rm pole}, 0,\varepsilon) + \Delta_0 F_2(\xi\to\infty) + \frac{2}{3} F_1(M_{\rm pole}, 0,\varepsilon)\log \biggl( \frac{\mu}{m} \biggr)\,\nonumber\\
=\,& -\! \frac{1}{2} \log^2 \biggl( \frac{\mu}{m } \biggr) + \frac{5}{12} \log \biggl(\frac{\mu}{m} \biggr) - \frac{89}{288}\,,\nonumber
\end{align}
independent of $M_{\rm pole}$ and in agreement with Ref.~\cite{Chetyrkin:1997un}. At this order, $\alpha_s$ in the previous expression can be evaluated with either $n_f+1$ or $n_f$ flavors, as the difference is $\mathcal{O}(\alpha_s^3)$.

\section{SCET Computations}\label{sec:SCET}
We turn our attention now to the computation of matrix elements which enter the factorization theorem of event shape distributions for $e^+e^-$ collisions, initiated by massless primary quarks:
\begin{align}\label{eq:factSCET}
\frac{1}{\sigma_0} \frac{{\rm d} \sigma}{{\rm d} e} =&\, Q^2 H_Q (Q, \mu)\!\int \!{\rm d} \ell J_{\tau} (Q^2 \tau - Q \ell, \mu) S_e (\ell, \mu)\,,\\[-0.1cm]
J_{\tau} (s, \mu) \equiv& \int_0^s \!{\rm d} s' J_n (s - s', \mu) J_n (s', \mu)\,,\nonumber
\end{align}
where $e=\tau$ (thrust) and $C$ (C-parameter), and where the hemisphere jet function appears also in the heavy jet mass factorization theorem. The factorized expression involves the product of the hard matching coefficient $H_Q(Q,\mu)$ times the convolution of jet $J_n (s, \mu)$ and soft $S_e (\ell, \mu)$ functions, whose natural scales will be denoted by $\mu_H$, $\mu_J$ and $\mu_S$, respectively, satisfying $\mu_H>\mu_J>\mu_S$. Since our formalism as it stands now solely applies to virtual massive bubbles, we will be able to compute only the corrections to the hemisphere jet function, which enters the factorized expressions for thrust, heavy jet mass and C-parameter in the massive scheme, also known as C-jettiness, see Refs.~\cite{Lepenik:2019jjk,Bris:2020uyb} for a discussion of massive schemes for heavy quarks, and Refs.~\cite{Salam:2001bd,Mateu:2012nk} for their effects on soft hadronization. It turns out that the hemisphere jet function can be computed as the discontinuity of a forward-scattering matrix element in which the bubble is virtual. Even though both results have been already computed in closed form, to the best of our knowledge, the small and large secondary mass expansions are not known. Furthermore, the renormalization group (RG) evolved jet function is not known in closed form, and in that respect our result in terms of expansions can be regarded as a new analytic result.

\subsection{Hard Matching Coefficient}\label{sec:hard}
In this section we compute the corrections to the Wilson coefficient appearing when matching QCD onto SCET due to a massive vector boson or a secondary massive quark bubble. The relevant hard scale in this case is $Q$, the center-of-mass energy. To write expressions as simple as possible, it will be useful to define the reduced masses for the vector boson $\hat m_g^2=m_g^2/Q^2$ and secondary quark $\hat m^2=m^2/Q^2$, where we include the squares to avoid making any assumption on the sign of $Q^2$.

\subsubsection{Review of QCD-SCET Matching for massless Quarks}\label{sec:massless}
Let us consider the QCD and SCET dijet currents, both defined in terms of bare fields, that we schematically denote by $\mathcal{O}_{\rm{QCD}}$ and $\mathcal{O}^{\rm bare}_{\rm{SCET}}$, respectively. For simplicity we consider only the vector current:
\begin{equation}\label{eq:QCD-SCET}
\mathcal{O}_{\rm{QCD}} = {\bar q}\gamma^\mu q\,,\qquad \mathcal{O}^{\rm bare}_{\rm{SCET}} = {\bar \chi}_n Y_n^\dagger \gamma^\mu Y_{\bar n} \chi_{\bar n}\,,
\end{equation}
where $Y_n$ and $Y_{\bar n}$ are ultrasoft Wilson lines that appear after BPS field redefinition~\cite{Bauer:2001yt}, and $\chi_{n}$ and $\chi_{\bar n}$ are jet fields, involving a collinear quark field and a collinear Wilson line. For simplicity, and since it plays no significant role, we ignore the Lorentz index in both currents. For later use, we symbolically define the soft $\mathcal{O}^{\rm bare}_{\rm soft}(\ell)\sim {\rm tr}[\overline Y_{\bar n}^\top Y_n\delta(\ell - Q\hat e)Y_n^\dagger \overline Y_{\bar n}^\ast]/N_c$ and collinear $\mathcal{O}^{\rm bare}_{\rm col}(s)\sim{\rm tr}[\bar n\!\!\!/\,\chi_n\delta(s-Q^2\hat e)\bar\chi_n]/(4N_c )$ operators that appear when squaring $\mathcal{O}^{\rm bare}_{\rm{SCET}}$, where the trace is over color (and also Dirac indices in the jet operator), $s$ is the invariant mass of the hemisphere, $e$ the event-shape variable, and $\hat e$ the operator that pulls out its value when acting on a state. For now we do not specify the number of active quark flavors (or, equivalently, the number of massive and massless gluons), although this will become important later on. While QCD vector current conservation implies its UV finiteness, the SCET operator needs renormalization, and the relation between bare and renormalized currents defines the renormalization constant $Z_{\rm{SCET}}$ (the renormalized current is also expressed in terms of bare fields)\footnote{For simplicity we omit the dependence of $Z_{\rm SCET}$ on $Q$ along with convolutions over the field labels.}
\begin{align}
\mathcal{O}_{\rm{SCET}}^{\rm ren}(\mu) =\, & Z_{\rm{SCET}}(\mu) \mathcal{O}^{\rm bare}_{\rm{SCET}}\,, \\
\log\bigl[Z_{\rm{SCET}}(\mu)\bigr] = \, & \delta Z^{\rm SCET}_{\rm nc} + L_\mu \delta Z^{\rm SCET}_{\rm cusp} = \sum_{n=1} \biggl[\frac{\alpha_s(\mu)}{4\pi}\biggr]^n
\Bigl(\delta Z^{\rm SCET}_{i,\rm nc} + L_\mu \delta Z^{\rm SCET}_{i,\rm cusp}\Bigr)\,,\nonumber\\
\delta Z^{\rm SCET}_{i,\rm cusp} = \, & \sum^{j=i}_1 \dfrac{\delta Z^{\rm SCET}_{i,j,\rm cusp}}{\varepsilon^j}\,,\nonumber\\
\delta Z^{\rm SCET}_{i,\rm nc} = \, &\sum^{j=i+1}_1 \dfrac{\delta Z^{\rm SCET}_{i,j,\rm nc}}{\varepsilon^j}\,,\nonumber
\end{align}
where $L_\mu = \log(- \mu^2/Q^2)$. Unlike the bare operator, the renormalized one depends on the renormalization scale $\mu$. Its dependence on this scale is given by the renormalization constant $Z_{\rm{SCET}}(\mu)$, from which one can compute the SCET anomalous dimension as \mbox{$\gamma_{\rm SCET}=-{\rm d}\!\log(Z_{\rm SCET})/{\rm d}\!\log(\mu)=\gamma_{\rm nc}^{\rm SCET} - L_\mu \Gamma_{\rm cusp}$}, which has cusp and non-cusp pieces. For completeness, we also define the QCD $\beta$-function coefficients:
\begin{align}
\Gamma_{\!\rm cusp} (\alpha_s)=\, &\sum_{n = 1} \Gamma_{\!n-1} \biggl(\frac{\alpha_s}{4 \pi} \biggr)^{\!\!n} \,,\\
\gamma_{\rm nc}^{\rm SCET} (\alpha_s)=\, &\sum_{n = 1} \gamma^{\rm SCET}_{n-1} \biggl(\frac{\alpha_s}{4 \pi} \biggr)^{\!\!n}\,,\nonumber\\
\beta_{\rm QCD} (\alpha_s) = \,& - 2 \alpha_s \sum_{n = 1} \beta_{n - 1} \biggl(\frac{\alpha_s}{4 \pi} \biggr)^{\!\!n}\, .\nonumber
\end{align}
In order to have UV-finite cusp and non-cusp anomalous dimensions, the following constraints must be satisfied:
\begin{align}\label{eq:UVcond}
\delta Z^{\rm SCET}_{j, i + 1,\rm cusp} =\,& -\! \frac{1}{j} \sum_{n = i}^{j - 1} n \beta_{j - n - 1}\delta Z^{\rm SCET}_{n, i,\rm cusp}\,,\\
\delta Z^{\rm SCET}_{j, i + 1,\rm nc} = \,&\frac{1}{j} \Biggl[ \delta Z^{\rm SCET}_{j, i,\rm cusp} - \sum_{n = \max (1, i - 1)}^{j - 1} n \beta_{j - n - 1}\delta Z^{\rm SCET}_{n, i,\rm nc} \Biggr],\nonumber
\end{align}
Once these are satisfied, the anomalous dimensions are obtained from the $1/\varepsilon$ terms:
\begin{equation}\label{eq:GammaSCET}
\Gamma_{\!n} = - 2 (n + 1) \delta Z^{\rm SCET}_{n + 1, 1,\rm cusp}\,,\qquad
\gamma^{\rm SCET}_n = 2 (n + 1) \delta Z^{\rm SCET}_{n + 1, 1,\rm nc}\, .
\end{equation}
One can use Eq.~\eqref{eq:UVcond} to obtain a simple closed from for the most divergent terms in the $Z$ factors which depend only on $\Gamma_0$:
\begin{equation}\label{eq:highest}
\delta Z^{\rm SCET}_{j ,j,\rm cusp} = -\frac{(- \beta_0)^{j - 1} \Gamma_{\!0}}{2 j}\,,\qquad
\delta Z^{\rm SCET}_{j, j + 1,\rm nc} = \delta Z^{\rm SCET}_{j ,j,\rm cusp} H_j\,.
\end{equation}
Solving the renormalization group evolution (RGE) equation one can relate renormalized operators at different scales summing up potentially large logarithms of their ratio: \mbox{$\mathcal{O}_{\rm{SCET}}^{\rm ren}(\mu_2)=U_{\rm SCET}(Q, \mu_2, \mu_1)\mathcal{O}_{\rm{SCET}}^{\rm ren}(\mu_1)$}. The renormalized matching coefficient is defined on renormalized operators:
\begin{equation}
\mathcal{O}_{\rm{QCD}} = C_{\rm{SCET}}^{\rm{ren}}(\mu) \mathcal{O}_{\rm{SCET}}^{\rm{ren}}(\mu) = Z_{\rm{SCET}}(\mu)C_{\rm{SCET}}^{\rm{ren}}(\mu) \mathcal{O}^{\rm{bare}}_{\rm{SCET}} \equiv C_{\rm{SCET}}^{\rm{bare}} \mathcal{O}^{\rm{bare}}_{\rm{SCET}}\,,
\end{equation}
hence the bare matching coefficient $C_{\rm{SCET}}^{\rm{bare}}=Z_{\rm{SCET}}C_{\rm{SCET}}^{\rm{ren}}$ relates $\mathcal{O}_{\rm{QCD}}$ to the bare SCET operator. To avoid large logarithms it is convenient to match QCD and SCET at the scale $\mu_H\sim Q$. From now on we drop the dependence of $\mathcal{O}_{\rm{SCET}}^{\rm ren}$ and $ C_{\rm{SCET}}^{\rm{ren}}$ on $\mu$. To compute $C_{\rm{SCET}}$ we consider the simplest matrix element: the quark form factor denoted by $\langle \mathcal{O}\rangle$. Since
$Z_{\rm{SCET}} =\langle \mathcal{O_{\rm SCET}} \rangle =\langle \mathcal{O_{\rm QCD}}\rangle= 1+\mathcal{O}(\alpha_s)$, taking the natural logarithm is convenient:
\begin{equation} \log (C_{\rm{SCET}}^{\rm{bare}}) = \log (Z_{\rm{SCET}}) + \log (C_{\rm{SCET}}^{\rm{ren}}) = \log (\langle \mathcal{O}_{\rm{QCD}} \rangle) - \log (\langle \mathcal{O}^{\rm{bare}}_{\rm{SCET}} \rangle) \,.
\end{equation}
Hence $\log (Z_{\rm{SCET}}) = -[\log (\langle \mathcal{O}^{\rm{bare}}_{\rm{SCET}}\rangle)]_{\rm{div}}$ and, unless otherwise stated, we adopt the $\overline{\rm{MS}}$ scheme and absorb in $\log (Z_{\rm{SCET}})$ only the $1 /\varepsilon^n$ UV-poles. Finally, the renormalized matching coefficient is computed as $\log (C_{\rm{SCET}}^{\rm{ren}}) = [\log(\langle \mathcal{O}_{\rm{QCD}} \rangle) - \log (\langle\mathcal{O}^{\rm{bare}}_{\rm{SCET}} \rangle)]_{\rm{fin}}$. The hard function appearing in the factorization theorem is simply $H(Q,\mu)=|C_{\rm SCET}^{\rm ren}(Q,\mu)|^2$ that evolves with the renormalization scale with the anomalous dimension $\gamma_H=\gamma_{\rm SCET}+\gamma_{\rm SCET}^\star$, to which we can associate a renormalization factor \mbox{$Z_H(\mu)=|Z_{\rm SCET}(\mu)|^2$} relating the bare and renormalized hard functions \mbox{$H_Q^{\rm bare}(Q)=Z_H(\mu) H(Q,\mu)$}. Note that the hard function's evolution is reversed as compared to that of the squared SCET operator since it is a matching coefficient, not a matrix element.

For massless partons, the QCD and SCET form factors are IR divergent, hence a regulator needs to be specified. However, the ratio (difference of logarithms) does not need any IR regularization, as both full theory and EFT are equal in the infrared. For massless partons and taking care of IR singularities with dimensional regularization (dimreg from now on), the SCET form factor beyond tree-level involves only scaleless integrals that vanish, so $\langle \mathcal{O}^{\rm bare}_{\rm SCET}\rangle=1$ and $\log(\langle \mathcal{O}^{\rm ren}_{\rm SCET}\rangle) = -\log (Z_{\rm{SCET}})$. Moreover, $C_{\rm SCET}^{\rm bare} = \langle \mathcal{O}_{\rm{QCD}} \rangle$ provided that dimreg is also chosen as IR regulator in the QCD form factor. Furthermore, $\log (Z_{\rm{SCET}}) = \log(\langle \mathcal{O}_{\rm{QCD}} \rangle)_{\rm div}$, where the QCD divergences are of IR origin, and $\log(C_{\rm{SCET}}^{\rm{ren}})=\log(\langle \mathcal{O}_{\rm{QCD}} \rangle)_{\rm fin}$. This is by far the simplest way to compute the Wilson coefficient, and was the strategy followed in Ref.~\cite{Gracia:2021nut} to carry out the one-loop computation with a modified massless gluon propagator.

The soft operator also needs renormalization, defining the soft renormalization factor as $\mathcal{O}_{\rm soft}^{\rm bare} = Z_S(\mu) \mathcal{O}^{\rm ren}_{\rm soft}(\mu)$.\footnote{We oversimplify the notation assuming multiplicative renormalization for the soft and collinear operators. A full-fledged analysis would of course require convolutions. If Fourier transforms are taken, renormalization is indeed multiplicative, but the renormalization factors depend on the Fourier variable through their cusp term.} The soft function is the vacuum matrix element of the soft operator $S^{\rm ren}_e(\ell, \mu)=\langle\mathcal{O}^{\rm ren}_{\rm soft}(\mu)\rangle=S^{\rm bare}_e(\ell)/Z_S(\mu)$, whose anomalous dimension $\gamma_S$ is set by $Z_S(\mu)$. To avoid large logarithms in the soft function one must set its renormalization scale to $\mu_S=Qe$. The jet function is the vacuum expectation value of the jet operator $J_n(s,\mu)\sim Z_{\rm SCET}(\mu)Z_S^{1/2}(\mu)\langle\mathcal{O}^{\rm bare}_{\rm col} \rangle=J_n^{\rm bare}(s)/Z_J(\mu)$, whose anomalous dimension $\gamma_J$ is set by $Z_J(\mu)$. One can also evolve $\mathcal{O}_{\rm col}^{\rm ren}$ with $\gamma_J$. To avoid large logarithms in the jet function its renormalization scale must be set to $\mu_J\sim Q\sqrt{e}$. From the relation $Z_H(\mu)Z_J^2(\mu)Z_S(\mu)=1$ stems the consistency condition among the anomalous dimensions $\gamma_H+2\gamma_J+\gamma_S=0$. Having discussed the running and matching in SCET, we can describe the structure of the factorization theorem for event shapes: first one matches QCD onto SCET at the scale $\mu_H$ and evolves $\mathcal{O}^{\rm ren}_{\rm SCET}(\mu_H)$ to some arbitrary $\mu<\mu_H$ (equivalently, evolves the hard function from $\mu$ to $\mu_H$). After inserting the measurement delta functions, $|\mathcal{O}^{\rm ren}_{\rm SCET}(\mu)|^2$ is split into $\mathcal{O}^{\rm ren}_{\rm col}(\mu)$ and $\mathcal{O}^{\rm ren}_{\rm soft}(\mu)$ which are evolved independently to $\mu_J$ and $\mu_S$, respectively. At these respective scales, the jet and soft functions are computed. With this procedure, only small logs appear in the matrix elements, and large logs of characteristic-scale ratios are summed up in the evolution. Note that the soft and jet functions' evolution is the same as their respective operators, since they are matrix elements (not matching coefficients).

For contributions of either massive vector bosons or secondary massive bubbles, the non-zero mass acts as an IR regulator and neither the QCD nor the SCET loop-level form factors vanish any longer if dimreg is used ---\,in fact, they are IR finite and no regulator is needed.\footnote{The computation with a massive vector boson is IR finite. For the massive bubble, the $\Pi^{\rm OS}$ insertion is also IR finite, but that proportional to $\Pi_0^{\rm \overline{MS}}$ needs regularization.} Furthermore, in either case the SCET form factors are finite only after soft bin subtractions are included, requiring regulators in individual Feynman diagrams due to rapidity divergences. With our computational strategy we bypass all problems at once: only the QCD Feynman diagrams contribute, soft bin subtractions identically vanish and there are no rapidity singularities. On the other hand, the calculation does not disentangle the QCD and SCET IR-finite contributions. As we will see, consistency conditions can be used to obtain these separately. For the one-loop computation with a shifted gluon propagator we have $\mathcal{Q}^2=-Q^2$, and the following result was found in Ref.~\cite{Gracia:2021nut}:
\begin{equation}
m_1 (h,\varepsilon) = - \frac{1}{2} \frac{\Gamma^2 (h - \varepsilon) \Gamma (1 - h +\varepsilon)}{\Gamma (3 + h - 2 \varepsilon)} \{ 2 - \varepsilon [3 + h^2 + h(2 - 3 \varepsilon) - \varepsilon (3 - 2 \varepsilon)] \}\,,
\end{equation}
from which we observe a double pole sits at $h=\varepsilon$.

\subsubsection{Formal Aspects of Matching with secondary Masses}
\begin{figure}[t]
\includegraphics[width=\textwidth]{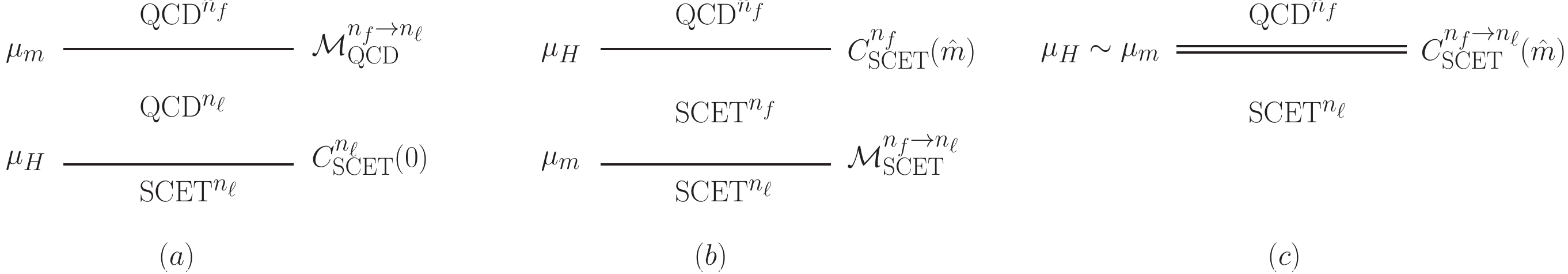}
\caption{Different paths to match QCD with $n_f$ active flavors to SCET with $n_\ell$ active flavors. The horizontal lines signal energy scales, as marked on its left, and delimit the regions of validity of the various EFTs. We also show to the right of each line the matching coefficient necessary to relate two consecutive EFTs. In panel (a) $\mu_m\gg \mu_H$ and hence the secondary quark mass is integrated out already in QCD before matching to SCET$^{(n_\ell)}$. In panel (b) $\mu_m\ll \mu_H$, such that the secondary quark is an IR scale in both QCD$^{(n_f)}$ and SCET$^{(n_f)}$. In panel (c) both scales are comparable, becoming necessary a direct matching between QCD$^{(n_f)}$ and SCET$^{(n_\ell)}$.}
\label{fig:scenarios}
\end{figure}

When considering massive quarks or gluons, one has to specify the hierarchy between $Q$ and the mass, as this determines if the latter is a UV or an IR scale.\footnote{We indicate the number of active flavors in QCD and SCET with a superscript. All operators in this section are renormalized.} For simplicity we restrict our discussion to secondary heavy quarks. If the mass is much smaller than $Q$ [\,panel (b) in Fig.~\ref{fig:scenarios}\,], then $m$ is an IR scale that should appear both on the QCD$^{(n_f)}$ and SCET$^{(n_f)}$ sides of the matching, and both theories will have this flavor as dynamic, hence using $\alpha_s^{(n_f)}$ with $n_f=n_\ell +1$ being $n_\ell$ the number of massless quarks. However, since Wilson coefficients are short-distance corrections they cannot depend on an IR scale. Since QCD$^{(n_f)}$ and SCET$^{(n_f)}$ should coincide in the limit $Q\to \infty$, which is equivalent to the mass tending to zero, and given that in a scheme in which the massive flavor is dynamic the massless limit is manifest, indeed the Wilson coefficient $C_{\rm SCET}^{(n_f)}(0)=\lim_{m\to0}\langle \mathcal{O}^{(n_f)}_{\rm QCD}\rangle/\langle\mathcal{O}^{(n_f)}_{\rm SCET}\rangle$ will not depend on the mass:\footnote{To avoid cluttering, all SCET operators and Wilson coefficients in this section are renormalized even though no superscript ``ren'' is shown.} it is simply the well-known massless matching coefficient with $n_f=n_\ell + 1$ active flavors. One can formally maintain $\hat m$-suppressed terms by simply not taking the $m\to0$ limit in the ratio of QCD and SCET form factors obtaining \mbox{$C_{\rm SCET}^{(n_f)}(\hat m^2)=\langle \mathcal{O}^{(n_f)}_{\rm QCD}\rangle/\langle\mathcal{O}^{(n_f)}_{\rm SCET}\rangle\equiv C_{\rm SCET}^{(n_f)}(0)+\Delta_0 C^{(n_f)}_{\rm SCET}(\hat m^2)$}. Moreover, since in mass-independent renormalization schemes such as $\overline{\rm MS}$ divergences do not depend on IR scales, the anomalous dimension of $C_{\rm SCET}^{(n_f)}(\hat m^2)$ is the same as that of $C_{\rm SCET}^{(n_f)}(0)$. This mass-dependent Wilson coefficient is used in Scenarios II, III and IV of Ref.~\cite{Pietrulewicz:2014qza} characterized by the condition $\mu_m<\mu_H$, where $\mu_m\sim m$ is the scale at which the secondary mass is integrated out. The computation we carry out with the Mellin-Barnes expansion gives access precisely to the quantity $C_{\rm SCET}^{(n_f)}(\hat m^2)$. If $\mu_m$ is larger than the jet or soft scales, one needs to match SCET$^{(n_f)}$ onto SCET$^{(n_\ell)}$, and the Wilson coefficient defined as
\begin{equation}\label{eq:scetMatch}
\mathcal{O}^{(n_f)}_{\rm SCET}(\mu)=\mathcal{M}^{(n_f\to n_\ell)}_{\rm SCET}(\mu)\mathcal{O}^{(n_\ell)}_{\rm SCET}(\mu)\,,
\end{equation}
is computed as the ratio of renormalized form factors in the two theories. To avoid large logarithms, it is convenient to match both operators at the scale $\mu_m$, and there is no running related to this matching condition. To carry out such computation we assume an IR regulator $\Delta$ other than dimreg is used. Furthermore, the UV poles do not depend on $\Delta$ and from the well known one-loop result ---\,or from the first line of Eq.~\eqref{eq:SCETmg}\,--- one can identify the $m_1$ and $m_2$ pieces of Eq.~\eqref{eq:F1}:
\begin{equation}
m_2 = \frac{1}{2}\,,\qquad
m_1 = \frac{3}{4} - \frac{1}{2} \log \biggl( - \frac{Q^2}{\mu^2} \biggr).
\end{equation}
With this at hand, using Eq.~\eqref{eq:limit} we obtain the matching coefficient at two-loop order
\begin{align}\label{eq:SCET2SCET}
\mathcal{M}^{(n_f\to n_\ell)}_{\rm SCET}(m,Q,\mu) =\,& \dfrac{\langle \mathcal{O}^{(n_f)}_{\rm SCET}\rangle}{\langle\mathcal{O}^{(n_\ell)}_{\rm SCET}\rangle} = 1 + \biggl[ \frac{\alpha_s(\mu)}{\pi} \biggr]^2
C_F T_F \mathcal{M}^{(2)}_{\rm SCET}(m,Q,\mu)\,,\\
\mathcal{M}^{(2)}_{\rm SCET}(m,Q,\mu) =\,& \Bigr[H^{\rm SCET}_2(m,Q,\mu,\varepsilon)\Bigl]_{\!\rm fin}
-\,\frac{1}{36} \log^3\biggl(\frac{\mu ^2}{m^2}\biggr)
-\frac{\pi ^2}{72} \log\biggl(\frac{\mu^2}{m^2}\biggr) \nonumber\\
&+\frac{\zeta_3}{18}-\biggl[\frac{3}{4} - \frac{1}{2} \log \biggl( - \frac{Q^2}{\mu^2} \biggr)\biggr]\biggl[\frac{1}{6} \log^2\biggl(\frac{\mu ^2}{m^2}\biggr)+\frac{\pi^2}{36}\biggr]\,,\nonumber
\end{align}
where $H^{\rm SCET}_2$ is the dispersive contribution of the SCET form factor, which is independent of IR regulators and, at the order we are working, $\alpha_s$ can be chosen with either $n_f$ or $n_\ell$ active flavors. In the event-shape factorization one uses $|\mathcal{M}^{(n_f\to n_\ell)}_{\rm SCET}|^2$. This matching condition is used in Ref.~\cite{Pietrulewicz:2014qza} if the renormalization scale is evolved to a scale smaller than $\mu_m$. At this point one can directly relate QCD$^{(n_f)}$ to SCET$^{(n_\ell)}$ and, if keeping all $\hat m$-suppressed terms, the relation reads: $\mathcal{O}^{(n_f)}_{\rm QCD}=C_{\rm SCET}^{(n_f)}(\hat m)\mathcal{M}^{(n_f\to n_\ell)}_{\rm SCET}\mathcal{O}^{(n_\ell)}_{\rm SCET}$.

If $m\gg Q$ (left panel in Fig.~\ref{fig:scenarios}), one should integrate out the massive secondary quark already in QCD$^{(n_f)}$ before matching to SCET$^{(n_\ell)}$ (that is, one matches QCD$^{(n_f)}$ onto QCD$^{(n_\ell)}$). Since the QCD quark form factor carries no anomalous dimension, after properly regulating the IR singularities in both theories it can be seen from Eq.~\eqref{eq:matching} that this amounts to removing the heavy quark from the QCD Lagrangian and using $\alpha_s^{(n_\ell)}$. The running of the strong coupling with $n_\ell$ active flavors sums up large logarithms of $\hat m^2 $. One can effectively keep all $\hat m$-suppressed terms considering the following matching coefficient: $\mathcal{O}^{(n_f)}_{\rm{QCD}} = \mathcal{M}^{(n_f\to n_\ell)}_{\rm QCD}\mathcal{O}^{(n_\ell)}_{\rm{QCD}}$, that at two loops reads
\begin{equation}
\mathcal{M}^{(n_f\to n_\ell)}_{\rm QCD}(\hat m,\mu) = \dfrac{\langle \mathcal{O}^{(n_f)}_{\rm QCD}\rangle}{\langle\mathcal{O}^{(n_\ell)}_{\rm QCD}\rangle}= 1 + \biggl[ \frac{\alpha_s(\mu)}{\pi} \biggr]^2 C_F T_F H^{\rm QCD}_2(\hat m^2)\,,
\end{equation}
where $H^{\rm QCD}_2$ is the dispersive contribution to the QCD quark form factor and, at the order we are working, $\alpha_s$ can be chosen with either $n_f$ or $n_\ell$ active flavors. Given that $H^{\rm QCD}_2(\hat m^2\to\infty)=0$, when strict hierarchies are considered the threshold condition is trivial. The matching coefficient between QCD$^{(n_\ell)}$ and SCET$^{(n_\ell)}$ is nothing less than the usual SCET Wilson coefficient with $n_\ell$ active flavors: $\mathcal{O}^{(n_\ell)}_{\rm QCD}=C_{\rm SCET}^{(n_\ell)}(0)\mathcal{O}^{(n_\ell)}_{\rm SCET}$. Hence, one can directly relate QCD$^{(n_f)}$ to SCET$^{(n_\ell)}$, and if keeping all $\hat m$-suppressed terms the relation reads: $\mathcal{O}^{(n_f)}_{\rm QCD}=\mathcal{M}^{(n_f\to n_\ell)}_{\rm QCD}C_{\rm SCET}^{(n_\ell)}(0)\mathcal{O}^{(n_\ell)}_{\rm SCET}$.

A more interesting situation occurs when $\hat m\simeq 1$ (right panel in Fig.~\ref{fig:scenarios}), that is, when both $m$ and $Q$ are UV scales, but comparable in size. In this case one matches QCD$^{(n_f)}$ directly to SCET$^{(n_\ell)}$ integrating out the hard scale and $m$ simultaneously, such that the Wilson coefficient, which we denote $\mathcal{O}^{(n_f)}_{\rm QCD}=C_{\rm SCET}^{(n_f\to n_\ell)}(\hat m^2)\mathcal{O}^{(n_\ell)}_{\rm SCET}$, depends on $Q$ and $m$. In this scenario, the mass appears on the QCD side, but not on the SCET one. The computation can be organized in two equivalent ways as follows:
\begin{equation}\label{eq:Cnfnl}
C_{\rm SCET}^{(n_f\to n_\ell)}(\hat m^2) = \dfrac{\langle \mathcal{O}^{(n_f)}_{\rm QCD}\rangle}{\langle\mathcal{O}^{(n_\ell)}_{\rm SCET}\rangle} =
\left\{\begin{array}{l} \dfrac{\langle \mathcal{O}^{(n_f)}_{\rm QCD}\rangle}{\langle\mathcal{O}^{(n_f)}_{\rm SCET}\rangle}
\dfrac{\langle \mathcal{O}^{(n_f)}_{\rm SCET}\rangle}{\langle\mathcal{O}^{(n_\ell)}_{\rm SCET}\rangle}
= \mathcal{M}^{(n_f\to n_\ell)}_{\rm SCET}C_{\rm SCET}^{(n_f)}(\hat m)\\[0.5cm]
\dfrac{\langle \mathcal{O}^{(n_f)}_{\rm QCD}\rangle}{\langle\mathcal{O}^{(n_\ell)}_{\rm QCD}\rangle}
\dfrac{\langle \mathcal{O}^{(n_\ell)}_{\rm QCD}\rangle}{\langle\mathcal{O}^{(n_\ell)}_{\rm SCET}\rangle}
= \mathcal{M}^{(n_f\to n_\ell)}_{\rm QCD}C_{\rm SCET}^{(n_\ell)}(0)
\end{array}\right.\,,
\end{equation}
implying the following consistency condition $\mathcal{M}^{(n_f\to n_\ell)}_{\rm QCD}C_{\rm SCET}^{(n_\ell)}(0) = \mathcal{M}^{(n_f\to n_\ell)}_{\rm SCET}C_{\rm SCET}^{(n_f)}(\hat m)$. Taking $\hat m^2\to \infty$ one simply has $\mathcal{M}^{(n_f\to n_\ell)}_{\rm SCET} = C_{\rm SCET}^{(n_\ell)}(0)/C_{\rm SCET}^{(n_f)}(\hat m^2\to\infty)$, where we have used the decoupling limit of $\mathcal{M}^{(n_f\to n_\ell)}_{\rm QCD}$. This gives a convenient way of computing $C_{\rm SCET}^{(n_f\to n_\ell)}(\hat m^2)$
\begin{equation}
C_{\rm SCET}^{(n_f\to n_\ell)}(\hat m^2) =C_{\rm SCET}^{(n_\ell)}(0) \dfrac{C_{\rm SCET}^{(n_f)}(\hat m)}{C_{\rm SCET}^{(n_f)}(\hat m^2\to\infty)}\,,
\end{equation}
making the decoupling limit manifest. The Wilson coefficient $C_{\rm SCET}^{(n_f\to n_\ell)}(\hat m^2)$ is used in Scenario~I of Ref.~\cite{Pietrulewicz:2014qza} characterized by the condition $\mu_m>\mu_H$.

\subsubsection{Variable-flavor number scheme}
The punchline of the previous section is that, as long as all subleading terms are kept, the three paths to reach SCET$^{(n_\ell)}$ from QCD$^{(n_f)}$ are equivalent and can be continuously described with a single setup. One simply chooses between the first or second lines of Eq.~\eqref{eq:Cnfnl} depending on the relative size of the hard scale and the mass. This was the basis of the variable-flavor number scheme (VFNS) setup presented in Ref.~\cite{Pietrulewicz:2014qza}, which, with minimal modifications, is sketched in Fig~\ref{fig:scenarioScheme}. If $\mu_H>\mu_J>\mu_S>\mu_m$ (scenario IV), one stops the matching sequence already at SCET$^{(n_f)}$ and computes the jet and soft functions with $\mathcal{O}_{\rm col}^{(n_f)}$ and $\mathcal{O}_{\rm soft}^{(n_f)}$. Hence, the secondary mass is an IR scale that enters those matrix elements, along with the hard function. Provided the common renormalization scale $\mu$ is chosen above $\mu_m$, all RG evolution proceeds with $n_f$ flavors as exposed in Sec.~\ref{sec:massless}. The opposite situation is Scenario I in which $\mu_m > \mu_H>\mu_J>\mu_S$: the EFT operator is $\mathcal{O}_{\rm SCET}^{(n_\ell)}$, hence all matrix elements, along with the hard function, are computed as massless. One needs to include the $|\mathcal{M}_{\rm QCD}^{(n_f\to n_\ell)}|^2$ matching coefficient, and provided $\mu<\mu_m$ all RG evolution involves $n_\ell$ flavors in the manner explained in Sec.~\ref{sec:massless}.

In scenario II, defined by $\mu_H>\mu_m>\mu_J>\mu_S$, the matching coefficient must be computed with mass effects $H(Q^2, m, \mu)=|C_{\rm SCET}^{(n_f)}(Q^2, m,\mu)|^2$. If the choice $\mu_m>\mu>\mu_J$ is made, $\mathcal{O}_{\rm SCET}^{(n_f)}$ is evolved with $n_f$ flavors from $\mu_H$ to $\mu_m$ where one has to add the additional matching coefficient $|\mathcal{M}^{(n_f\to n_\ell)}_{\rm SCET}|^2$, and keep evolving $\mathcal{O}_{\rm SCET}^{(n_\ell)}$ with $n_\ell$ flavors from $\mu_m$ to $\mu$. At this scale, the jet and soft operators defined from fields in $\mathcal{O}_{\rm SCET}^{(n_\ell)}$ individually evolve with $n_\ell$ flavors from $\mu$ to $\mu_J$ and $\mu_S$, respectively, scales at which the jet and soft functions, with $n_\ell$ massless quarks are computed. The last situation to discuss is scenario III, defined by $\mu_H>\mu_J>\mu_m>\mu_S$, that also involves the hard function $|C_{\rm SCET}^{(n_f)}(Q^2, m,\mu)|^2$, and where, provided the common scale satisfies the condition $\mu_J>\mu>\mu_m$, the operator $\mathcal{O}_{\rm SCET}^{(n_f)}$ is evolved from $\mu_H$ to $\mu$ with $n_f$ active flavors. The jet operators are evolved with $n_f$ flavors between $\mu$ and $\mu_J$. Since the jet function is computed with the collinear fields defined in SCET$^{(n_f)}$, the secondary quark is still an active degree of freedom. Since $\mu_m>\mu_S$, to avoid large logarithms it is convenient to integrate the secondary quark from the ultrasoft Lagrangian. Since it is a copy of QCD, one only needs to use $\alpha_s^{(n_\ell)}$. Finally, one needs to match $\mathcal{O}_{\rm soft}^{(n_f)}$ onto $\mathcal{O}_{\rm soft}^{(n_\ell)}$, relation that defines the soft matching coefficient:
\begin{figure}[t]\centering
\includegraphics[width=1\textwidth]{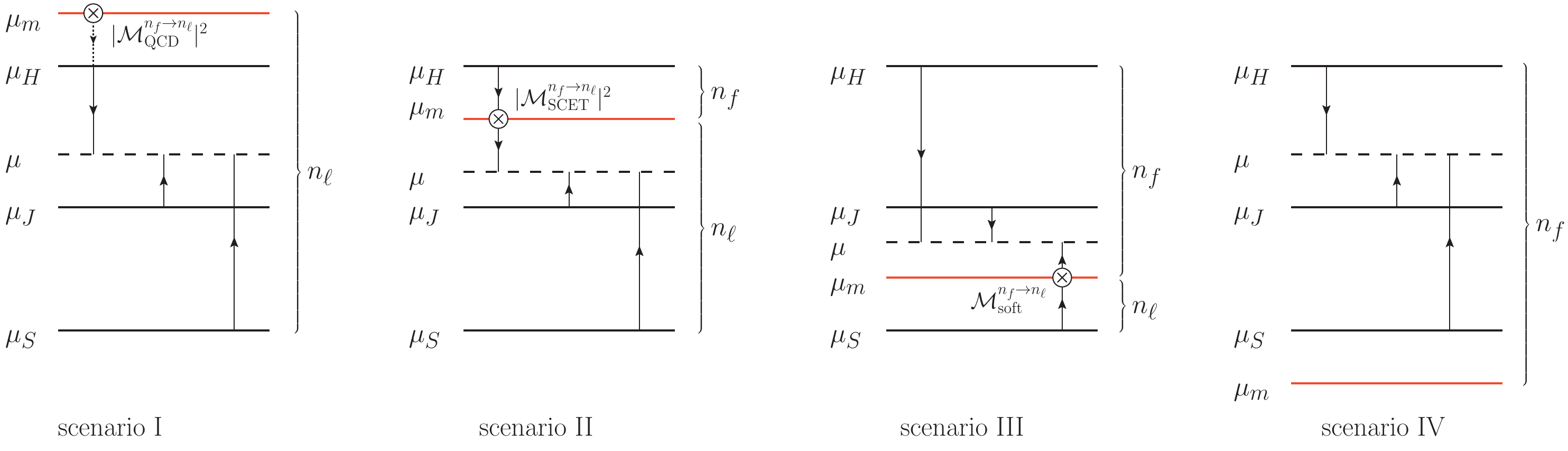}
\caption{Different VFNS scenarios defined by the relative size of the mass-mode scale $\mu_m$ (red horizontal line) and the SCET scales (hard, jet, and soft, solid black horizontal lines). The common renormalization scale $\mu$ is shown as a dashed horizontal line, the matching conditions, when necessary, appear as a cross, and the evolution from the characteristic scales to $\mu$ are indicated with vertical arrows. The dotted vertical arrow signifies that there is no evolution in the $n_\ell$-flavor QCD current. We indicate with curly brackets the number of active flavors participating in the running of each matrix element.
\label{fig:scenarioScheme}}
\end{figure}
\begin{equation}
\mathcal{O}_{\rm soft}^{(n_f)} = \mathcal{M}_{\rm soft}^{(n_f\to n_\ell)} \mathcal{O}_{\rm soft}^{(n_\ell)}\,.
\end{equation}
There is no running associated to the coefficient $\mathcal{M}_{\rm soft}^{(n_f\to n_\ell)}$, and it contains small logarithms provided the matching is performed at the scale $\mu_m$. Hence $\mathcal{O}_{\rm soft}^{(n_f)}(\mu)$ is evolved to $\mu_m$ with $n_f$ flavors, and after matching, $\mathcal{O}_{\rm soft}^{(n_\ell)}$ is evolved from $\mu_m$ to $\mu_S$ with $n_\ell$ flavors. The soft function is computed with $\mathcal{O}_{\rm soft}^{(n_\ell)}$ and hence is insensitive to this massive quark. All in all, the factorization theorem will involve the hard and jet functions with $n_f$ flavors and mass effects, a massless soft function with $n_\ell$ flavors and the soft matching condition $\mathcal{M}_{\rm soft}^{(n_f\to n_\ell)}$.

For other choices of the common renormalization scale $\mu$ we might also need to match $\mathcal{O}_{\rm col}^{(n_f)}$ onto $\mathcal{O}_{\rm col}^{(n_\ell)}$ (the collinear Lagrangian is simply a boosted copy of the QCD one, so once more the strong coupling is the only modification), defining the jet matching coefficient
\begin{equation}
\mathcal{O}_{\rm col}^{(n_f)} = \mathcal{M}_{\rm jet}^{(n_f\to n_\ell)} \mathcal{O}_{\rm col}^{(n_\ell)}\,.
\end{equation}
From Eq.~\eqref{eq:scetMatch} one has the condition $\mathcal{M}^{(n_f\to n_\ell)}_{\rm soft}(\mu)|\mathcal{M}^{(n_f\to n_\ell)}_{\rm jet}(\mu)|^2=|\mathcal{M}^{(n_f\to n_\ell)}_{\rm SCET}(\mu)|^2$. The jet matching coefficient can also be determined comparing the jet functions computed in the $n_f$ and $n_\ell$ theories, approach followed in this article.

\begin{figure*}[t!]
\subfigure[]
{\includegraphics[width=0.49\textwidth]{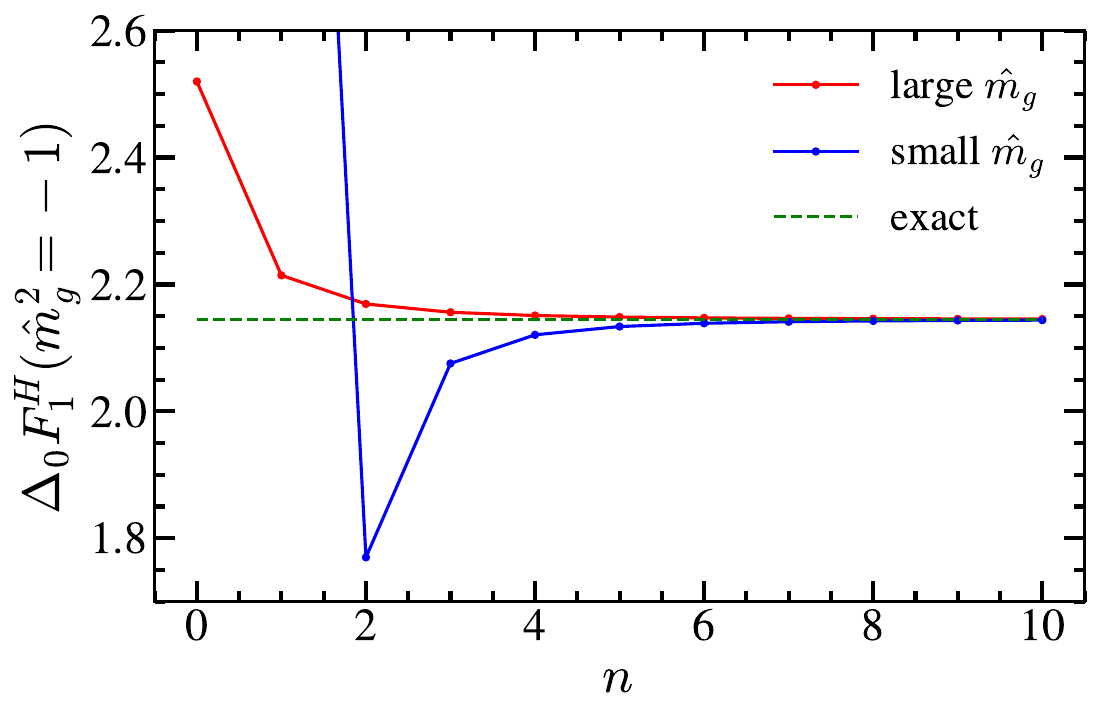}
\label{fig:SCETMg1}}
\subfigure[]{\includegraphics[width=0.462\textwidth]{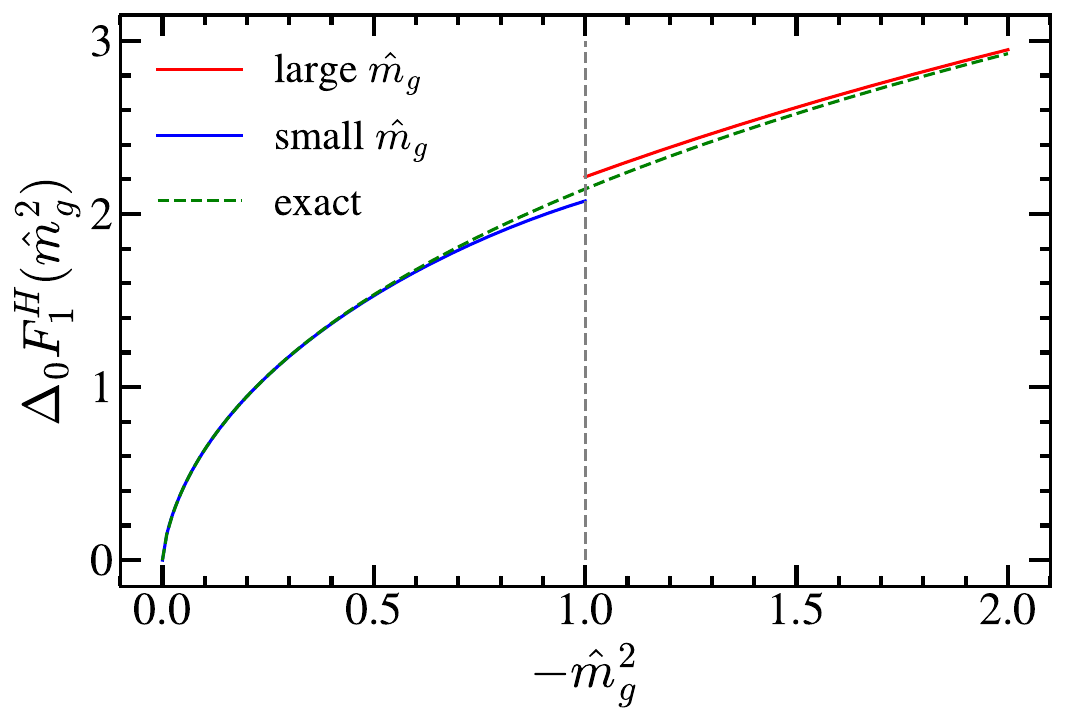}
\label{fig:SCETMg}}
\caption{Gluon mass correction to the massless SCET Wilson coefficient in its exact form (dashed green), small- (blue) and large-mass (red) expansions. Left panel: $\Delta_0 F_1^H$ at the boundary between the mass expansions $\hat m_g^2 = -1$, as a function of the expansion order $n$ of each series. Right panel: Dependence of $\Delta_0 F_1^H(\hat m_g^2)$ with the reduced gluon mass, including $3$ and $2$ non-zero terms in the small- and large-mass expansions, respectively.}
\label{fig:MgSCET}
\end{figure*}
\subsubsection{Massive vector Boson}
In this and the following section we label results with the superscript $H$, since from this coefficient the hard factor can be obtained. The Mellin transform reads:
\begin{equation}
\mathcal{M}^H_1(h,\hat{m}_g^2,0) = - \frac{\Gamma^2(h) \Gamma^2(1-h)}{h (h + 1) (h + 2)} (-\hat{m}_g^2)^{- h}\xrightarrow[| h | \gg 1]{} - \frac{\pi^2\! \csc^2 (\pi h)}{h^3} (-\hat{m}_g^2)^{- h}\,.
\end{equation}
From this result we can read off the convergence radius: $\hat m_g$ smaller or larger than $1$ call for small- or large-mass expansions. This result can be confirmed using the Cauchy's root test on the general terms shown in Eq.~\eqref{eq:SCETMg}. In fact, we have used Cauchy's root test in every single series presented in this article, and confirmed that the convergence radius coincides with that obtained inspecting the large $h$ behavior of the Mellin transform. As shown in Fig.~\ref{fig:SCETMg1}, at the boundary one can use either expansion, and both series converge equally fast.
The rest of relevant results before showing the expansions are
\begin{align}\label{eq:SCETmg}
\delta Z_1^{\rm SCET}\biggl(\frac{Q^2}{\mu^2},\varepsilon\biggr)&= -C_F\biggl[\frac{2}{\varepsilon^2} + \frac{1}{\varepsilon}\bigl( 2 L_\mu + 3 \bigr)\biggr]\,,\\
F^H_{1,\rm ren}\biggl(0,\frac{Q^2}{\mu^2}\biggr) &= - \frac{1}{4}\! L_\mu^2 -\frac{3}{4}\! L_\mu + \frac{\pi^2}{24} - 2\,,\nonumber\\
\Delta_0^{\infty} F_1^H(\hat m_g^2) & = \frac{1}{4}\! \log^2 (- \hat{m}_g^2) + \frac{3}{4}\! \log (-\hat{m}_g^2) + \frac{\pi^2}{6} + \frac{7}{8} \,,\nonumber
\end{align}
where we have displayed the one-loop massless limit already renormalized, which agrees with the well-known result of Ref.~\cite{Fleming:2007qr}. The divergences shown in the second line make clear the result does not correspond to the full-theory computation, since a massive vector boson yields a UV- and IR-finite result and does not need any regulator. From these we obtain the one-loop cusp and non-cusp anomalous dimensions using Eq.~\eqref{eq:GammaSCET}:
\begin{equation}\label{eq:Cusp1loop}
\Gamma_{\!0} =\ 4 C_F= \frac{16}{3}\,,\qquad
\gamma_{0}^{\rm SCET} =-6C_F=-8\,.
\end{equation}
Furthermore, the condition $\delta Z_{1,1,\rm cusp}^{\rm SCET} = \delta Z_{1,2,\rm nc}^{\rm SCET}=-\Gamma_{\!0}/2$ is verified. The two expansions for the correction to the massless limit are computed easily noting that there are double poles at all negative and positive integer values of $h$, except for $h=-1$ and $h=-2$ that are triple. Once again, the factor containing all gamma functions in $\mathcal{M}_1$ is symmetric under reversing the sign of $h$, thence a symmetry between the infinite sums will be manifest. We found
\begin{align}\label{eq:SCETMg}
\Delta_0 F_1^H(\hat m_g^2) = \,& \Delta_0^\infty F_1^H - \sum_{n = 1} G^H_n(\hat m_g^2)
= \sum_{n = 3} G^H_{-n}(\hat m_g^2) - \hat{m}_g^2 \biggl[ \frac{1}{2} \log^2 (- \hat{m}_g^2) + \frac{\pi^2}{3} + 1 \biggr] \nonumber \\
& - \hat{m}_g^4 \biggl[ \frac{1}{4} \log^2 (- \hat{m}_g^2) - \frac{3}{4} \log (- \hat{m}_g^2) + \frac{\pi^2}{6} + \frac{7}{8} \biggr] \nonumber \,,\\
G^H_{n}(\hat m_g^2) = \, & \frac{(- \hat{m}_g^2)^{- n}}{n (n + 1) (n + 2)} \biggl[ \log(- \hat{m}_g^2) + \frac{3 n^2 + 6 n + 2}{n (n + 1) (n + 2)} \biggr]\,.
\end{align}
The infinite sum can be carried out analytically:
\begin{align}\label{eq:anSumF}
\Delta_0 F_1^H(\hat m_g^2) = \,& \frac{\pi^2}{12}
- \frac{1}{2} \bigl( 1 + \hat{m}_g^2 \bigr)^{2}\, \text{Li}_2 \bigl( 1 + \hat{m}_g^2 \bigr) -\frac{\hat{m}_g^2}{2} \bigl[\log (- \hat{m}_g^2) + 1\bigr] \\
& - \frac{\hat{m}_g^2}{12}\bigl(2+ \hat{m}_g^2 \bigr) \bigl[ \pi^2 + 3 \log^2 (- \hat{m}_g^2)\bigr] . \nonumber
\end{align}
For $\hat m_g > 0$ the expression above develops an imaginary term. To take the real part, relevant to obtain the hard factor, one simply has to make the following replacements:
\begin{align}
\text{Li}_2 \bigl( 1 + \hat{m}_g^2 \bigr) \to\, & \frac{\pi ^2}{6}-\text{Li}_2\bigl(-\hat{m}_g^2\bigr)-\log (1 + \hat{m}_g^2) \log (\hat{m}_g^2)\,,\\
\log^n (- \hat{m}_g^2) \to\, & {\rm Re}\bigl\{\bigl[\,\log (\hat{m}_g^2) + i\pi\bigr]^n\bigr\}.\nonumber
\end{align}
The results for the exact result and expansions are shown in Fig.~\ref{fig:MgSCET}. The respective approximations can be made arbitrarily precise adding more and more terms. Obtaining expansions for the hard matching coefficient poses no difficulty, but to avoid cluttering we do not show them.

At this point we can split the Wilson coefficient in the QCD and SCET terms. In order for that, we use that the QCD form factor must vanish in the decoupling limit $m_g\to \infty$. Hence $F_1^{\rm QCD}(\hat m_g^2)=\Delta_\infty F_1^H(\hat m_g^2) = \Delta_0 F_1^H(\hat m_g^2) - \Delta_0^\infty F_1^H(\hat m_g^2)$, and using the result in Eq.~\eqref{eq:anSumF} we find full agreement with Refs.~\cite{Kniehl:1988id,Hoang:1995ex,Hoang:1995fr}. Note that the $\hat m_g^2\to 0$ limit of $F_1^{\rm QCD}$ diverges, signaling the need for an IR regulator. Finally, we can obtain the bare SCET form factor which contains all UV divergences but is IR finite. It takes the following simple form $F_1^{\rm SCET} = F_1^{\rm QCD} - F_1^H = - \Delta_0^\infty F_1^H - F_1^H(0,Q,\mu) = -F_1^H(\hat m_g^2\to\infty,Q,\mu)$, and corresponds to the contribution of the residue at $h = \varepsilon$:
\begin{align}
F_1^{\rm SCET}\!\biggl(\hat m_g^2,\frac{Q^2}{\mu^2},\varepsilon\biggr) \! =\,& \frac{\Gamma (\varepsilon)}{2} \biggl(\frac{\mu^2 e^{\gamma_E}}{m_g^2} \biggr)^{\!\!\varepsilon} \biggl[ H_{1 -\varepsilon} + \log \bigl( - \hat m_g^2\bigr) + \pi \cot (\pi\varepsilon) + \frac{1 + \varepsilon - \varepsilon^2}{(1 - \varepsilon) (2 -\varepsilon)} \biggr]\nonumber\\
\equiv\, & F_{1,{\rm ren}}^{\rm SCET}\biggl(\hat m_g^2,\frac{Q^2}{\mu^2}\biggr) - \frac{\delta Z_1^{\rm SCET}\!\Bigl(\frac{Q^2}{\mu^2},\varepsilon\Bigr)}{4C_F}+ \mathcal{O}(\varepsilon)\,,\nonumber\\
F_{1,{\rm ren}}^{\rm SCET}\biggl(\hat m_g^2,\frac{Q^2}{\mu^2}\biggr) \!=\,& -\! \frac{1}{4} \!\log^2 \bigl( - \hat m_g^2\bigr) - \frac{3}{4} \!\log\bigl( - \hat m_g^2\bigr)
+ \frac{1}{4}L_\mu^2+ \frac{3}{4}L_\mu - \frac{5 \pi^2}{24} + \frac{9}{8} \,,
\end{align}
where the harmonic number with a non-integer argument can be expressed in terms of the digamma function $\psi ^{(0)}$ ---\,the derivative of the logarithm of the gamma function\,--- as follows: $H_{1- \varepsilon}=\psi ^{(0)}(2-\varepsilon )+\gamma_E$. One can also relate the cotangent to digamma functions: $\pi \cot (\pi\varepsilon)=\psi ^{(0)}(1-\varepsilon )-\psi ^{(0)}(\varepsilon)$. Our unexpanded expression coincides with Eq.~(321) of Ref.~\cite{SimonThesis}, and our expanded result agrees with Ref.~\cite{Chiu:2009yx}.

We close the section computing the matching coefficient between QCD/SCET with massive and massless gluons (operators labeled with a superscript $n_g$) and QCD/SCET with only massless gluons (operators labeled with a superscript $n_\ell$), defined by the relation $\mathcal{O}_{\rm{th}}^{(n_g)} = \mathcal{M}_{\rm th}^{(n_g\to n_\ell)}\mathcal{O}_{\rm{th}}^{(n_\ell)}$ on renormalized operators, where th$\,=\,$QCD or SCET. Since the contribution of massless gluons is the same in both theories, the matching coefficient is simply the massive vector boson contribution to the QCD/SCET form factor (the latter renormalized), since the strong coupling is the same in both theories at one-loop:
\begin{align}
\mathcal{M}^{(n_g\to n_\ell)}_{\rm QCD}(\hat m_g^2,\mu) =\,& 1 + \frac{\alpha_s(\mu)}{\pi} C_F F_1^{\rm QCD}(\hat m_g^2)\,,\\
\mathcal{M}_{\rm SCET}^{(n_g\to n_\ell)}(m_g, Q^2,\mu) =\,& 1+ \frac{\alpha_s(\mu)}{\pi}C_F F_{1,{\rm ren}}^{\rm SCET}\biggl(\hat m_g^2,\frac{Q^2}{\mu^2}\biggr)\,,\nonumber
\end{align}
where
$\mathcal{M}_{\rm SCET}^{(n_g\to n_\ell)}$ agrees with Eq.~(29) of Ref.~\cite{Gritschacher:2013pha}.

\subsubsection{Secondary massive bubble}
\begin{figure*}[t!]
\subfigure[]
{\includegraphics[width=0.49\textwidth]{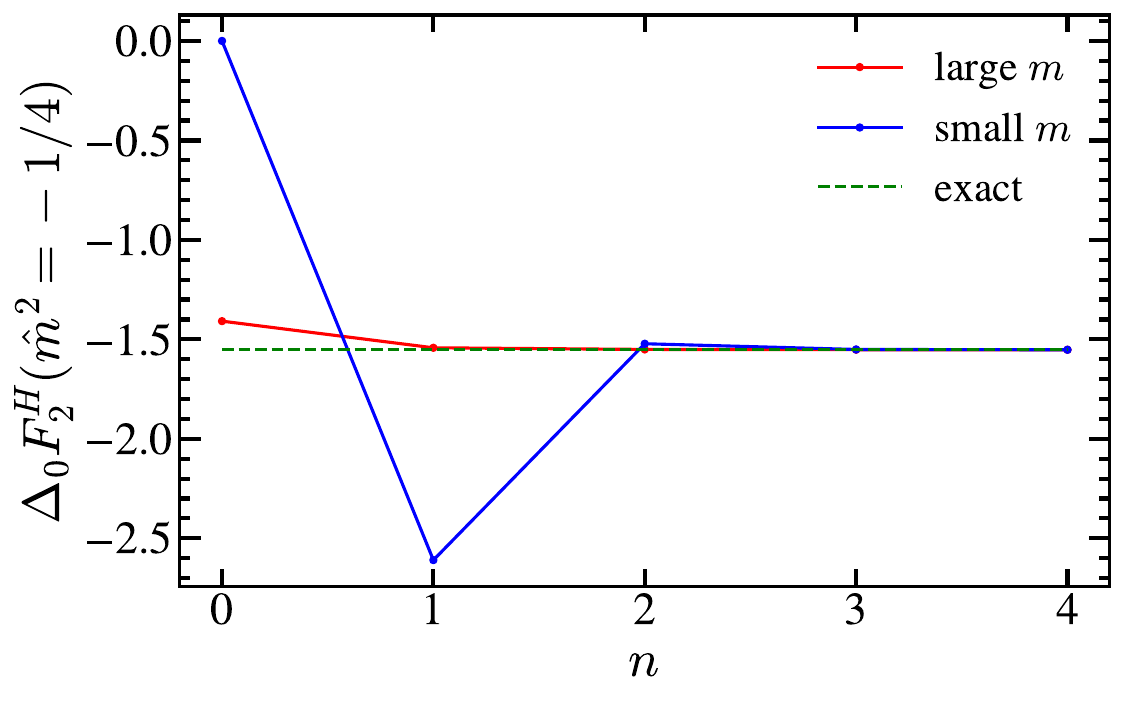}
\label{fig:SCETSec1}}
\subfigure[]{\includegraphics[width=0.49\textwidth]{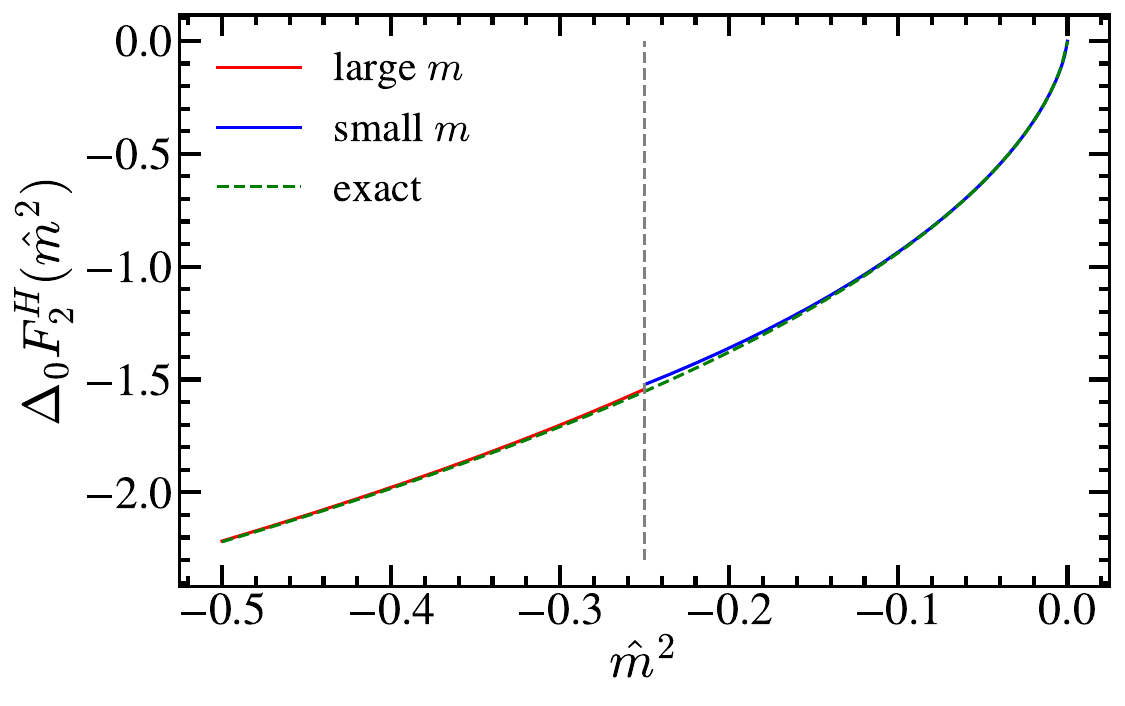}
\label{fig:SCETSec}}
\caption{Secondary mass correction to the massless SCET Wilson coefficient in its exact form (dashed green), small- (blue) and large-mass (red) expansions. Left panel: $\Delta_0 F_2^H$ at the boundary between the mass expansions $\hat m^2 = -1/4$, as a function of the expansion order $n$ of each series. Right panel: Dependence of $\Delta_0 F_2^H(\hat m^2)$ with the reduced secondary quark mass, including $2$ non-zero terms in each expansion.}
\label{fig:SecSCET}
\end{figure*}

The relevant results for the QCD to SCET matching coefficient with massive corrections are in this case
\begin{align}\label{eq:HSec}
\mathcal{M}^H_2(h,\hat m^2,0) &\, = - \bigl( -\hat m^2 \bigl)^{\!-h}
\frac{4 (h + 1) \Gamma^2(1- h)\Gamma^4(h)}{\Gamma (2 h + 5)}
\xrightarrow[| h | \gg 1]{} - \frac{\pi^{5 / 2} \csc^2 (\pi h)}{4 h^{9 / 2}} \bigl( -4\hat m^2 \bigl)^{\!-h}\,,\nonumber\\
\delta Z_{2,n_f}^{\rm SCET}\biggl(\frac{Q^2}{\mu^2},\varepsilon\biggr)&\,= -16C_FT_F \biggl[\frac{1}{8 \varepsilon^3} +\frac{1}{\varepsilon^2} \biggl( \frac{1}{12} L_\mu +\frac{1}{18} \biggr)
- \frac{1}{\varepsilon} \biggl( \frac{5}{36}L_\mu+\frac{\pi ^2}{48}+\frac{65}{432}\biggr)\biggr]\,,\nonumber\\
F^H_{2,\rm ren}\biggl(0,\frac{Q^2}{\mu^2}\biggr) &\,= \frac{1}{36}L_\mu^3+\frac{19}{72}L_\mu^2+\biggl(\frac{209}{216}+\frac{\pi ^2}{36}\biggr)
L_\mu+\frac{\zeta_3}{36}+\frac{23 \pi ^2}{432}+\frac{4085}{2592}\,,\nonumber\\
\Delta_0^\infty F_2^H(\hat m^2) & \,= -\frac{1}{36}L_m^3-\frac{19}{72}L_m^2-\biggl(\frac{265}{216}+\frac{\pi^2}{36}\biggr) L_m+\frac{\zeta_3}{3}-\frac{19 \pi ^2}{216}-\frac{3355}{1296}\,.
\end{align}
with $L_m=\log(-\hat m^2)$. Our results for the massless limit and the renormalization factor agree with Ref.~\cite{Matsuura:1987wt}. The latter satisfies Eqs.~\eqref{eq:UVcond} and \eqref{eq:highest} with the substitution $\beta_0\to \beta_0^{(n_f)}$. Using Eq.~\eqref{eq:GammaSCET} we obtain the two-loop anomalous dimension
\begin{align}\label{eq:Cusp2loop}
\Gamma^{(n_f)}_{\!1} =\,& -\!\frac{80}{9}C_F T_F = -\frac{160}{27}\,,\\
\gamma_{1,n_f}^{\rm SCET} =\,&64 C_FT_F\biggl(\frac{\pi ^2}{48}+\frac{65}{432}\biggr) = \frac{520}{81}+\frac{8 \pi ^2}{9}\,.\nonumber
\end{align}
From the first line of Eq.~\eqref{eq:HSec} it becomes clear that the small- and big-mass expansions converge for $4m^2\leq|Q|^2$ and $4m^2\geq|Q|^2$, respectively. Fast convergence at $4m^2=|Q|^2$ is expected from any of the two expansions, as can be seen in Fig.~\ref{fig:SCETSec1}. After removing the massless limit one can set $\varepsilon=0$ and examine the pole structure of the Mellin transform. On the positive real axis there are double poles at all integer values of $h$. On the negative real axis one has a quartic pole at $h = - 2$ and triple poles at $h = - n$ with $n\neq 2$. We can use the inverse mapping theorem to obtain the corresponding expansions for $\Delta_0 F_2$:
\begin{align}
\Delta_0 F_2^H(\hat m^2) =\,& \Delta^\infty_0 F_2^H(\hat m^2) + 4 \sum_{n = 1} \frac{(n + 1) [(n-1)!]^2 }{ (2 n + 4)!} ( -\hat m^2)^{-n} \biggl[ 2 (H_{n - 1} - H_{2 n + 1}) \\
&- L_m - \frac{7 + 4n}{(n + 2) (2 n + 3)} \biggr]\nonumber\\
=\,& \hat m^4\biggl(L_m^2 -\frac{1}{6} L_m^3 -\frac{33 + 2 \pi^2}{12} L_m + \frac{13}{4}+ \frac{\pi^2}{3} + 2 \zeta_3 \biggr)\nonumber\\
& + 2 \sum_{n = 1}^{n\neq 2}
\frac{ (2 n - 2)! \bigl( -\hat m^2\bigr)^n}{(n - 2) (2 n - 3) (n!)^2} \biggl[ \frac{2(4 n - 7)}{(n - 2) (2 n - 3)}\biggl( H_{2 n - 2} - H_n + \frac{1}{2} L_m \biggr) \nonumber\\
& - 2 \biggl( H_{2 n - 2} - H_n + \frac{1}{2} L_m \biggr)^{\!\!2} + 2 H_{2 n - 2}^{(2)} - H_n^{(2)} - \frac{\pi^2}{6} - \frac{37 - 42 n + 12 n^2}{(n - 2)^2 (2 n - 3)^2} \biggr] ,\nonumber
\end{align}
where $H^{(2)}_k=\sum_{i=1}^k n^{-2}$ is the harmonic number of order $2$. The infinite sum for the large-mass expansion can be summed up and we obtain the following result:
\begin{align}
\Delta_0 F_2^H(\hat m^2) =\, & \biggl( \frac{23 r^2}{72} + \frac{5}{24} \biggr) r \biggl[ \text{Li}_2 \biggl( \frac{r - 1}{r + 1} \biggr) - \text{Li}_2 \biggl( \frac{r + 1}{r - 1} \biggr) \biggr]
+ \biggl( \frac{55 r^2}{72} + \frac{25}{54} \biggr) L_m \\
& + \biggl( \frac{5}{48} - \frac{r^4}{16} + \frac{r^2}{8} \biggr) \biggl[ \text{Li}_3 \biggl( \frac{r - 1}{r + 1} \biggr) + \text{Li}_3 \biggl( \frac{r + 1}{r - 1} \biggr) - 2 \zeta_3 \biggr]
+ \frac{119 r^2}{72} \nonumber\\
&-\frac{L_m^3}{36}-\frac{19 L_m^2}{72}-\biggl(\frac{265}{216}+\frac{\pi
^2}{36}\biggr) L_m+\frac{\zeta_3}{3}-\frac{19 \pi ^2}{216}-\frac{119}{72}\,, \nonumber
\end{align}
with $r=\sqrt{1+4\hat m^2}$, in agreement with Ref.~\cite{Kniehl:1989kz,Hoang:1995fr}. In the equation above all terms are manifestly real for $0>\hat m^2>-1/4$. For $\hat m^2 < - 1/4$ some terms develop imaginary parts, but $\Delta_0 F_2^H(\hat m) $ is still real-valued. To have every term explicitly real in this case we simply make the following substitutions:
\begin{align}
r \biggl[ \text{Li}_2 \biggl( \frac{r - 1}{r + 1} \biggr) - \text{Li}_2 \biggl( \frac{r + 1}{r - 1} \biggr) \biggr] \rightarrow\, &
- 2 \sqrt{-1-4\hat m^2} \, {\rm Cl}_2 \!\biggl[ \arccos \biggl( \frac{r^2 + 1}{r^2 - 1} \biggr) \biggr], \\
\text{Li}_3 \biggl( \frac{r - 1}{r + 1} \biggr) + \text{Li}_3 \biggl( \frac{r + 1}{r - 1} \biggr) \rightarrow\, &
2 \,{\rm Cl}_3 \!\biggl[ \arccos \biggl( \frac{r^2 + 1}{r^2 - 1} \biggr) \biggr] ,
\nonumber
\end{align}
where the Clausen functions are defined as infinite sums: ${\rm Cl}_2(\alpha) = \sum_{k = 1} \sin (k \alpha)/k^2$ and ${\rm Cl}_3(\alpha) = \sum_{k = 1} \cos (k \alpha)/k^3$.
For $\hat m^2>0$ a genuine imaginary part is generated. To obtain the real part (which is most relevant to compute the hard factor), one has to do the following replacements:
\begin{align}
\text{Li}_2 \biggl( \frac{r - 1}{r + 1} \biggr) - \text{Li}_2 \biggl( \frac{r + 1}{r - 1} \biggr) \to \,& 2 \text{Li}_2 \biggl( \frac{r - 1}{r + 1} \biggr) + \frac{1}{2} \log^2 \biggl(\frac{r - 1}{r + 1} \biggr)
- \frac{\pi^2}{3} \,,\\
\text{Li}_3 \biggl( \frac{r - 1}{r + 1} \biggr) + \text{Li}_3 \biggl( \frac{r + 1}{r - 1} \biggr) \to \,& 2 \text{Li}_3 \biggl( \frac{r - 1}{r + 1} \biggr) - \frac{1}{6} \log^3 \biggl(\frac{r + 1}{r - 1} \biggr) + \frac{\pi^2}{3} \log \biggl( \frac{r + 1}{r - 1}\biggr)\,, \nonumber\nonumber
\end{align}
along with $L_m^n \to {\rm Re}[\,\log(\hat m^2) + i \pi]^n$. Obtaining the expansions for the SCET hard factor is trivial from the results quoted in this section and to avoid cluttering these will not be explicitly shown.

A comparison of the small- and large-mass expansions to the summed-up result is shown in Fig.~\ref{fig:SecSCET}, where one can observe that both expansions converge very fast, especially for large masses, where including only two terms is enough to achieve sub-percent accuracy everywhere the sum converges. The dispersive contribution to the two-loop QCD form factor is IR-finite and given by $H^{\rm {QCD}}_{2}(\hat m^2)=F^H_2(\hat m^2) - F^H_2(\hat m^2\to\infty)$. The dispersive contribution to the SCET form factor is also IR finite and obtained as the following limit: $F^{\rm {SCET}}_{2, \rm{disp}}=-F_2^H(\hat m^2 \to \infty)$, which is simply the pole at $h=\varepsilon$ of the $H_2$ contribution,
\begin{align}
H^{\rm {SCET}}_{2}\!\biggl(\hat m^2,\frac{Q^2}{\mu^2}\varepsilon\biggr)\!=\, &\frac{1}{8 \varepsilon^3} + \frac{1}{\varepsilon ^2}\biggl(\frac{L_\mu}{4}-\frac{L_m}{6}+\frac{1}{18}\biggr)+
\frac{1}{\varepsilon}\biggl[\frac{L_\mu^2}{4}+\frac{L_m^2}{12}+L_\mu \biggl(\frac{1}{9}-\frac{L_m}{3}\biggr) \\
&-\frac{L_m}{4}-\frac{\pi^2}{144}-\frac{65}{432}\biggr]
+\frac{L_\mu^3}{6}+L_\mu \biggl(\frac{L_m^2}{6}-\frac{L_m}{2}-\frac{\pi ^2}{72}-\frac{65}{216}\biggr)
\nonumber\\
& +\frac{7 L_m^2}{18}+L_\mu^2 \biggl(\frac{1}{9}-\frac{L_m}{3}\biggr)+L_m\biggr(\frac{121}{216}+\frac{\pi ^2}{36}\biggl)-\frac{5 \zeta_3}{12}+\frac{\pi ^2}{18}+\frac{875}{864}\,.\nonumber
\end{align}
Our result agrees with Ref.~\cite{SimonThesis}. We can obtain the matching SCET decoupling coefficient introducing the finite part of the result above in Eq.~\eqref{eq:SCET2SCET}:
\begin{align}\label{eq:SCETMatching}
\mathcal{M}^{(2)}_{\rm SCET}\!\biggl(\hat m^2,\frac{Q^2}{\mu^2}\biggr)\!\!=\,& \frac{L_\mu^3}{18}+\frac{1}{36}L_m^3+\frac{19}{72}L_m^2-L_\mu^2 \biggl(\frac{L_m}{12}+\frac{1}{72}\biggr)-L_\mu
\biggl(\frac{L_m}{4}+\frac{65}{216} +\frac{\pi ^2}{24} \biggr)\nonumber \\
& +\biggl(\frac{121}{216}+\frac{\pi ^2}{24}\biggr) L_m-\frac{13 \zeta_3}{36}+\frac{875}{864}+\frac{5 \pi ^2}{144}\,,
\end{align}
in agreement with Ref.~\cite{Pietrulewicz:2014qza} if setting $\mu_m=\mu_H$.

\subsection{Jet function}\label{sec:JetSCET}
In this section we present results for the SCET single-hemisphere jet function, which appears in the factorization theorems for $2$-jettiness and $C$-jettiness, modified versions of thrust and C-parameter which are designed to enhance the sensitivity to quark masses. The momentum-space jet function contains distributions: Dirac delta and plus functions. Some of these might get obscured when expanding in big and small masses, therefore we compute the cumulative jet function defined as
\begin{equation}
\Sigma_n^J(s_c,\mu) = \int_0^{s_c} {\rm d}s\,J_n(s,\mu)\,,
\end{equation}
to obtain those, and provide the expansions for the non-distributional terms of the differential jet function. For either secondary massive quarks or massive vector bosons, the jet function has real and virtual radiation contributions. The virtual contains only distributions that become singular at $s=0$, while the real radiation has only non-distributional terms. The virtual correction is easy to obtain since for large (gluon or secondary quark) masses one cannot radiate a massive particle any more. Hence, the expansion for large masses will be given by the residue of a single pole sitting at $h=\varepsilon$. The non-distributional terms (which are proportional to a Heaviside theta function) are simply obtained as the sum of residues on the real non-positive axis, from which one must subtract the radiative tails of the plus distributions coming from the virtual diagrams.

The renormalization of the jet function takes place through the convolution of a $Z$ factor, which splits the bare result in its divergent part and the renormalized jet function:
\begin{align}\label{eq:ZJet}
J_n^{\rm bare}(s) =\, & \int {\rm d}s' Z_J(s'-s,\mu) J_n(s',\mu)\,, \\
Z_J(s,\mu) \doteq\, & \delta(s) + \frac{\alpha_s(\mu)}{\pi}C_F\delta Z_1^J + \biggl[\frac{\alpha_s(\mu)}{\pi}\biggr]^2C_FT_F\delta Z_2^J \,,\nonumber\\
\delta Z_i^J =\, & \delta Z_{i,\rm nc}^J \delta(s) + \frac{\delta Z_{i,\rm cusp}^J}{\mu^2}\biggl[\dfrac{\mu^2}{s}\biggr]_{+} \nonumber\,,\\
\delta Z_{i,\rm cusp}^J =\, & \sum_{j=1}^i \frac{\delta Z_{i,j,\rm cusp}^J}{\varepsilon^j}\,,\nonumber\\
\delta Z_{i,\rm nc}^J =\, & \sum_{j=1}^{i+1} \frac{\delta Z_{i,j,\rm nc}^J}{\varepsilon^j}\,.\nonumber
\end{align}
The renormalized jet function obeys an RGE equation which takes the form of a convolution, where the anomalous dimension is also a distribution:
\begin{align}\label{eq:JetGamma}
\mu \dfrac{\rm d}{{\rm d}\mu} J_n(s,\mu) \, = & \int {\rm d}s' \gamma_J(s-s') J_n(s',\mu)\,,\\
\gamma_J(s,\mu) \, = & -\! \frac{2\Gamma_{\rm cusp}}{\mu^2}\biggl[\dfrac{\mu^2}{s}\biggr]_{+} + \gamma^J_{\rm nc} \delta(s)\,,\nonumber\\
\gamma_{\rm nc}^J \, = &\sum_{n = 1} \gamma^J_n \biggl[\frac{\alpha_s (\mu)}{4 \pi} \biggr]^n .\nonumber
\end{align}
To derive $\gamma_J(s,\mu)$ from $Z_J(s,\mu)$ it should be noted that the derivative of the plus distribution appearing in the third line of Eq.~\eqref{eq:ZJet} with respect to $\log(\mu)$ equals $-2\delta(s)$. Assuming that the cancellation of UV-divergent terms identical to those appearing in Eq.~\eqref{eq:UVcond} takes place, the anomalous dimensions are proportional to the $1/\varepsilon$ terms in the $Z$ factor:
\begin{align}\label{eq:JetGammaZ}
\Gamma_0 =\, & -\!4 C_F\delta Z_{1,1,\rm cusp}^{J}\,, &\Gamma_1^{(n_f)} &= - 32 C_FT_F\delta Z_{2,1,\rm cusp}^{J}\,,\\
\gamma_{0}^J =\, & 8 C_F\delta Z_{1,1,\rm nc}^{J}\,, &\gamma_{1,n_f}^{J} &= 64 C_FT_F\delta Z_{2,1,\rm nc}^{J}\,,\nonumber
\end{align}

For the one-loop computation of $\Sigma_n^J$ with a shifted gluon propagator we have $\mathcal{Q}^2=s_c$, and the following result was found in Ref.~\cite{Gracia:2021nut}:
\begin{equation}
m_1 (h,\varepsilon) = \frac{1}{2} \frac{\Gamma (2 - \varepsilon)}{(h - \varepsilon) \Gamma (1-h)\Gamma (3 + h - 2 \varepsilon)}
\biggl[ 5 - \varepsilon + \frac{2 (2 - h)}{h -\varepsilon} \biggr] ,
\end{equation}
from which we observe a double pole sits at $h=\varepsilon$. We label quantities related with the differential and cumulative jet functions with $J$ and $\Sigma_J$ superscripts, respectively. The (dimensionless) Fourier transform of the SCET jet function is defined as
\begin{equation}\label{eq:jet-fourier}
\tilde{J}_n (y, \mu) = \!\int_{0}^{\infty} \!\!\text{d} s\, e^{- i s y}J_n ( s, \mu) \,.
\end{equation}

\subsubsection{Massive vector Boson}
To simplify the notation we define the dimensionless and positive-definite variable \mbox{$\bar s_g= s/m_g^2$}, which will be used in the differential and cumulative versions. The relevant results for the jet function are:
\begin{align}
\mathcal{M}^{\Sigma_J}_1(h,\bar{s}_g,0) & = \frac{(3 h + 4)\bar s_g^h}{2 h^3 (2 + h) (1 + h)} \xrightarrow[| h | \gg 1]{} \frac{3\bar s_g^h}{2 h^4} \,,\\
\delta Z_J^1(s,\mu,\varepsilon) & = \biggl(\frac{1}{\varepsilon^2} + \frac{3}{4\varepsilon}\biggr) \delta(s)- \frac{1}{\varepsilon}\frac{1}{\mu^2} \biggl[ \frac{\mu^2}{s} \biggr]_+\,,\nonumber\\
F^J_{1,\rm ren}(0,s,\mu) &\,=\frac{1}{\mu^2} \biggl[ \frac{\mu^2}{s} \log\biggl(\frac{s}{\mu^2}\biggr) \biggr]_+ - \frac{3}{4} \frac{1}{\mu^2} \biggl[ \frac{\mu^2}{s} \biggr]_+
+\biggl( \frac{7}{4}- \frac{\pi^2}{4} \biggr)\delta(s) \,,\nonumber\\
m_g^2\Delta^\infty_0 F^J_1(s, m_g) &\,= - \biggl[ \frac{\log(\bar s_g)}{\bar s_g} \biggr]_+
+ \frac{3}{4} \biggl[ \frac{1}{\bar s_g} \biggr]_+- \frac{5}{8} \delta(\bar s_g)\,.\nonumber
\end{align}
Using Eq.~\eqref{eq:JetGammaZ} on the second line of the previous equation we recover the one-loop cusp anomalous dimension shown in Eq.~\eqref{eq:Cusp1loop}, along with the non-cusp jet anomalous dimension coefficient $ \gamma_1^J = 6C_F = 8$.

The Mellin-Barnes transform for the differential jet function is trivially obtained applying a derivative: $\mathcal{M}^{J}_1(h,\bar{s}_g,0) = h \mathcal{M}^{\Sigma_J}_1(h,\bar{s}_g,0)/s$.
Interestingly, after setting $\varepsilon=0$ there is a finite number of poles on each side of the real axis. The virtual contribution is simply the pole at $h=\varepsilon$, and prior to renormalization we find
\begin{align} F_1^{J,\rm virt} = \,& \frac{ \Gamma (\varepsilon) }{2 } \biggl( \frac{\mu^2 e^{\gamma_E}}{m_g^2} \biggr)^{\!\!\varepsilon}
\biggl\{ \biggl[ 2\pi \cot (\pi \varepsilon ) - 2\log \biggl( \frac{\mu^2}{m_g^2} \biggr) +\frac{3-\varepsilon^2}{(2-\varepsilon) (1-\varepsilon)} \biggr] \delta (s) - \frac{2}{\mu^2}\!
\biggl[ \frac{\mu^2}{s} \biggr]_{+} \biggr\}\nonumber\\
= \,&\delta Z_J^1(s,\mu,\varepsilon) + F^{J,\rm virt}_{1,\rm ren}(s,m_g,\mu) +\mathcal{O}(\varepsilon)\,,\nonumber \\
F^{J,\rm virt}_{1,\rm ren} = \,& \frac{1}{8} \biggl[ 6 \log \biggl( \frac{\mu^2}{m_g^2} \biggr) - 4 \log^2 \biggl( \frac{\mu^2}{m_g^2} \biggr)
+ 9 - 2 \pi^2 \biggr] \delta (s)- \log \biggl( \frac{\mu^2}{m_g^2} \biggr) \frac{1}{\mu^2} \biggl[ \frac{\mu^2}{s} \biggr]_+ \,.
\end{align}
The unexpanded result agrees with Eq.~(365) of Ref.~\cite{SimonThesis}\footnote{There is indeed a typo in that equation: there should be a minus sign in front of $H_{1-d/2}$.} and $F^{J,\rm virt}_{1,\rm ren}$ in the expanded expression reproduces Eq.~(33) of Ref.~\cite{Gritschacher:2013pha} up to a global factor of $2$ that accounts for the fact that in that article the jet function accounts for the two hemispheres combined. The real-radiation part is then obtained simply as the sum of residues of $\mathcal{M}^{J}_1(h,\bar{s}_g,\varepsilon)$ to the left of the origin minus the radiative tail of the virtual contribution. This can be computed as the sum of the residues at $h = 0$ (double), $- 1$ and $- 2$ (simple) setting $\varepsilon = 0$ before computing them:
\begin{equation}
F_1^{J,\rm real}(\bar s_g) = \frac{1}{s}\biggl[ \frac{(1 - \bar s_g) (1 + 3 \bar s_g)}{4 \bar s_g^2} + \log (\bar s_g) \biggr] \theta (\bar s_g - 1)\,,
\end{equation}
again in agreement with Eq.~(34) of Ref.~\cite{Gritschacher:2013pha} (up to the factor of two already mentioned). The one-loop correction to jet matching coefficient relating the jet functions in the theories with and without massive vector bosons is given by $F_{1,\rm ren}^{J,\rm virt}$, in agreement with Eq.~(37) of Ref.~\cite{Gritschacher:2013pha} (once again we account for the factor of two and the overall minus sign present due to the difference in the definition of the matching coefficient)
\begin{align}
J_n^{(n_g)}(s)=\,&\int_0^s {\rm d}s'\mathcal{M}^{(n_g\to n_\ell)}_J(s-s') J_n^{(n_\ell)}(s')\,,\\
\mathcal{M}^{(n_g\to n_\ell)}_J=\,&\delta(s)+\frac{\alpha_s(\mu)}{\pi} C_FF^{J,\rm virt}_{1,\rm ren}\,,\nonumber
\end{align}
The gluon mass correction to the massless jet function can be written as
\begin{equation}\label{eq:TildeJ}
\Delta_0 J_1(s,m_g^2) \equiv \frac{1}{s} \tilde J_1(\bar s_g)\,,
\end{equation}
From that, we define the following RG-evolved jet function, which is the relevant object to carry out large-log resummation in differential and cumulative cross sections:
\begin{figure*}[t!]
\subfigure[]
{\includegraphics[width=0.49\textwidth]{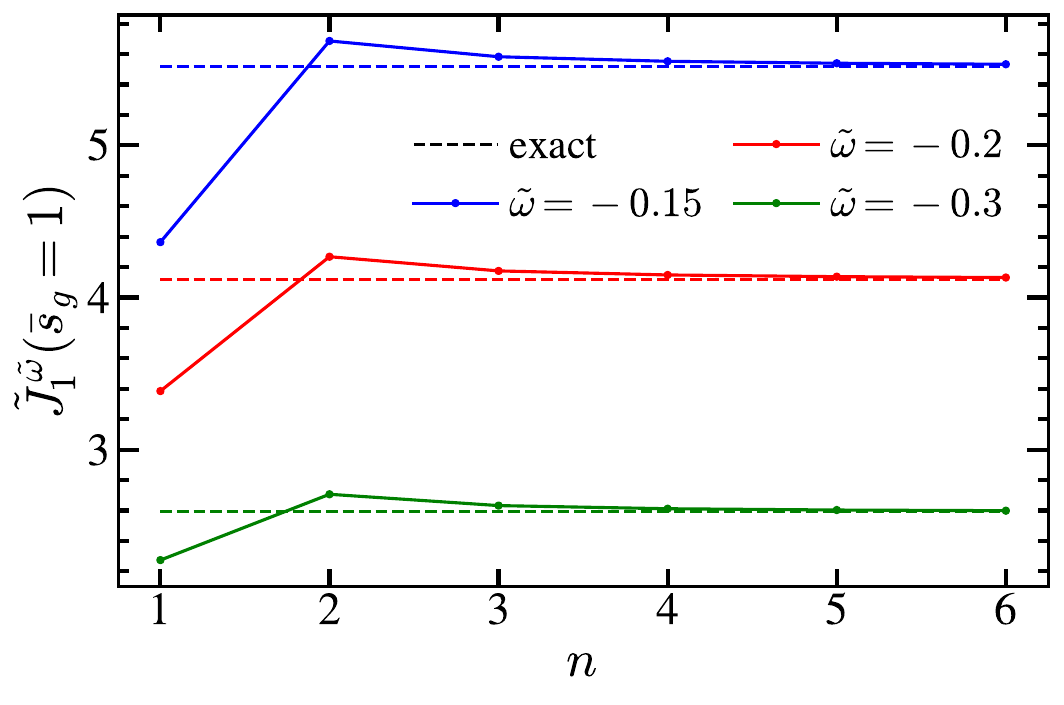}
\label{fig:JetMg1}}
\subfigure[]{\includegraphics[width=0.49\textwidth]{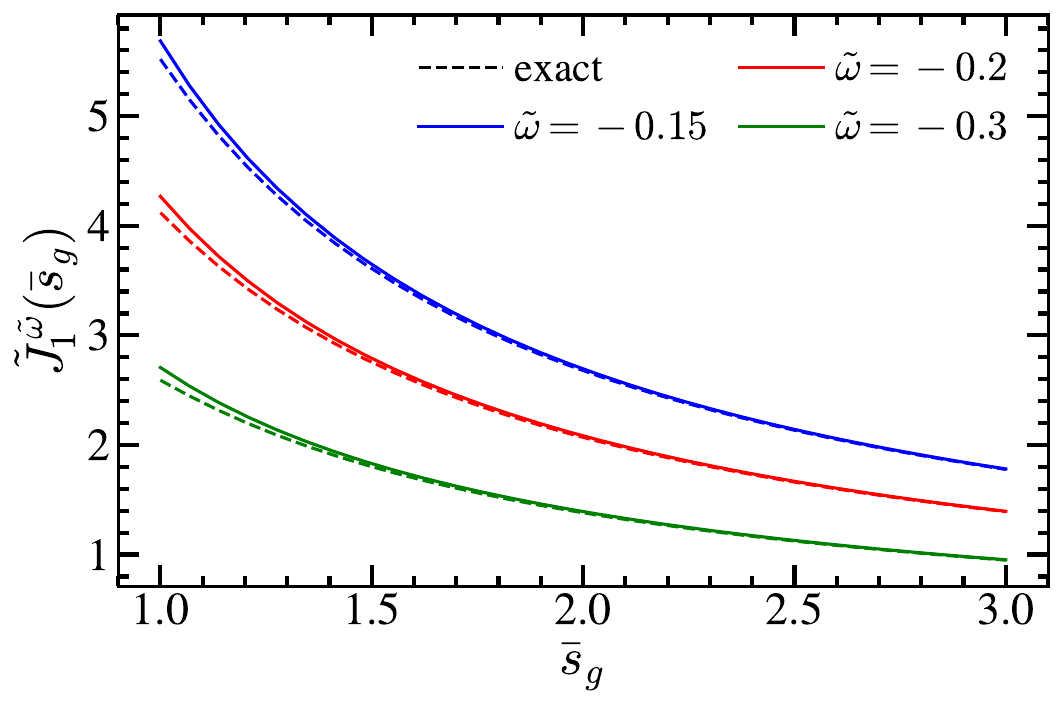}
\label{fig:JetMg}}
\caption{Gluon mass correction to the massless RG-evolved jet function in its exact form (dashed), and expansions for small masses (solid lines) for three values of $\tilde \omega$: $-0.15$ (blue), $-0.2$ (red), and $-0.3$ (green). Left panel: small mass expansion of $\tilde J_1^{\tilde\omega}$ at threshold $\bar s_g=1$ as a function of the expansion order $n$. Right panel: Dependence of $\tilde J_1^{\tilde\omega}$ with $\bar s_g$, including $2$ non-zero terms in the small-mass expansion.}
\label{fig:MgJet}
\end{figure*}
\begin{align}\label{eq:RGEvolved}
\tilde J_1^{\tilde \omega}(\bar s_g) =\ \int_0^{\bar s_g} \frac{{\rm d}s'}{s'}
\biggl(1-\frac{s'}{\bar s_g}\biggr)^{\!-1-\tilde \omega} \tilde J_1(s')
= \int_{c - i \infty}^{c + i \infty} \frac{{\rm d} h}{2 \pi i} \frac{(3 h + 4)\Gamma(h)\bar s_g^h}{2 h^2 (2 + h) (1 + h)(-\omega )_h}\,,
\end{align}
with $-1<c<0$ and where $(a)_n=\Gamma(a+n)/\Gamma(a)$ is the Pochhammer symbol. The convergence regions are identical as for the fixed-order case. Closing to the right one encounters only the triple pole at $h = 0$. Closing to the left there are double poles at $h = - 1, - 2$ and simple poles at $h = - k$ with $k \geqslant 3$. All in all, we find
\begin{align}
\tilde J_1^{\tilde \omega}(\bar s_g) =\, & \dfrac{1}{2\bar s_g}\bigl[L_g (\tilde \omega+1)+5 \tilde \omega+4\bigr] +
\dfrac{1}{2}\sum_{n=3}\frac{1}{\bar s_g^n}\frac{(4-3 n) (\tilde \omega+1)_n}{n^2(n-2) (n-1) n!} \\
& + \dfrac{1}{8\bar s_g^2}\bigl[L_g(\tilde \omega+1)_2
+2\tilde \omega (\tilde \omega+2) +1\bigr] = -\frac{L_g^2}{2}+\frac{3 L_g}{4}+\frac{\psi ^{(1)}(-\tilde \omega)}{2}-\frac{\pi ^2}{12}-\frac{5}{8} \nonumber\,,
\end{align}
where $L_g=\log(\bar s_g) - \gamma_E - \psi^{(0)}(-\tilde \omega)$ and the trigamma function $\psi ^{(1)}$ is the derivative of the digamma $\psi^{(0)}$. The series can be summed up and we find a closed form for the RG-evolved jet function:
\begin{align}
\tilde J_1^{\tilde \omega}(\bar s_g) =\, & \dfrac{1}{2\bar s_g}\bigl[L_g (\tilde \omega+1)+5 \tilde \omega+4\bigr]
+ \dfrac{1}{8\bar s_g^2}\bigl[L_g(\tilde \omega+1)_2 +2\tilde \omega (\tilde \omega+2) +1\bigr]\\
& +\frac{(\tilde \omega+1)_3}{216 \bar s_g^3}
\bigl[4 \, _4F_3\bigl(1,1,3,\tilde \omega+4;4,4,4;\bar s_g^{-1}\bigr)-9 \,
_3F_2\bigl(1,1,\tilde \omega+4;4,4;\bar s_g^{-1}\bigr)\bigr]\,.\nonumber
\end{align}
Results for the exact result and the small-mass expansion are shown in Fig.~\ref{fig:MgJet}. Nice convergence is achieved for any value of $\bar s_g$ (in particular, at threshold) for the various values of the resummation parameter $\tilde \omega$ we have tested.

For completeness, we present results for the Fourier transform of the mass corrections to the massless one-loop jet function. To obtain the Mellin transform we use the following integral
\begin{equation}
\int_0^\infty \frac{{\rm d}s}{s} e^{-i s y} \biggl(\frac{s}{m_g}\biggr)^{\!\!h} = \Gamma (h) (i m_g y)^{-h}\,.
\end{equation}
With this result one immediately finds the Fourier transform of $\Delta_0 J_1(s,m_g^2) $:
\begin{align}\label{eq:FourJMg}
\Delta_0 \tilde J_1(y_g) =\, & \int_{c - i \infty}^{c + i \infty} \frac{{\rm d} h}{2 \pi i} \frac{(3 h + 4)\Gamma(h)(i y_g)^{-h}}{2 h^2 (2 + h) (1 + h)} \\
=\, & iy_g\biggl(\frac{L_y}{2}-\frac{5 }{2}\biggr) + y_g^2\biggl( \frac{L_y}{8}-\frac{1}{4}\biggr) -\frac{1}{2}\sum_{n=3} \frac{(3 n-4) (-i y_g)^n}{(n-2) (n-1) n^2\, n!}\nonumber\\
= \, &\frac{1}{8}e^{-i y_g} (5-i y_g) -i y_g \, _3F_3(1,1,1;2,2,2;-i y_g)-\biggl(\frac{y_g^2}{8}+\frac{i y_g}{2}+\frac{3}{4}\biggr) \Gamma (0,i y_g) \nonumber\\
&- \frac{3}{4} L_y -\frac{5}{8}\,,\nonumber
\end{align}
with $y_g=m_g y$, $-1 <c < 0$, and $L_y = \log(i y_g) + \gamma_E$. In the resummed expression, on the third line, $\Gamma (0,i y_g)$ is the incomplete gamma function, whose integral expression for a complex second argument can be found in Ref.~\cite{Gracia:2023qdy}. Using Cauchy's root test is simple to check that the series converges in the entire complex $y_g$ plane.
For large $h$, the Mellin transform appearing in Eq.~\eqref{eq:FourJMg} behaves as $\propto [-ih/(y\,e)]^h/h^{7/2}$, hence the contour integral in the complex $h$ place can be closed only towards the negative real axis no matter what the value of $y$ is. We will encounter the same behavior in the rest of Fourier-transformed jet functions discussed in the remainder of this article, but will not repeat the argument: only expansions for small masses shall be presented, whose convergence radius will be the entire complex plane. In all cases we have double checked with Cauchy's root test that the converge radius is indeed $\infty$.

\subsubsection{Secondary massive bubble}
\begin{figure*}[t!]
\subfigure[]
{\includegraphics[width=0.482\textwidth]{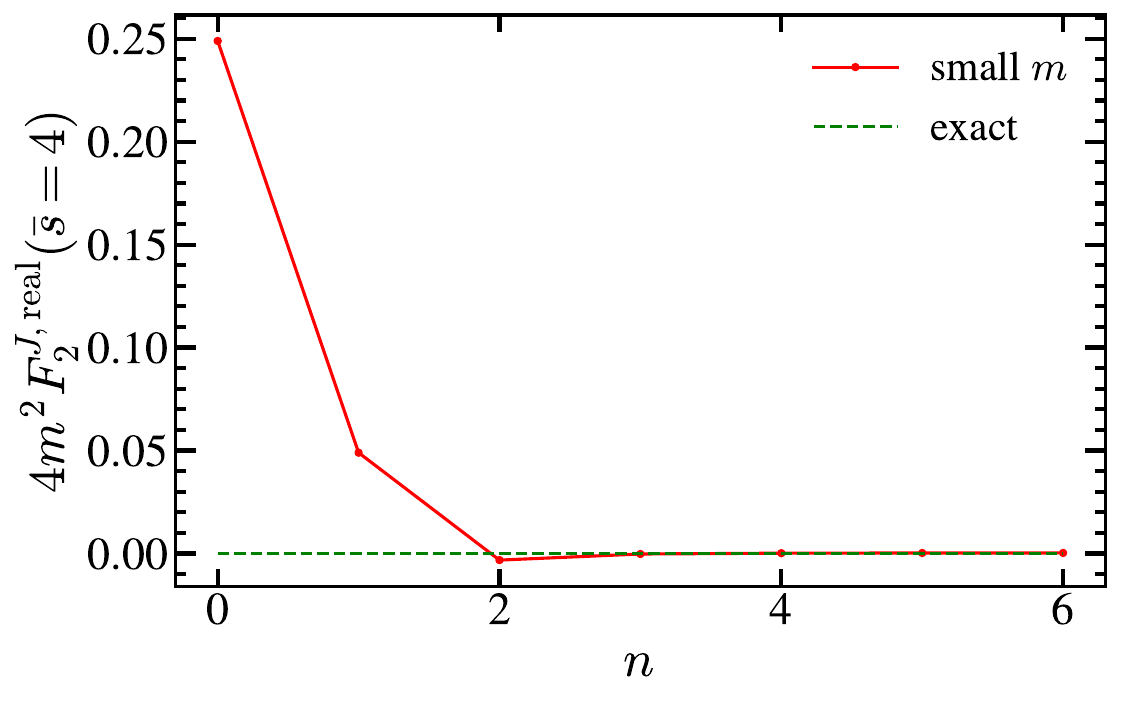}
\label{fig:JetSec1}}
\subfigure[]{\includegraphics[width=0.498\textwidth]{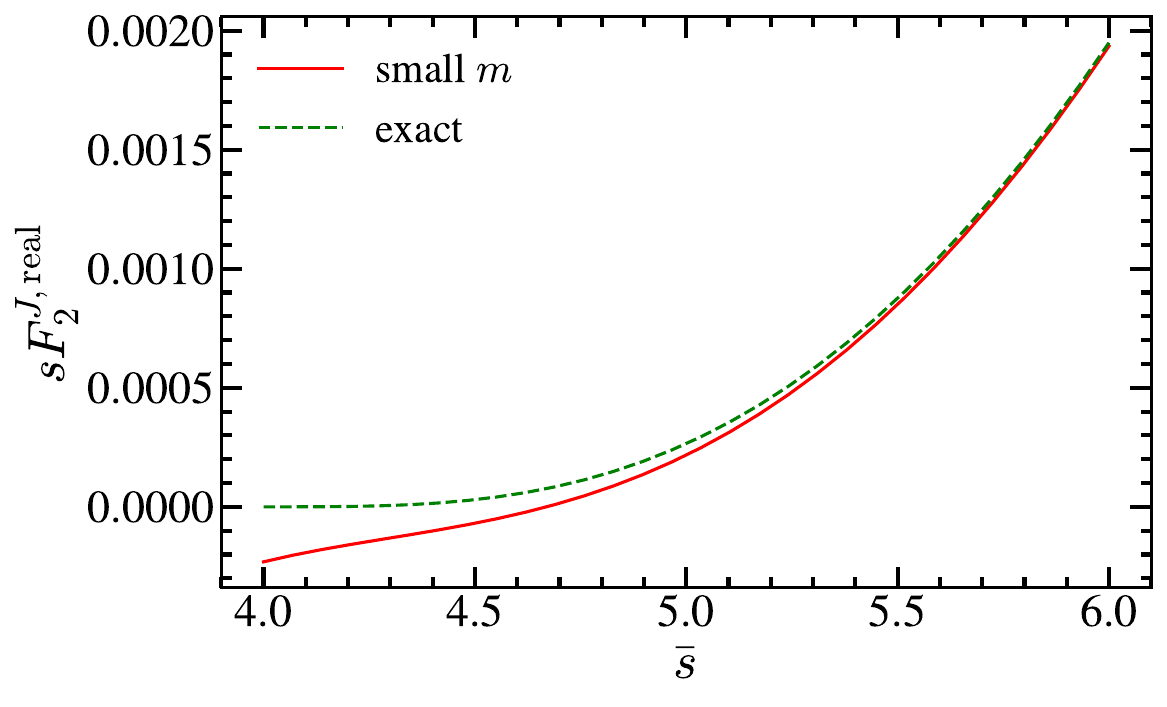}
\label{fig:JetSec}}
\caption{Secondary mass corrections to the jet function: real radiation part. We show exact results as green dashed lines, and the expansions for small masses as red solid lines. Left panel: small mass expansion of $4m^2 F^{J,\rm real}_2$ at threshold $\bar s=4$ as a function of the expansion order $n$. Right panel: Dependence of $s F^{J,\rm real}_2$ with $\bar s$, including $6$ non-zero terms in the expansion.}
\label{fig:SecJet}
\end{figure*}

We proceed in the same way as in the previous section, switching between cumulative and differential to identify distributions and virtual corrections. To simplify expressions as much as possible, we define $\bar s = s/m^2$. In this case we also compute the expansion for the differential jet function prior to showing results for its RG evolution. The most relevant expressions before we discuss any expansion are the following:
\begin{align}\label{eq:CumJetSec}
\mathcal{M}^{\Sigma_J}_2(h,\bar s_c,0) =\, & \frac{2 (3 h + 4) (h + 1) \Gamma^2 (h)}{h^2 \Gamma (2 h + 5)} \bar s_c^h \xrightarrow[| h | \gg 1]{} \frac{3 \sqrt{\pi}}{8} \frac{1}{h^{11 / 2}} \biggl( \frac{\bar s_c}{4} \biggr)^{\!\!h} \,,\\
\delta Z^J_{2}(s,\mu,\varepsilon) = \,& \biggl[\frac{1}{4 \varepsilon^3} - \frac{1}{72\varepsilon^2} - \dfrac{1}{\varepsilon}\biggl(\frac{\pi^2}{72} + \frac{121}{432}\biggr) \biggr]\delta(s)
+ \biggl(\frac{5}{18\varepsilon}-\frac{1}{6\varepsilon^2}\biggr)
\dfrac{1}{\mu^2}\biggl[\dfrac{\mu^2}{s}\biggr]_+\,,\nonumber\\
F^J_{2, \rm fin}(0,s,\mu,\varepsilon) =\, & \biggl(\frac{\zeta_3}{9} + \frac{17 \pi^2}{108} - \frac{4057}{2592}\biggr)\delta(s) - \biggl( \frac{\pi^2}{18} - \frac{247}{216} \biggr) \dfrac{1}{\mu^2}\biggl[\dfrac{\mu^2}{s}\biggr]_+ \nonumber\\
& - \frac{29}{36} \dfrac{1}{\mu^2}\biggl[\dfrac{\mu^2}{s}\log\biggl(\frac{s}{\mu^2}\biggr)\biggr]_+
+ \frac{1}{6} \dfrac{1}{\mu^2}\biggl[\dfrac{\mu^2}{s}\log^2\biggl(\frac{s}{\mu^2}\biggr)\biggr]_+ \,,\nonumber\\
m^2\Delta^\infty_0 F^J_2(\bar s,m) = \, & -\!\frac{1}{6} \biggl[\dfrac{\log^2(\bar s)}{\bar s}\biggr]_++\frac{29}{36}\biggl[\dfrac{\log(\bar s)}{\bar s}\biggr]_+-\biggl(\frac{359}{216}-\frac{\pi^2}{18}\biggr) \biggl[\dfrac{1}{\bar s}\biggr]_+ \nonumber\\
& + \biggl(\frac{4325}{1296}-\frac{2 \zeta_3}{3}-\frac{29 \pi ^2}{216} \biggr)\delta(\bar s) \,.\nonumber
\end{align}
The jet anomalous dimension is obtained from the second line using Eq.~\eqref{eq:JetGammaZ}. We recover the two-loop cusp anomalous dimension of Eq.~\eqref{eq:Cusp2loop} and the corresponding non-cusp piece, and observe the cancellation of UV-divergent terms $1/\varepsilon^n$:
\begin{equation}
\gamma_{1,n_f}^J = -\,C_FT_F \biggl(\frac{484}{27}+\frac{8 \pi ^2}{9}\biggr) = -\biggl(\frac{968}{81}+\frac{16 \pi ^2}{27} \biggr)\,.
\end{equation}
From the first line of Eq.~\eqref{eq:CumJetSec} we see that the expansion for small masses converges if $\bar s > 4$, that is, above threshold. Much as happened for the massive vector boson, below threshold one only has the contribution from the virtual diagrams, captured by the residue of the only pole with $h>0$, sitting at $h=\varepsilon$ [\,after removing the massless limit and setting $\varepsilon=0$, the pole moves to $h=0$, making it of multiplicity $4$, and one is left with the last line of Eq.~\eqref{eq:CumJetSec}\,]. We can split the virtual and real-radiation contributions again, and compare to known results. In fact, the virtual terms of the mass corrections to the two-loop massless jet function are given by $\Delta^\infty_0 F^J_2(\bar s)$, in agreement with Ref.~\cite{Pietrulewicz:2014qza}.\footnote{Despite appearances, Eq.~(41) of Ref.~\cite{Pietrulewicz:2014qza} is $\mu$-independent as it should: the dependence on the renormalization scale is entirely contained in the massless jet function.} The non-distributional (or real-radiation) part is the sum of poles on the negative real axis minus the radiative tail, which is simply the sum of residues corresponding to all poles with $h \leqslant 0$ having set $\varepsilon = 0$. The extra factor of $h$ in $\mathcal{M}^{J}_2(h,\hat m,0)$ makes the pole at $h=0$ triple. The pole at $h=-2$ is double, while the rest of poles sitting at negative integer values of $h$ are all simple. The result quoted below is valid only for $\bar s > 4$, since otherwise it identically vanishes, and the series is convergent in its whole domain of validity, as can be observed in Fig.~\ref{fig:JetSec}:
\begin{align}
s F^{J,\rm real}_2 =\, & \frac{359}{216}-\frac{\pi^2}{18} + \frac{1}{6}\log^2(\bar s) -\frac{29}{36}\log(\bar s) - \frac{\log (\bar s)+1}{2 \bar s^2}
+\!\sum_{n=1}^{n\neq2} \frac{\bar s^{-n}(3n-4)(2 n-2)!}{ (2n-3)(n - 2)n (n!)^2} \nonumber\\
=\, & \frac{359}{216}-\frac{\pi^2}{18} + \frac{1}{6}\log^2(\bar s) -\frac{29}{36}\log(\bar s) - \frac{1}{\bar s}
+ \frac{1}{\bar s^3}\biggl[{}_3F_2\biggl(1,1,\frac{3}{2};3,4;\frac{4}{\bar s}\biggr)\nonumber\\
& -\frac{7}{9} \,_3F_2\biggl(1,1,\frac{3}{2};4,4;\frac{4}{\bar s}\biggr)\!+\frac{4}{27} \,_4F_3\biggl(1,1,\frac{3}{2},3;4,4,4;\frac{4}{\bar s}\biggr)\biggr]
- \frac{\log (\bar s)+1}{2 \bar s^2}\,,\!\!
\end{align}
where to get to the last equality we have summed up the infinite series. This result is equivalent to Eq.~(42) of Ref.~\cite{Pietrulewicz:2014qza}, which is expressed in terms of logarithms and a dilogarithm. As can be seen in Fig.~\ref{fig:JetSec1}, indeed our result for $\Delta_0 F_{2,\rm real}^J$ exactly vanishes at $\bar s = 4$. The jet matching condition is obtained using Eq.~\eqref{eq:matching}, and the result we have obtained
\begin{align}
J_n^{(n_f)}(s)=\, &\int_0^s {\rm d}s'\mathcal{M}^{(n_f\to n_\ell)}_J(s-s') J_n^{(n_\ell)}(s')\,,\\
\mathcal{M}_J^{(n_f\to n_\ell)} =\, & \delta(s) + \biggl[\frac{\alpha_s(\mu)}{\pi}\biggr]^2 T_F\,C_F \mathcal{M}^{(2)}_J(m,s,\mu)\,,\nonumber\\
\mathcal{M}^{(2)}_J(m,s,\mu)=\,& \biggl[ \frac{29}{72} \log ^2\biggl(\frac{\mu^2}{m^2}\biggr)-\frac{1}{18} \log^3\biggl(\frac{\mu^2}{m^2}\biggr)
-\biggl(\frac{233}{216}+\frac{\pi^2}{36}\biggr) \log \biggl(\frac{\mu^2}{m^2}\biggr)-\frac{5 \zeta_3}{9} \nonumber\\
& +\frac{5 \pi ^2}{216}+\frac{1531}{864}\biggr] \delta(s) +
\biggl[\frac{5}{9} \log \biggl(\frac{\mu^2}{m^2}\biggr)-\frac{1}{6} \log ^2\biggl(\frac{\mu^2}{m^2}\biggr)-\frac{14}{27}\biggr]
\frac{1}{\mu^2} \biggl[\frac{\mu^2}{ s}\biggr]_+, \nonumber
\end{align}
is in agreement with Eq.~(46) of Ref.~\cite{Pietrulewicz:2014qza} if setting $\mu_m=\mu_J$ after reversing the sing and accounting for the factor of $2$ explained already. In the previous equation, the strong coupling can be evaluated with either $n_f$ or $n_\ell$ active flavors.

We discuss next the RG-evolved hemisphere jet function, defined as in Eqs.~\eqref{eq:TildeJ} and \eqref{eq:RGEvolved}, and considering once more only the evolution of the correction to the massless result. From the inverse Mellin transform we find:
\begin{equation}
\tilde J_2^{\tilde \omega}(\bar s) = \int_0^{\bar s} \frac{{\rm d}\bar s'}{\bar s'}
\biggl(1-\frac{\bar s'}{\bar s}\biggr)^{\!-1-\tilde \omega} \tilde J_2(\bar s')
= \int_{c - i \infty}^{c + i \infty} \frac{{\rm d} h}{2 \pi i} \frac{2(h+1)(3 h + 4)\Gamma^3(h)\bar s^h}{h\Gamma(2h+5)(-\tilde\omega )_h}\,,
\end{equation}
where the convergence radius does not depend on $\tilde\omega$. Once again, closing to the right for $\bar s < 4$ one picks only the multiplicity-4 pole at $h=0$ corresponding to the virtual radiation contribution. For $\bar s> 4$ one closes to the left, finding an infinite number of poles sitting at integer negative values of $h$: double at $h=-2$, simple otherwise. Defining the logarithm $L=\log(\bar s) -\gamma_E - \psi ^{(0)}(-\tilde \omega)$ we obtain the following results:
\begin{align}
\tilde J_2^{\tilde \omega}(\bar s) =\, & \frac{29 L^2}{72}-\frac{L^3}{18}+\frac{L}{216} \big[36 \psi ^{(1)}(-\tilde \omega)+6 \pi ^2-359\bigr]\\
& +\frac{1}{1296} \big[72 \psi ^{(2)}(-\tilde \omega) -522 \psi ^{(1)}(-\tilde \omega)-720 \zeta_3-87 \pi ^2+4325\bigr] \nonumber \\
=\, & -\!\frac{1}{8\bar s^2}\biggl\{\!\biggl[L^2-\frac{\pi ^2}{6}-\psi ^{(1)}(-\tilde\omega)\biggr] (\tilde\omega+1)_2
+L [\tilde\omega(5\tilde\omega+11)+4]+\tilde\omega (7\tilde\omega+11)+1\!\biggr\} \nonumber\\
& + \sum_{n=1}^{n\neq 2}\frac{(3 n-4) (2 n-2)! (\tilde\omega+1)_n}{\bar s^n n(n-2)(2 n-3) (n!)^3} \biggl[L + 3 H_n-2 H_{2 n-2}+\psi^{(0)}(-\tilde\omega)-\psi ^{(0)}(-n-\tilde\omega)\nonumber\\
& +\frac{12 n^3-45 n^2+56 n-24}{n(n-2) (2 n-3) (3 n-4)}\biggr]. \nonumber
\end{align}
The infinite sum can be carried out and one obtains an analytic expression in terms of MeijerG functions. We find it more convenient to carry out the truncated sum, adding as many terms as necessary to achieve the desired numerical accuracy. For efficient computer implementations, it is convenient to express $\psi ^{(0)}(-n-\tilde\omega)$ in terms of $\psi ^{(0)}(-\tilde\omega)$ as follows:
\begin{figure*}[t!]
\subfigure[]
{\includegraphics[width=0.49\textwidth]{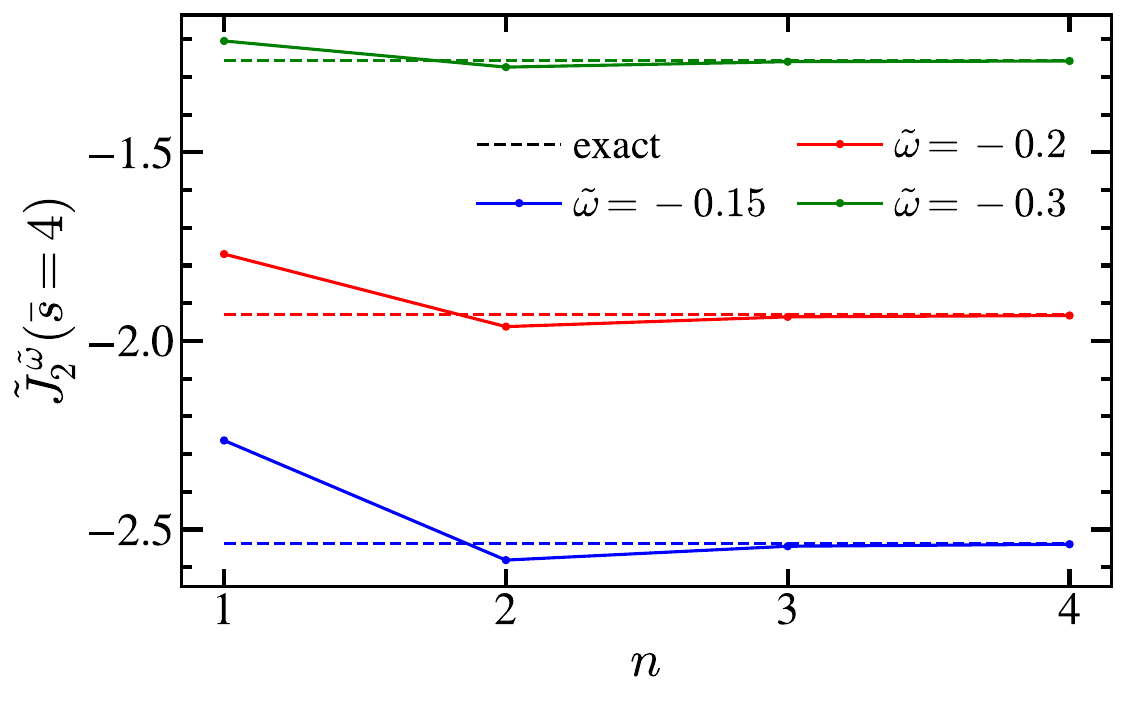}
\label{fig:JetSecRG1}}
\subfigure[]{\includegraphics[width=0.49\textwidth]{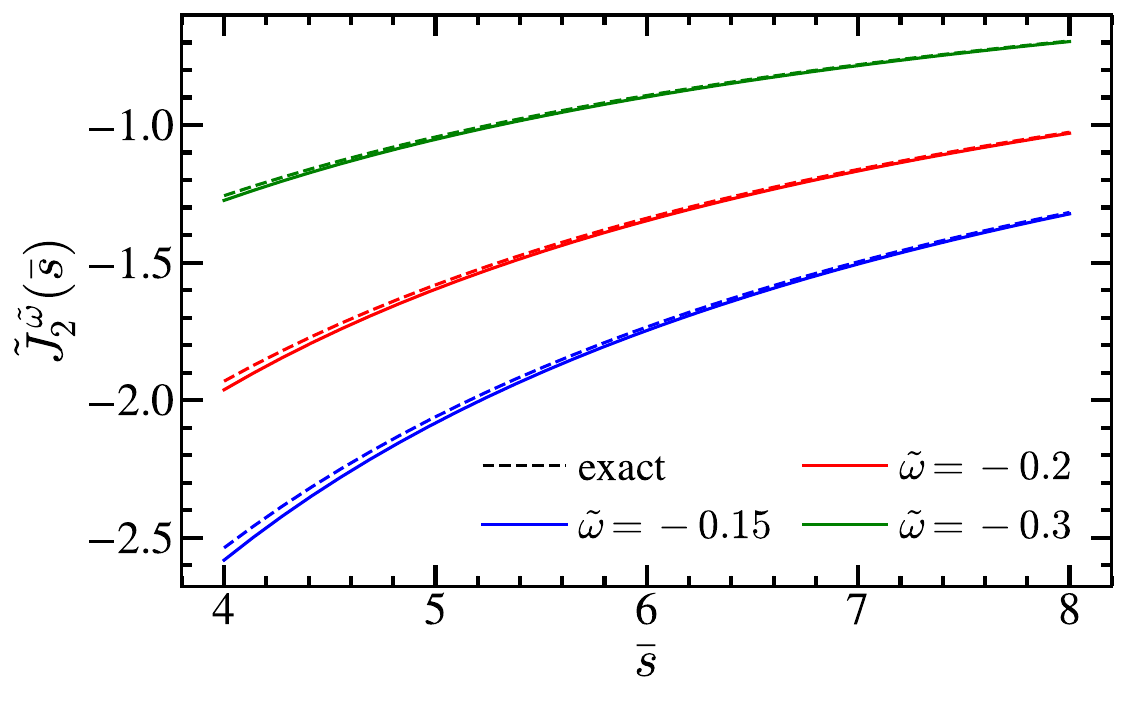}
\label{fig:JetSecRG}}
\caption{Secondary mass correction to the massless RG-evolved jet function in its exact form (dashed), and expansions for small masses (solid lines) for three values of $\tilde \omega$: $-0.15$ (blue), $-0.2$ (red), and $-0.3$ (green). Left panel: small mass expansion of $\tilde J_2^{\tilde\omega}$ at threshold $\bar s=4$ as a function of the expansion order $n$. Right panel: Dependence of $\tilde J_2^{\tilde\omega}$ with $\bar s$, including two non-zero terms in the small-mass expansion.}
\label{fig:SecJetRG}
\end{figure*}
\begin{equation}
\psi ^{(0)}(-n-\tilde\omega)=\psi ^{(0)}(-\tilde\omega) + \sum_{i=1}^{n}\dfrac{1}{i+\tilde\omega}\,.
\end{equation}
In Fig.~\ref{fig:SecJetRG} we show the good convergence of the small-mass expansion, and how it agrees with the single term corresponding to the virtual radiation contribution at $\bar s = 4$. The agreement of the two series is a strong cross-check on our results.

Obtaining the expansion of the Fourier transform of the secondary-mass correction to the two-loop jet function is straightforward, picking up the poles with negative values of $h$, and we find for the Fourier transform of $\Delta_0 J_2(s,m) $:
\begin{align}
\Delta_0 \tilde J_2(x_m) =\, & \int_{c - i \infty}^{c + i \infty} \frac{{\rm d} h}{2 \pi i}\frac{2 (h+1) (3 h+4) \Gamma^3(h) (i x_m)^{-h}}{h \Gamma (2 h+5)} \\
=\, & x_m^2\biggl[\frac{L_x^2}{8}-\frac{5 L_x}{8}-\frac{\pi ^2}{48}+\frac{7}{8}\biggr] -\sum_{n=1}^{n\neq2} \frac{(3 n-4) (2 n-2)! (-i x_m)^n }{(n-2) n (2 n-3) (n!)^3}\nonumber\\
&\times \biggl[L_x+2 H_{2 n-2}-3 H_n+\frac{24-12 n^3+45 n^2-56
n}{n(n-2)(2 n-3) (3 n-4)}\biggr]\,,\nonumber
\end{align}
with $x_m=m y$, $-1 <c < 0$, and $L_x = \log(i x_m) + \gamma_E$.

Before closing the section on SCET computations, the following comment is in order. If applying the strict EFT philosophy, when the secondary quark is no longer active (that is, if $\mu_m>\mu_J$), the secondary quark simply and plainly does not participate in the jet function. The matching condition $\mathcal{M}_J^{(n_f\to n_\ell)}$ accounts for the discrepancy of the two EFTs in the UV. Since the two theories are required to agree for $m\to\infty$, the matching condition takes into account only the virtual radiation, as real emissions cannot occur in this limit. One can however make the transition between the two scenarios smoother by including the mass-suppressed real-radiation contribution not accounted for in $\mathcal{M}_J^{(n_f\to n_\ell)}$, which will naturally become increasingly small as $\bar s$ decreases. This ``improved'' $n_\ell$ flavored jet function is simply:
\begin{equation}
J_n^{(n_\ell)}(s,\mu,m) = J_n^{(n_\ell)}(s,\mu,0) + \biggl[\frac{\alpha^{(n_\ell)}_s(\mu)}{\pi}\biggr]^2 C_F T_F \Delta_\infty J_n(\bar s,m)\,,
\end{equation}
that agrees with the EFT result $J_n^{(n_\ell)}(s,\mu,0)$ in the limit $m\to \infty$. In Ref.~\cite{Pietrulewicz:2014qza}, this mass-modified jet function is obtained from the $n_f$ jet function computation using an OS renormalization factor $Z_J^{\rm OS}$. Even if different in spirit, the results are of course equivalent.

\section{bHQET computations}\label{sec:bHQET}
We consider now the situation of jets produced by boosted massive primary quarks (the most relevant scenario is for unstable tops) resulting from $e^+e^-$ collisions. While for jets whose invariant mass is larger than the quark mass (specifically, for $m_J^2-m_t^2\sim m_t^2$) SCET can be used to write down a factorized cross section in a way analogous to the case of primary massless quarks, when the jet mass is very close to the heavy quark mass, that is for $m_J^2-m_t^2\sim m_t \Gamma_t \ll m_t^2$, SCET must be matched onto a boosted version of HQET, dubbed bHQET \cite{Fleming:2007qr,Fleming:2007xt}. In such situation, the relevant degrees of freedom on top of the heavy quarks are ultracollinear particles, which are soft in the rest frame of the tops $k^\mu_{\rm c,rest}\sim \mathcal{O}(\Gamma_t)$ and large-angle soft radiation which scales in the center-of-mass frame as $k^\mu_{\rm s,CM}\sim \mathcal{O}(m_t\Gamma_t/Q)$, with smaller virtuality than the ultracollinear radiation. When matching the SCET dijet operator to the corresponding bHQET current, an additional matching coefficient $C_m$ needs to be taken into account. Furthermore, the heavy-quark fields, along with the ultracollinear Wilson lines, define a new jet function, which will be referred to as the bHQET jet function $B_n(\hat s,\mu)$:
\begin{align}\label{eq:factbHQET}
\frac{1}{\sigma_0} \frac{{\rm d}\sigma_{\rm bHQET}}{\text{d} e} =&\, Q^2 H (Q,\mu_m) H_m\! \biggl( \!m, \frac{Q}{m}, \mu_m, \mu \!\biggr) \!\!\int \!\!{\rm d} \ell\,
B_{\tau}\! \biggl( \!\frac{Q^2 \tau - Q \ell}{m} - 2m, \mu \!\biggr) S_{e}(\ell, \mu)\,,\\
B_{\tau} (\hat{s}, \mu) =&\, m\! \int_0^{\hat{s}}\! \text{d} \hat{s}' B_n (\hat{s}- \hat{s}', \mu) B_n (\hat{s}', \mu)\,.\nonumber
\end{align}
The jet function does not contain top loops, since the non-relativistic HQET propagators have only a pole and such loops identically vanish. The soft function is identical to its SCET counterpart, except for the fact that there is no soft field for the primary heavy quark left in the bHQET Lagrangian, hence top quarks cannot be produced and do not appear in closed loops.

In this article we consider mass corrections from secondary quarks (say bottom quarks) and massive vector bosons to the matching between SCET and bHQET (where the top quarks are primarily produced) and the bHQET hemisphere jet function (for the top quark). In this context, we also need a VFNS framework analogous to that developed in Ref.~\cite{Pietrulewicz:2014qza} for SCET. Even though the formal aspects of that setup will be discussed in a forthcoming publication, we present in this section the results for the most relevant computations, carried out with the Mellin-Barnes strategy which is the main focus of this article. This also includes the hard mass matching condition defined as $\mathcal{O}_{\rm bHQET}^{(n_f)}=\mathcal{M}_{\rm bHQET}^{(n_f\to n_\ell)}\mathcal{O}_{\rm bHQET}^{(n_\ell)}$, being $\mathcal{O}_{\rm bHQET}$ dijet current operators in bHQET, resulting from integrating the secondary massive bubble at a scale smaller than the primary mass, along with the matching condition for the bHQET jet function $\mathcal{M}_B^{(n_f\to n_\ell)}$.

\subsection{Hard matching coefficient}\label{sec:Hm}
In first place we compute $1$- and $2$-loop quantum corrections to the Wilson coefficient relating the dijet operators defined in SCET and bHQET caused by a massive vector boson or a quark bubble of massive secondary quarks. The relevant hard scale is the mass $M$ of the primary quark. In this section we use $\mu_M\sim M$ to denote the scale of the primary mass at which SCET is matched onto bHQET, and leave $\mu_m\sim m$ as the scale associated to the secondary mass. Likewise, $M$ and $m$ will denote the primary and secondary masses. Finally, $n_f+1$ will be used to denote EFTs in which both primary and secondary massive quarks are active, while $n_\ell = n_f - 1$ signifies that none of those is active any more. Operators labeled with $n_f$ ($n_\ell +1$) have the primary (secondary) quark active while the secondary (primary) has been integrated out.

For the one-loop computation with a shifted gluon propagator we have $\mathcal{Q}^2=M^2$, and the following result was found in Ref.~\cite{Gracia:2021nut}:

\begin{align}
h_1^h (h, \varepsilon) =\, & \frac{1}{2} \frac{(1 - h) \Gamma (1 - h + \varepsilon) \Gamma (1 + 2 h - 2 \varepsilon)}{\Gamma (3 + h - 2 \varepsilon) (h - \varepsilon)^2}\\
& \times \{ 2 + h^2 (3 - 2 \varepsilon) + 4 h (1 - \varepsilon)^2 - \varepsilon [5 - (5 - 2 \varepsilon) \varepsilon] \}\,,\nonumber
\end{align}
again depicting a double pole at $h=\varepsilon$

\subsubsection{Review of SCET-bHQET Matching for massless Quarks}\label{sec:Cmassless}
In this case we deal with the SCET and bHQET dijet currents, both defined in terms of bare fields, schematically denoted by $\mathcal{O}^{\rm bare}_{\rm{SCET}}$ and $\mathcal{O}^{\rm bare}_{\rm{bHQET}}$. Restricting ourselves to vector currents only we have\footnote{For massless primary quarks, the vector and axial-vector form factors are identical up to $\mathcal{O}(\alpha_s^2)$ in QCD or SCET, even in the presence of massive secondary quarks. For massive primary quarks the axial and vector form factors differ in full QCD by mass-suppressed corrections, but are still identical in SCET and bHQET at leading power. Hence, our discussion remains general.}
\begin{equation}
\mathcal{O}^{\rm bare}_{\rm{bHQET}} = {\bar h}_{v_+} W_n Y_n^\dagger \gamma^\mu Y_{\bar n} W^\dagger_{\bar n} h_{v_-}\,,
\end{equation}
where $W_n$ and $W_{\bar n}$ are ultracollinear Wilson lines which are identical to those defined in SCET. The soft Wilson lines have already been defined after Eq.~\eqref{eq:QCD-SCET}. The spin structure can be simplified due to heavy-quark spin symmetry, but since it plays no role in our discussion we do not show the simplified form. The SCET current was already shown in Eq.~\eqref{eq:QCD-SCET}. The bHQET dijet current needs multiplicative renormalization through a $Z$-factor, defining the renormalized operator
\begin{equation}
\mathcal{O}_{\rm{bHQET}}^{\rm ren}(\mu) = Z_{\rm{bHQET}}(\mu) \mathcal{O}^{\rm bare}_{\rm{bHQET}} \,.
\end{equation}
The renormalized bHQET operator depends on the renormalization scale $\mu$ at which it is renormalized, and the dependence with this scale is set by the anomalous dimension $\gamma_{\rm bHQET}$, calculable through $Z_{\rm{bHQET}}(\mu)$ in the usual fashion. Such anomalous dimension is of regular nature, meaning it does not contain a cusp part.

The matching condition between SCET and bHQET is defined on renormalized operators and reads
\begin{align}
\mathcal{O}^{(n_f),\rm ren}_{\rm{SCET}}(\mu) =\, & C_{\rm bHQET}^{(n_f\to n_\ell)}\mathcal{O}^{(n_\ell),\rm ren}_{\rm{bHQET}}(\mu)\,,\qquad
\mathcal{O}^{(n_f),\rm bare}_{\rm{SCET}} = C_{\rm bHQET}^{(n_f\to n_\ell),\rm bare} \mathcal{O}^{(n_\ell),\rm bare}_{\rm{bHQET}}\,,\\
C_{\rm bHQET}^{(n_f\to n_\ell),\rm bare} =\,& \frac{Z_{\rm{bHQET}}(\mu)}{Z_{\rm{SCET}}(\mu)} C_{\rm bHQET}^{(n_f\to n_\ell)}\,,\nonumber
\end{align}
hence $C_{\rm bHQET}^{(n_f\to n_\ell),\rm bare}$ relates the SCET and bHQET bare currents. To avoid large logarithms, one must match both EFTs at the scale $\mu_M\sim M$.\footnote{We consider the primary quark mass in the pole scheme.} The simplest matrix element that can be used to compute this Wilson coefficient is the quark form factor which we again denote by $\langle \mathcal{O} \rangle$. For simplicity, in the rest of this subsection we drop the number of flavors along with the dependence on $\mu$. Taking logarithms is once more convenient since all factors involved equal $1$ at lowest order. Hence
\begin{align}
\log(C_{\rm bHQET}^{\rm bare}) =\,& \log(Z_{\rm{bHQET}}) - \log(Z_{\rm{SCET}}) + \log(C_{\rm bHQET}^{\rm ren}) \\
= \, &\log(\langle \mathcal{O}^{\rm bare}_{\rm SCET}\rangle) - \log(\langle \mathcal{O}^{\rm bare}_{\rm bHQET}\rangle)\,.\nonumber
\end{align}
We can assume the SCET renormalization factor $Z_{\rm{SCET}}$ is already known from previous computations. Both matrix elements are IR divergent and a regulator has to be specified. Since the infrared physics in both EFTs is identical, the matching coefficient is free from IR singularities and the regulator choice does not affect the final result. Different choices, however, might simplify or complicate the computations. If a regulator other than dimreg is chosen one has $\log(Z_{\rm{bHQET}}) = -[\,\log(\langle \mathcal{O}^{\rm bare}_{\rm bHQET}\rangle)]_{\rm div}$, where again we adopt the $\overline{\rm MS}$ prescription to absorb UV singularities. Let us discuss in some detail the simplest choice to carry out the computation: using dimreg to regulate IR divergences. In this situation, for massless secondary quarks one has $\langle \mathcal{O}^{\rm bare}_{\rm bHQET}\rangle = 1$ to all orders. Hence $\log(C_{\rm bHQET}^{\rm bare}) = \log(\langle \mathcal{O}^{\rm bare}_{\rm SCET}\rangle)$ where all $1/\varepsilon^n$ divergences are of UV origin, such that $\log(C_{\rm bHQET}^{\rm ren}) = \log(\langle \mathcal{O}_{\rm SCET}\rangle)_{\rm fin}$. The downside is that $Z_{\rm{bHQET}}$ must be determined indirectly, but this is not complicated: $\log(Z_{\rm bHQET}) = \log(\langle \mathcal{O}^{\rm bare}_{\rm SCET}\rangle)_{\rm div} + \log(Z_{\rm SCET})$. The divergent structure $\log(Z_m) \equiv\log(\langle \mathcal{O}^{\rm bare}_{\rm SCET}\rangle)_{\rm div}$ resembles that of a quantity with cusp anomalous dimension:
\begin{align}
\log(Z_m) =\,& \delta Z^{m}_{\rm cusp} \log\biggl(\frac{\mu^2}{M^2}\biggr) + \delta Z^{m}_{\rm nc}\,,\\
\delta Z^{m}_{\rm cusp} =\, & \sum_{n=1} \biggl[\frac{\alpha_s(\mu)}{4\pi}\biggr]^n \delta Z_{n,\rm cusp}^{m}\,,\qquad
&\delta Z^{m}_{\rm nc}& =\sum_{n=1} \biggl[\frac{\alpha_s(\mu)}{4\pi}\biggr]^n \delta Z_{n,\rm nc}^{m}\,,\nonumber\\
\delta Z_{n,\rm cusp}^{m} =\, & \sum_{i=1}^n \frac{\delta Z_{n,i,\rm cusp}^{m}}{\varepsilon^i}\,,\qquad
&\delta Z_{n,\rm nc}^{m}& = \sum_{i=1}^{n+1} \frac{\delta Z_{n,i,\rm nc}^{m}}{\varepsilon^i}\,,\nonumber
\end{align}
where $\alpha_s(\mu)\equiv \alpha_s^{(n_\ell)}(\mu)$, that is, the primary quark is not active. This applies to all series appearing in this section unless otherwise stated. These coefficients also obey the consistency conditions laid out in Eq.~\eqref{eq:UVcond}. Since the bHQET current operator is free from cusp anomalous dimensions, the sum \mbox{$\log(Z_{\rm bHQET}) =\log(Z_m)+\log(Z_{\rm SCET})$} cannot contain a logarithm of $\mu$. This can only occur if $\delta Z^{m}_{\rm cusp}= - \delta Z^{\rm SCET}_{\rm cusp}$. Moreover, that equality, along with Eq.~\eqref{eq:UVcond}, implies that $\delta Z_{n,n+1,\rm nc}^{m} = - \delta Z_{n,i,\rm nc}^{\rm SCET}$. All in all, we have
\begin{align}
\log(Z_{\rm bHQET}) =\, & \delta Z^{\rm SCET}_{\rm cusp} \log(-\hat M^2) + \delta Z^{\rm bHQET}_{\rm nc}\,,\\
\delta Z^{\rm bHQET}_{\rm nc} =\, & \delta Z^{m}_{\rm nc} +\delta Z^{\rm SCET}_{\rm nc} = \sum_{n=1} \biggl[\frac{\alpha_s(\mu)}{4\pi}\biggr]^n\delta Z^{\rm bHQET}_{n,\rm nc} \,,\nonumber\\
Z^{\rm bHQET}_{n,\rm nc} =\, & \sum_{i=1}^n \dfrac{\delta Z^{m}_{n,i,\rm nc} + \delta Z^{\rm SCET}_{n,i,\rm nc}}{\varepsilon^i}
\equiv \sum_{i=1}^n \dfrac{\delta Z^{\rm bHQET}_{n,i,\rm nc}}{\varepsilon^i}\,,\nonumber
\end{align}
where we have introduced the reduced mass $\hat M=M/Q$ in the first line. Adding up the relation in the second line of Eq.~\eqref{eq:UVcond} applied to $\delta Z_{n,i,\rm nc}^{m}$ and $\delta Z_{n,i,\rm nc}^{\rm SCET}$, and taking into account that $\delta Z_{i-1,i,\rm nc}^{\rm bHQET}=0$, it is found that the coefficients $\delta Z_{n,i,\rm nc}^{\rm bHQET}$ obey the relation in the first line of that equation, crucial to obtain a UV-finite bHQET anomalous dimension. We find:
\begin{align}
\gamma_{\rm bHQET}(\alpha_s) =\, & -\! \frac{{\rm d}Z_{\rm bHQET}}{{\rm d}\log\mu}=-\Gamma_{\rm cusp}(\alpha_s) \log(-\hat M^2) + \sum_{n=1}\gamma_n^{\rm bHQET} \biggl(\frac{\alpha_s}{4 \pi} \biggr)^{\!\!n} \,,\\
\gamma^{\rm bHQET}_n =\, & 2 (n+1)\delta Z^{\rm bHQET}_{n,1,\rm nc}\,.\nonumber
\end{align}
This result can be used to RG-evolve the bHQET dijet current operator. At this point, a discussion along the lines of that presented at the end of Sec.~\ref{sec:massless} could be added, but we refrain doing so to avoid repetition.

\subsubsection{Formal aspects of bHQET matching}\label{sec:formal}
A presentation analogous to that laid out in Sec.~\ref{sec:hard} can be carried out for the effects of secondary masses on the matching between SCET and bHQET. The same discussion applies to the case of a massive vector boson, and to avoid repetition we focus on the secondary bubble only. If $\mu_m\gg \mu_M$,\footnote{We assume in this section $Q>\mu_m$.} one first integrates out the secondary quark by matching SCET$^{(n_f+1)}$ onto SCET$^{(n_f)}$ (both theories with a massive, active, primary quark, whose mass $M$ is an IR scale) obtaining the matching condition \mbox{$\mathcal{M}_{\rm SCET}^{(n_f+1\to n_f)}(M,m,Q,\mu) = \langle \mathcal{O}_{\rm SCET}^{(n_f+1)}\rangle/\langle \mathcal{O}_{\rm SCET}^{(n_f)}\rangle$}. Since IR dynamics is the same in both theories, its $M\to 0$ limit is well defined and has been already computed, see Eq.~\eqref{eq:SCETMatching}. Subsequently, the hard shell of the primary massive quark is integrated out by matching SCET$^{(n_f)}$ onto bHQET$^{(n_\ell)}$ through the ``ordinary'' massless Wilson coefficient $C^{(n_f\to n_\ell)}_m(0,\mu)=\langle \mathcal{O}_{\rm SCET}^{(n_f)}\rangle/\langle \mathcal{O}_{\rm bHQET}^{(n_\ell)}\rangle$. On the other hand, if $\mu_m \ll \mu_M$, the secondary mass is an infrared scale both in SCET$^{(n_f+1)}$ and bHQET$^{(n_\ell+1)}$, hence the matching coefficient relating both theories should not depend on it. In this case, the hard shell of the primary quark is integrated out first by matching SCET$^{(n_f+1)}$ onto bHQET$^{(n_\ell + 1)}$, obtaining $C^{(n_f+1\to n_\ell +1)}_m(0,\mu) = \lim_{m\to 0} \langle \mathcal{O}_{\rm SCET}^{(n_f+1)}\rangle/\langle \mathcal{O}_{\rm bHQET}^{(n_\ell+1)}\rangle$. In a way analogous to Scenario II in SCET, we can keep the formally power suppressed mass effects in the matching condition simply not taking the limit of vanishing secondary quark mass: $C^{(n_f+1\to n_\ell +1)}_m(m,\mu) = \langle \mathcal{O}_{\rm SCET}^{(n_f+1)}\rangle/\langle \mathcal{O}_{\rm bHQET}^{(n_\ell+1)}\rangle$. This is the IR-finite quantity that shall be computed in this section. Finally, if the masses of the primary and secondary quarks are comparable, both are UV scales that do not live in bHQET and one integrates the hard scales $\mu_m$ and $\mu_M$ simultaneously, ``jumping'' two quark flavors at once:
\begin{equation}
\!\!\! C^{(n_f+1\to n_\ell)}_m \!= \dfrac{\langle \mathcal{O}_{\rm SCET}^{(n_f+1)}\rangle}{\langle \mathcal{O}_{\rm bHQET}^{(n_\ell)}\rangle}=\!
\left\{\begin{array}{l} \!\! \dfrac{\langle \mathcal{O}^{(n_f+1)}_{\rm SCET}\rangle}{\langle\mathcal{O}^{(n_\ell + 1)}_{\rm bHQET}\rangle}
\dfrac{\langle \mathcal{O}^{(n_\ell + 1)}_{\rm bHQET}\rangle}{\langle\mathcal{O}^{(n_\ell)}_{\rm bHQET}\rangle}
= \mathcal{M}^{(n_\ell + 1\to n_\ell)}_{\rm bHQET}C_{\rm bHQET}^{(n_f + 1\to n_\ell+1)}(m)\\[0.5cm]
\dfrac{\langle \mathcal{O}^{(n_f+1)}_{\rm SCET}\rangle}{\langle\mathcal{O}^{(n_f)}_{\rm SCET}\rangle}
\dfrac{\langle \mathcal{O}^{(n_f)}_{\rm SCET}\rangle}{\langle\mathcal{O}^{(n_\ell)}_{\rm bHQET}\rangle}
= \mathcal{M}^{(n_f+1\to n_f)}_{\rm SCET}(M)C_{\rm bHQET}^{(n_f\to n_\ell)}(0)
\end{array}\right.\!\!,
\end{equation}
hence, we can obtain the bHQET matching condition as
\begin{equation}\label{eq:bHQETMatchGeneric}
\mathcal{M}^{(n_\ell + 1\to n_\ell)}_{\rm bHQET}(M, Q, m, \mu) =\mathcal{M}^{(n_f+1\to n_f)}_{\rm SCET}(0, Q, m, \mu)
\frac{C_{\rm bHQET}^{(n_f\to n_\ell)}(M,0,\mu)}{C_{\rm bHQET}^{(n_f + 1\to n_\ell+1)}(M\to 0,m,\mu)}\,,
\end{equation}
where we use the fact that $\mathcal{M}^{(n_\ell + 1\to n_\ell)}_{\rm bHQET}(M,Q,m,\mu)$ depends on $M$ through logarithms only, and hence it coincides with its $M\to 0$ expression. Furthermore, we can set $M=0$ in $\mathcal{M}^{(n_f+1\to n_f)}_{\rm SCET}$ since the limit, as already argued, is smooth. Finally, we consider the expansion \mbox{$C_{\rm bHQET}^{(n_\ell + 1\to n_\ell)}(M\to 0,m)$} since the limit of massless primary quark is not analytic and involves logarithms. Since the non-logarithmic dependence is through ratios of masses, this limit coincides with taking $m\to \infty$. From the result above we can obtain
\begin{equation}\label{eq:SCETMatchGeneric}
\! \mathcal{M}^{(n_f+1\to n_f)}_{\rm SCET}(M,Q, m, \mu) = \mathcal{M}^{(n_f+1\to n_f)}_{\rm SCET}(0, Q, m, \mu)
\dfrac{C_{\rm bHQET}^{(n_f + 1\to n_\ell+1)}(M,m,\mu)}{C_{\rm bHQET}^{(n_f + 1\to n_\ell+1)}(M,m\to \infty,\mu)}.
\end{equation}
From this result it is clear that the dependence of $\mathcal{M}^{(n_f+1\to n_f)}_{\rm SCET}(M,Q,m,\mu)$ on $\mu$ is $M$ independent, and the $M\to 0$ limit is trivially satisfied. These relations can be used to isolate the SCET and bHQET matrix elements due to a massive vector boson, which are well-defined by themselves (that is, there is no need to specify an IR regulator) since $m_g$ regulates infrared and collinear divergences, or the dispersive contribution from a secondary massive bubble, also IR finite.

\subsubsection{Massive vector Boson}
In this section we use again the dimensionless variable $\xi_g=m_g/M$ defined in Sec.~\ref{sec:MSglue}. The most relevant results to be presented before turning to the small- and large-$m_g$ expansions are the following:
\begin{align}\label{eq:MgbHQET}
\mathcal{M}_1^{m}(h, 0,\xi_g ) \!=\, & \frac{[2+h (3 h + 4)]\Gamma^2(2 - h) \Gamma (2 h)
\xi_g^{-2 h}}{h^2 (1-h^2) (h + 2)} \xrightarrow[| h | \gg 1]{} - \frac{3 \pi^\frac{3}{2}\!
\csc^2 (\pi h) }{2 \sqrt{h}}\biggl(\frac{\xi_g}{2}\biggr)^{\!-2 h} ,\nonumber\\
\delta Z_1^{m}\!\biggl(0, \frac{M}{\mu}, \varepsilon\biggr) =\, & C_F\biggl[\frac{2}{\varepsilon^2} + \frac{1}{\varepsilon} \biggl( 2 L_M + 1 \biggr)\!\biggr],\nonumber\\
F_{1,\rm ren}^{m}\!\biggl(0, \frac{M}{\mu}\biggr) =\, & 1 + \frac{\pi^2}{24} + \frac{1}{4} L_M^2+ \frac{1}{4}L_M\,,\nonumber\\
\Delta_0^\infty F_1^m (\xi_g)=\, & \frac{1}{8} - \frac{\pi^2}{4} - \log^2 ( \xi_g) -\frac{1}{2} \log ( \xi_g),
\end{align}
with $L_M\equiv \log(\mu^2/M^2)$. The form of the $1/\varepsilon^n$ terms on the second line makes clear the result cannot correspond to a SCET computation, since in this EFT the primary mass is an infrared scale and UV divergences can only depend on UV physics. The result in the second line agrees with the known one-loop result, computed for the first time in Ref.~\cite{Fleming:2007xt}, see Eq.~(131) therein. The one-loop $Z_{\rm bHQET}$ factor is obtained adding $\delta Z_1^{m}$ and $\delta Z_1^{\rm SCET}$, from where we can also compute the one-loop anomalous dimension:
\begin{equation}\label{eq:ZbHQETMg}
\delta Z_1^{\rm bHQET} = -\frac{2C_F}{\varepsilon}\bigl[\log(-\hat M^2) + 1\bigr]\,,\qquad
\gamma_1^{\rm bHQET} = - 4C_F\bigl[\log(-\hat M^2) + 1\bigr]\,,
\end{equation}
both in agreement with Ref.~\cite{Fleming:2007xt}, see Eqs.~(129) and (131) therein . The result in the last line of Eq.~\eqref{eq:MgbHQET} comes from the triple pole at $h=0$. From the first line of Eq.~\eqref{eq:MgbHQET} one can see that the small and large gluon mass expansions will converge for $m_g$ smaller or larger that $2M$, respectively. On the negative real axis there are double poles at $h=-1$ and $-2$, and simple poles at all other negative integer and half-integer values of $h$. On the positive real axis one finds a triple pole at $h=1$ and double poles at all other integer values of $h$. All in all, we find
\begin{align}
\Delta_0 F_1^m(\xi_g) = \, & -\xi_g^{2} \biggl[ \log ( \xi_g) + \frac{1}{4}\biggr]
+ \frac{3}{16}\xi_g^{4} \bigl[ 4 \log ( \xi_g) - 3 \bigr] \\
&+\sum_{n=1}^{n \neq 2, n \neq 4} \bigl( - \xi_g\bigr)^{n}\, \frac{(n + 2) (3 n^2 - 8 n +8) \Gamma^2 \bigl( \frac{n}{2} + 1 \bigr) }{(n - 4) (n - 2) n^2\, n!}\nonumber\\
=\,& \Delta^\infty_0 F_1^m + \frac{3}{2} \xi_g^{-2}+ \sum_{n = 2} \xi_g^{-2 n}\,
\frac{ (2n-1)!}{ [(n + 2)!]^2} \biggl\{ \frac{8 + 22 n + 20 n^2 + 9 n^3 - 2 n^4 - 3 n^5}{n}
\nonumber\\
&- 2 (n^2 - 1) (n + 2) (3 n^2 + 4 n + 2) \bigl[ \log (\xi_g) + H_{n-1} - H_{2 n-1} \bigr] \!\biggr\} \nonumber.
\end{align}
The term linear in the gluon mass (caused by the pole at $h=-1/2$) is directly related to the $u=1/2$ renormalon found in Ref.~\cite{Gracia:2021nut}. The expansion for small $m_g$ can be summed up and we find the following analytic result:
\begin{align}
\Delta_0 F_1^m(\xi_g) = \, &\frac{10 \xi_g^2 - 4 - 3\xi_g^4}{4 r_g} \xi_g\log \biggl( \frac{\xi_g+r_g}{2} \biggr) - \log^2 \biggl( \frac{\xi_g + r_g}{2}\biggr)\\
&- \!\frac{\pi^2 + 3 \xi_g^2 + \xi_g^2(4 - 3\xi_g^2) \log (\xi_g)}{4}\,, \nonumber
\end{align}
with $r_g = \sqrt{\xi_g^2 - 4}$. The expression above has all terms manifestly real for $\xi_g \geq 2$. For $\xi_g < 2$ the result is also real-valued, but to have all terms manifestly real one has to make the following replacements:
\begin{align}
r_g \,&\to \hat r_g = \sqrt{4-\xi_g^2}\,,\\
\log^2\biggl( \frac{\xi_g+r_g}{2} \biggr) \,&\to -\arctan^2\biggl(\frac{\hat r_g}{\xi_g}\biggr)\,,\nonumber \\
\frac{1}{r_g}\log\biggl( \frac{\xi_g+r_g}{2} \biggr) \,&\to \frac{1}{\hat r_g}\arctan\biggl(\frac{\hat r_g}{\xi_g}\biggr)\,.\nonumber
\end{align}
At the limiting value $\xi_g=2$ one gets of course a finite result, $\Delta_0 F_1^m(2) = 8\log (2)-6-\pi ^2/4$.

\begin{figure*}[t!]
\subfigure[]
{\includegraphics[width=0.494\textwidth]{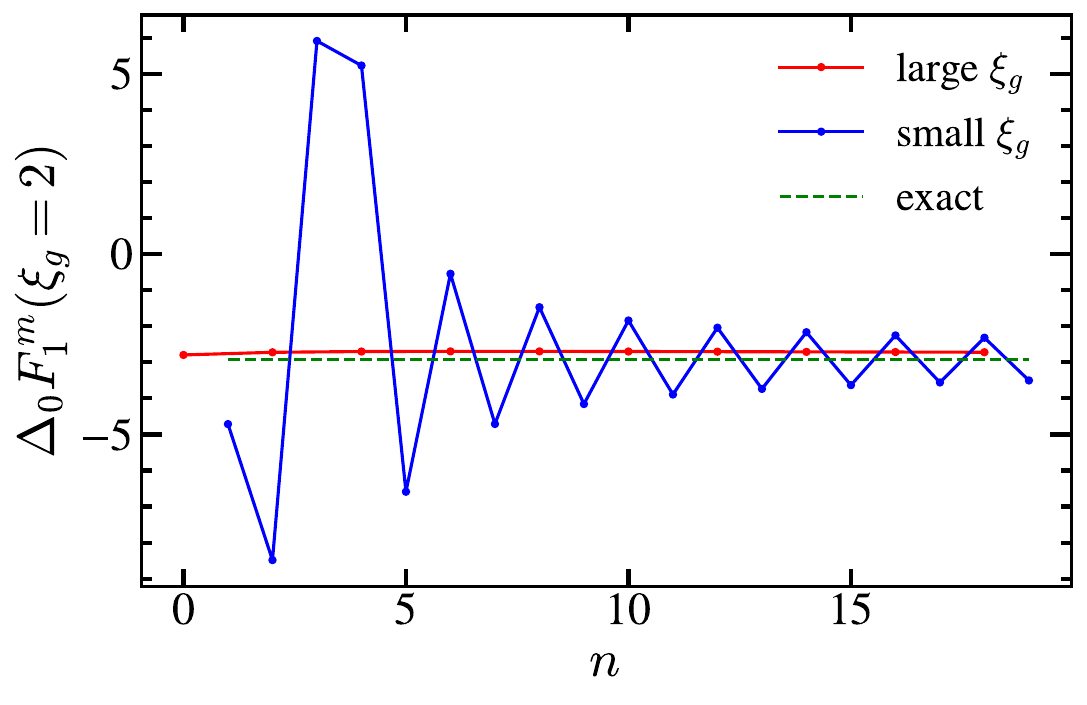}
\label{fig:HmMg1}}
\subfigure[]{\includegraphics[width=0.486\textwidth]{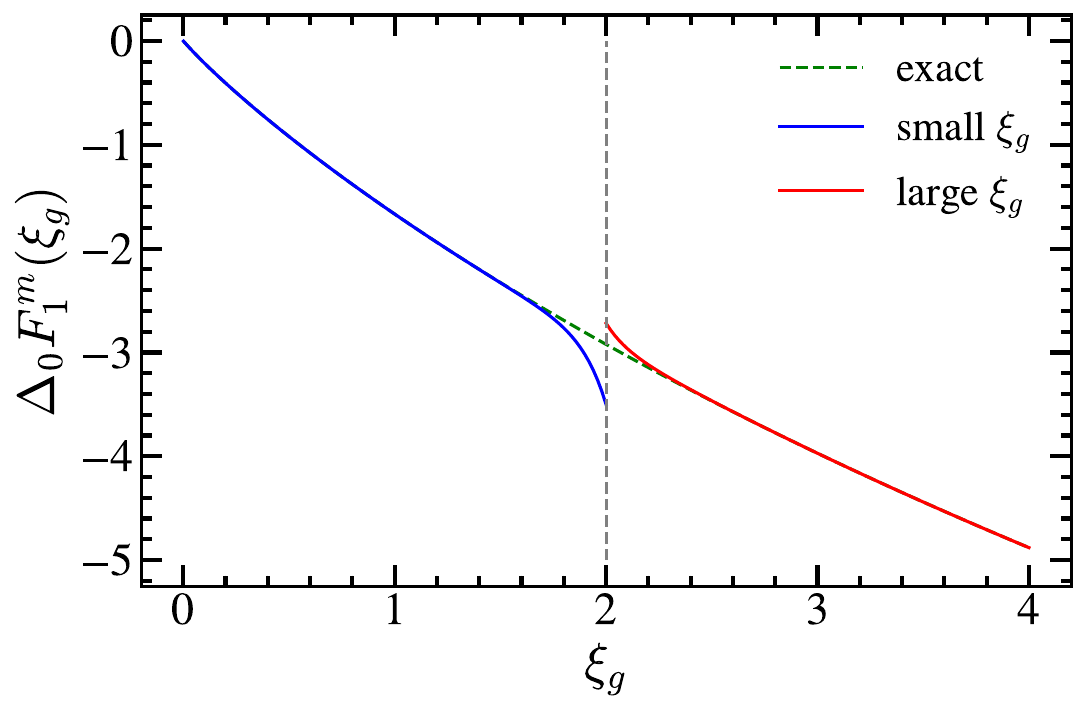}
\label{fig:HmMg}}
\caption{Gluon mass correction to the massless Wilson coefficient relating the dijet operators in SCET and bHQET in its exact form (dashed green), small- (blue) and large-mass (red) expansions. Left panel: $\Delta_0 F_1^m$ at the boundary between the mass expansions $\xi_g = 2$, as a function of the expansion order $n$ of each series. Right panel: Dependence of $\Delta_0 F_1^m(\xi_g)$ with the reduced gluon mass, including $20$ and $10$ non-zero terms in the small- and large-mass expansions, respectively.}
\label{fig:MgHm}
\end{figure*}

In Fig.~\ref{fig:MgHm} we study the convergence of the small and large gluon mass series expansions. The small mass expansion has an oscillatory behavior, possibly related to the $u=1/2$ renormalon, and converges at a slower pace than its large-mass counterpart. We observe that the latter needs many terms to approach the exact value at the limiting value $\xi_g=2$, hence it is also slowly convergent.

Next we compute the matching condition $\mathcal{M}_{\rm bHQET}^{(n_g\to n_\ell)}$ at one loop, which coincides with the bHQET massive vector boson form factor $F_1^{\rm bHQET}$ and is obtained from the $m_g\to \infty$ limit of the SCET to bHQET Wilson coefficient, see Eq.~\eqref{eq:bHQETMatchGeneric}:
\begin{align}
\mathcal{M}_{\rm bHQET}^{(n_g\to n_\ell)} =\, & 1 + C_F \frac{\alpha_s(\mu)}{\pi} F_1^{\rm bHQET}\biggl(\hat M, \frac{m_g}{\mu},\varepsilon\biggr) ,\\
F_1^{\rm bHQET}\!\biggl(\hat M, \frac{m_g}{\mu},\varepsilon\biggr) =\, & F_{1,\rm ren}^{\rm SCET}(0,Q,m_g,\mu) - F_{1,\rm ren}^{m}\!\biggl(0, \frac{M}{\mu}\biggr)
- \Delta^\infty_0 F_1^m(\xi_g)- \frac{\delta Z_1^{\rm bHQET}}{4C_F} \nonumber\\
=\, & \biggl[\frac{1}{2\varepsilon} + \log \Bigl(\frac{\mu }{m_g}\Bigr)\biggr]\bigl[\log(-\hat M^2) + 1\bigr] \,.\nonumber
\end{align}
The presence of powers of $\log(\mu/m_g)$ makes clear that the gluon mass is acting as a regulator for IR divergences. The $\mu$ dependence of the result above is consistent with the $m_g$-independent $Z$-factor depicted in Eq.~\eqref{eq:ZbHQETMg}, in which cusp terms are absent. The dependence on the renormalization scale also agrees with that of Eq.~(130) in Ref.~\cite{Fleming:2007xt}, where off-shellness was used as a regulator. We finish this section by computing the primary-mass-corrected matching condition $\mathcal{M}_{\rm SCET}^{(n_g\to n_\ell)}$ at one loop using Eq.~\eqref{eq:SCETMatchGeneric}, which coincides with the SCET form factor for massive primary quarks and a massive virtual gluon:
\begin{align}
\mathcal{M}_{\rm SCET}^{(n_g\to n_\ell)} =\, & 1 + C_F \frac{\alpha_s(\mu)}{\pi} F_1^{\rm SCET}(M,Q, m_g,\mu) \,,\\
F_1^{\rm SCET}(M,Q,m_g,\mu) =\, & F_{1,\rm ren}^{\rm SCET}(0,Q,m_g,\mu) + \Delta_0 F_{1}^m(\xi_g)
- \Delta^\infty_0 F_1^m(\xi_g)\,. \nonumber
\end{align}
Obviously, the result in the second line can also be computed adding $F_1^{\rm bHQET}$ and $F_{1,\rm ren}^m$.

\subsubsection{Secondary massive bubble}
In this section we will write the relevant theoretical expressions in terms of the dimensionless parameter $\xi=m/M$ defined in Sec.~\ref{sec:MSbSec}. The most relevant results prior to discussing any expansions are
\begin{align}\label{eq:SecMassUnExp}
\mathcal{M}^{m}_2(h,\xi,0) & = \frac{ (1-h)(3 h^2+4 h+2)\Gamma^2(1-h) \Gamma^2(h)\xi^{-2h}}{2 h^2 (h+2) (2 h+1) (2 h+3)}
\xrightarrow[| h | \gg 1]{} -\frac{3 \pi ^2 \csc ^2(\pi h)}{8 h^2}\xi^{-2h},\nonumber\\
\delta Z_{2,n_f}^{m}\!\biggl(0, \frac{M}{\mu}, \varepsilon\biggr)\!\! &\,= 16C_FT_F\biggl[\frac{1}{8 \varepsilon^3}+\frac{1}{12\varepsilon^2}\biggl(L_M-\frac{1}{3}\biggr)-\frac{1}{\varepsilon}\biggl(\frac{5}{36} L_M+\frac{\pi ^2}{48}+\frac{5}{432}\biggr)\biggr],\nonumber\\
F^{m}_{2,\rm ren}\!\biggl(0, \frac{M}{\mu}\biggr)\!\! &\,= -\frac{1}{36}L_M^3-\frac{13}{72} L_M^2- L_M\biggl(\frac{\pi ^2}{18}+\frac{77}{216} \biggr)-\frac{13 \zeta_3}{36}-\frac{37 \pi ^2}{432}-\frac{1541}{2592}\,,\nonumber\\
\Delta_0^\infty F_2^{m}(\xi )\! & \,= \frac{2}{9}L_\xi^3+\frac{13}{18}L_\xi^2+\biggl(\frac{133}{108}+\frac{\pi ^2}{9}\biggr)\!L_\xi+\frac{1747}{1296}+\frac{13 \pi ^2}{108}\,.
\end{align}
with $L_\xi = \log(\xi)$. From the first line we see that the boundary between expansions is $\xi=1$, value at which both converge due to the $h^{-2}$ overall dependence of $\mathcal{M}^{m}_2(h,\xi,0)$ for large $h$. The divergent pieces and massless part agree with the results found in Ref.~\cite{Hoang:2015vua}. From the UV singularities we can obtain the $n_f$ piece of the two-loop bHQET renormalization factor, simply adding $\delta Z_{2,n_f}^{m}$ and $\delta Z_{2,n_f}^{\rm SCET}$ from Eq.~\eqref{eq:HSec}, and from that result obtaining the corresponding anomalous dimension is trivial
\begin{align}
\delta Z_{2,n_f}^{\rm bHQET} =\,& C_FT_F\bigl[\log(-\hat M^2)+1\bigr]\biggl(-\frac{4}{3}\frac{1}{\varepsilon^2}
+\frac{20}{9}\frac{1}{\varepsilon}\biggr)\,,\\
\gamma_{2,n_f}=\,& \frac{80}{9}C_FT_F\bigl[\log(-\hat M^2)+1\bigr] =
\frac{160}{27}\bigl[\log(-\hat M^2)+1\bigr] \,.\nonumber
\end{align}
The $Z$ factor agrees with Eq.~(3.6) of Ref.~\cite{Hoang:2015vua} and the anomalous dimension reproduces the corresponding piece in Eq.~(2.19) of the same reference.

\begin{figure*}[t!]
\subfigure[]
{\includegraphics[width=0.494\textwidth]{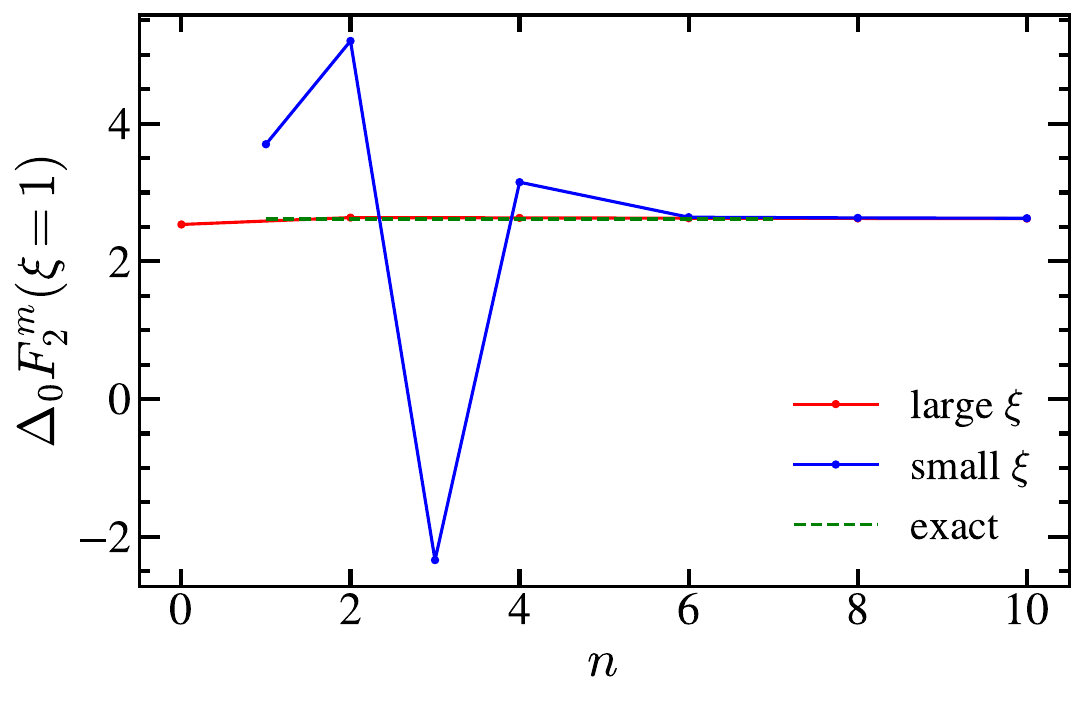}
\label{fig:HmSec1}}
\subfigure[]{\includegraphics[width=0.486\textwidth]{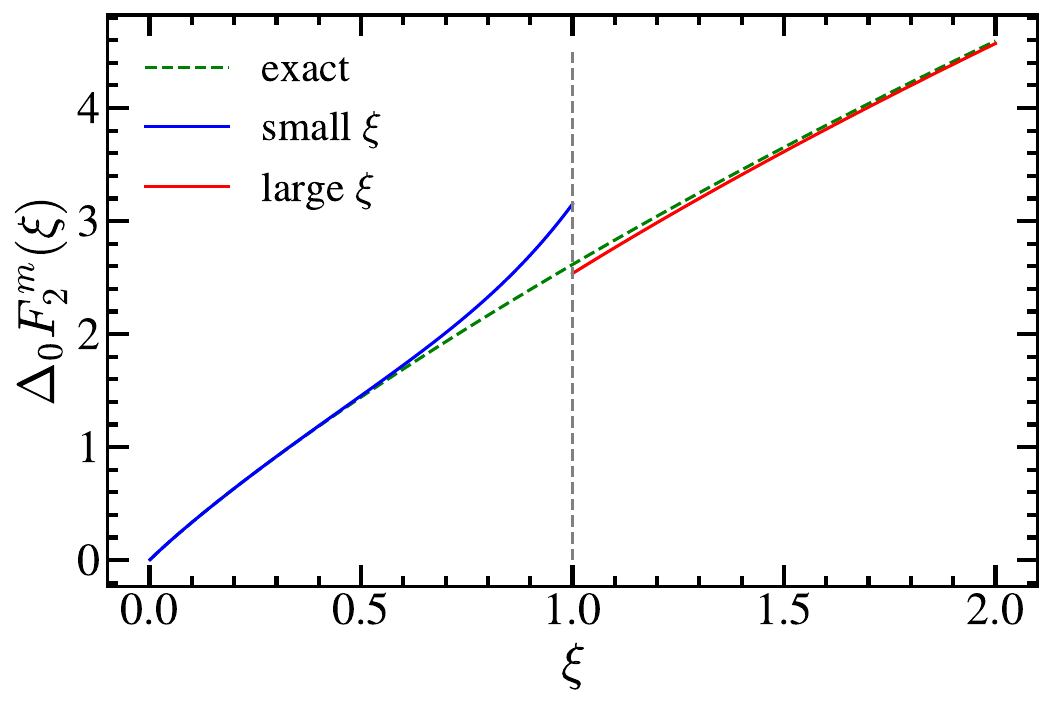}
\label{fig:HmSec}}
\caption{Secondary mass correction to the two-loop massless Wilson coefficient relating the dijet operators in SCET and bHQET in its exact form (dashed green), small- (blue) and large-mass (red) expansions. Left panel: $\Delta_0 F_2^m$ at the boundary between the mass expansions $\xi = 1$, as a function of the expansion order $n$ of each series. Right panel: Dependence of $\Delta_0 F_2^m(\xi)$ with the secondary quark mass, including $4$ and $1$ non-zero terms in the small- and large-mass expansions, respectively.}
\label{fig:MgSec}
\end{figure*}
Let us study now the different series for the mass corrections: there are double poles at all integer values of $h$ except for $h=1$, which is simple, and $h=-2$ with multiplicity equal to $3$. Additionally, simple poles sit at $h=-1/2,\,-3/2$. Since the factor containing gamma functions in $\mathcal{M}^{m}_2(h,\xi,0)$ is invariant under the replacement $h\to-h$, one expects some symmetry in the infinite sums appearing in the expansions for small and large mass. Indeed, after applying the converse mapping theorem, we get the following expansions:
\begin{align}
\Delta_0 F_2^m(\xi) =\,& \Delta^\infty_0 F_2^m(\xi) -\sum_{n=1}A_n(\xi)\\
=\,&\frac{3 \pi ^2 \xi }{8} -\frac{55 \pi ^2 \xi ^3}{72} + \xi ^4 \biggl(\frac{3}{2}L_\xi^2 -3 L_\xi+\frac{\pi ^2}{4}+\frac{145}{48}\biggl)+\sum_{n=1}^{n\neq 2}A_{-n}(L_\xi)\,, \nonumber\\
A_n(\xi) =\, & \frac{\xi ^{-2n}}{n^2 (n+2) (2 n+1) (2 n+3)}\Biggl[(2+4n+3n^2)(n-1)L_\xi \nonumber\\
&+\frac{3 \bigl(8 n^6+20 n^5-45 n^3-68 n^2-42 n-8\bigr)}{2 n (n+2) (2 n+1) (2 n+3)}\Biggr].\nonumber
\end{align}
For the limiting case $\xi=1$ one finds $\Delta_0 F_2^m(1) = \zeta_3/3+83/36-\pi ^2 / 108$. The series for large $\xi$ can be summed up and we find the following expression, in which all terms are manifestly real for $\xi>1$:
\begin{align}\label{eq:F2Mexact}
\Delta_0 F_2^m=\,& \frac{\pi ^2 \xi ^4}{8}-\frac{55 \pi ^2 \xi ^3}{108}+\frac{83 \xi ^2}{36}+\frac{13}{18}L_\xi^2+
\!\biggl(\frac{14 \xi ^2}{9}+\frac{\pi ^2}{6}\biggr)L_\xi +\frac{\pi ^2 \xi }{4} -\frac{10}{9}L_\xi^3 + \frac{13 \pi^2}{216}\\&
+\frac{1}{3} \biggl[\text{Li}_3\biggl(\frac{1}{\xi ^2}\biggr)+2L_\xi^2 \log \bigl(\xi ^2-1\bigr)\biggr]\!
+\biggl(\frac{13}{36}-\frac{3 \xi^4}{4}-\frac{L_\xi}{3}\biggr) \text{Li}_2\biggl(1-\frac{1}{\xi ^2}\biggr)\nonumber\\
& +\biggl(\frac{3 \xi }{4}-\frac{55 \xi ^3}{36}\biggr) \biggl[L_\xi \log (\xi +1)-\frac{1}{2} \log ^2(\xi +1)+\text{Li}_2(1-\xi )-\text{Li}_2\biggl(\frac{\xi }{\xi +1}\biggr)\biggr]\nonumber\,.
\end{align}
For $\xi<1$ the expression is still real, but the individual pieces on the term in square brackets in the second line become complex. To have each term manifestly real, the following substitution should be implemented:
\begin{equation}
\text{Li}_3\biggl(\frac{1}{\xi ^2}\biggr)+2L_\xi^2 \log \bigl(\xi ^2-1\bigr) \to \text{Li}_3\bigl(\xi ^2\bigr) + 2 L_\xi^2\log \bigl(1-\xi ^2\bigr)+\frac{4}{3}L_\xi^3-\frac{2}{3} \pi ^2 L_\xi\,.
\end{equation}
In Fig.~\ref{fig:MgSec}, a comparison of the exact expression and both expansions is shown at various orders. We see that at $\xi = 1$ both expansions converge very quickly, although the large-mass one is accurate already at very low orders, and the small-mass series exhibits an oscillatory behavior related to the poles at half-integer values: after the third power of $\xi$ is included, the sign alternating behavior disappears. Both expansions excel reproducing the exact result in their respective convergence domains, but the large-mass expansion is particularly fast at approaching the all-order result.

From these results we can compute the bHQET flavor matching condition, that is, the Wilson coefficient relating bHQET dijet operators in two consecutive EFTs, the UV one with an active secondary massive quark, the IR one where the secondary quark has been integrated out. Using Eq.~\eqref{eq:bHQETMatchGeneric} we obtain
\begin{align}
\mathcal{M}^{(n_\ell + 1\to n_\ell)}_{\rm bHQET}(\hat M, m, \mu) = \,& 1 + \biggl[\frac{\alpha_s(\mu)}{\pi}\biggr]^2C_FT_F \mathcal{M}^{(2)}_{\rm bHQET}\biggl(\hat M,\frac{m}{\mu}\biggr) \,,\\
\mathcal{M}^{(2)}_{\rm bHQET}\biggl(\hat M,\frac{m}{\mu}\biggr) = \,& \mathcal{M}^{(2)}_{\rm SCET}(m,Q,\mu) - \frac{2}{3} F^m_{1,\rm ren}\!\biggl(0, \frac{M}{\mu}\biggr) \log\biggl(\frac{\mu}{m}\biggr)\nonumber\\
&- F^m_{2,\rm ren}\!\biggl(0, \frac{M}{\mu}\biggr) - \Delta^\infty_0 F_2^m(\xi)\nonumber\\
= \,& \frac{1}{3}\Bigl[\log\bigl(-\hat M^2\bigr)+1\Bigr]
\biggl[ \log ^2\biggl(\frac{\mu}{m}\biggr)-\frac{5}{3} \log \biggl(\frac{\mu}{m}\biggr)+\frac{7}{9}\biggr]
\,,\nonumber
\end{align}
where at this order, $\alpha_s$ can be used with either $n_\ell+1$ or $n_\ell$ running flavors. Using Eq.~\eqref{eq:SCETMatchGeneric} we obtain the following result for the SCET flavor condition corrected for the effects of a non-zero primary quark mass $M$, which is an active degree of freedom in both EFTs, but where the secondary quark appears only in the SCET with $n_f+1$ active flavors
\begin{align}
\mathcal{M}^{n_f + 1\to n_f}_{\rm SCET}(M, Q, m, \mu) = \,& 1 + \biggl[\frac{\alpha_s(\mu)}{\pi}\biggr]^2C_FT_F\mathcal{M}^{(2)}_{\rm SCET}(M,m,Q,\mu) \,,\\
\mathcal{M}^{(2)}_{\rm SCET}(M,m,Q,\mu) = \,& \Delta_0 F^m_2(\xi) + \mathcal{M}^{(2)}_{\rm SCET}(m,Q,\mu) - \Delta^\infty_0 F_2^m(\xi)\,,\nonumber
\end{align}
where at the order we are working, the strong coupling can be evolved with either $n_f+1$ or $n_f$ flavors. Next we compute the dispersive contribution to the SCET and bHQET form factors separately. For bHQET we compute the finite and divergent parts of the dispersive contribution, which can be combined in a concise form:
\begin{align}
H_{2}^{\rm bHQET}\biggl(\hat M, \frac{m}{\mu},\varepsilon\biggr) =\,& \mathcal{M}^{(2)}_{\rm bHQET}\!\biggl(\hat M,\frac{m}{\mu}\biggr) + \frac{\delta Z_{1}^{\rm bHQET}}{12 C_F} \biggl[\! \biggl( \frac{\mu^2e^{\gamma_E}}{m^2} \biggr)^{\!\!\varepsilon} \Gamma (\varepsilon) -\frac{1}{\varepsilon}\biggr]-\!\frac{\delta Z_{2,n_f}^{\rm bHQET}}{16 C_F T_F} \nonumber\\
=\, & \bigl[\log(-\hat M^2) + 1\bigr]\biggl\{\frac{1}{12 \varepsilon ^2}+\frac{1}{\varepsilon}\biggl[\frac{1}{3} \log \biggl(\frac{\mu }{m}\biggr)-\frac{5}{36}\biggr] \nonumber\\
& + \frac{2}{3} \log^2\biggl(\frac{\mu}{m}\biggr)-\frac{5}{9} \log \biggl(\frac{\mu}{m}\biggr)+\frac{\pi ^2}{72}+\frac{7}{27}\biggr\}\,.
\end{align}
We proceed in the same way for the SCET dispersive contribution with massive primary and secondary quarks and find
\begin{align}
H_{2}^{\rm SCET}\!\biggl(\hat M, \frac{m}{\mu},\varepsilon\biggr) =\,& \mathcal{M}^{(2)}_{\rm SCET}(M,m,Q,\mu) + \frac{\delta Z_{1}^{\rm SCET}}{12 C_F} \biggl[\! \biggl( \frac{\mu^2e^{\gamma_E}}{m^2} \biggr)^{\!\!\varepsilon} \Gamma (\varepsilon) -\frac{1}{\varepsilon}\biggr]-\!\frac{\delta Z_{2,n_f}^{\rm SCET}}{16 C_F T_F} \nonumber\\
=\, & \mathcal{M}^{(2)}_{\rm SCET}(M,m,Q,\mu) + \frac{1}{8 \varepsilon ^3} +\frac{1}{\varepsilon^2}\biggl[\frac{1}{12} \log (-\hat m^2)-\frac{1}{6} \log
\biggl(\frac{\mu }{m}\biggr)+\frac{1}{18}\biggr] \nonumber \\
&-\frac{1}{\varepsilon}\biggl\{\biggl[\frac{1}{3} \log (-m^2)+\frac{7}{9}\biggr] \log \biggl(\frac{\mu}{m}\biggr)
+\frac{5}{36} \log (-\hat m^2)+\log ^2\biggl(\frac{\mu }{m}\biggr)\nonumber \\
&+\frac{5 \pi ^2}{144}+\frac{65}{432}\biggr\} - \frac{\pi ^2}{72}
\log (-\hat m^2)-\frac{8}{9} \log ^3\biggl(\frac{\mu }{m}\biggr)-\frac{\pi ^2}{18} \log \biggl(\frac{\mu }{m}\biggr)\nonumber\\
& +\frac{\zeta_3}{18}-\frac{\pi ^2}{48}-\biggl[\frac{1}{3} \log (-\hat m^2)+\frac{1}{2}\biggr] \log ^2\biggl(\frac{\mu }{m}\biggr)\,.
\end{align}

We finish the section by, starting from our results, computing the contribution of a secondary massive bubble to the SCET-bHQET Wilson coefficient, for the case in which the particle running in the bubble is also the primary quark. This contribution does not coincide with $F_2^m(M,M,\mu,\varepsilon)$ since the massive bubble with a primary quark contributes to the SCET part, but does not appear on the bHQET since in this EFT the mass $M$ is not an IR scale. The renormalized correction can be expressed as\footnote{One-loop quantities refer to the physical case of a massless gluon, even if no argument is shown.}
\begin{align}
F_{2,M}^{m}(M) =\, & \biggr\{H^{\rm SCET}_{2}\!\biggl(\hat M, \frac{M}{\mu},\varepsilon\biggr) -\frac{1}{3}F_1^{\rm SCET} \biggl[\! \biggl( \frac{\mu^2e^{\gamma_E}}{M^2} \biggr)^{\!\!\varepsilon} \Gamma (\varepsilon) -\frac{1}{\varepsilon} -\log\biggl(\frac{\mu^2}{M^2}\biggr)\!\biggr]\!\biggr\}_{\!\rm fin}\\
=\, & F_{2,\rm ren}^m(M,M,\mu) + H_{2,\rm ren}^{\rm bHQET}\biggl(\hat M, \frac{M}{\mu}\biggr)
+\frac{1}{3} F_{1,\rm ren}^{m} L_M + \frac{\varepsilon}{3}\frac{\delta Z_{1}^{\rm bHQET}}{4 C_F} \biggl( \frac{1}{2}L_M^2+\frac{\pi^2}{12}\biggr)
\nonumber\\
=\, &\frac{1}{18}L_M^3-\frac{1}{72} L_M^2- L_M\biggl(\frac{\pi ^2}{24}+\frac{65}{216} \biggr)-\frac{\zeta_3}{36}-\frac{41 \pi ^2}{432}+\frac{5107}{2592}
\nonumber\\
& + \log(-\hat M^2)\biggl(\frac{1}{12} L_M^2-\frac{5}{18} L_M+\frac{7}{27}\biggr)\,.\nonumber
\end{align}
To arrive to the second expression we have used that since the term in square brackets on the first line is $\mathcal{O}(\varepsilon)$,
one can replace $F_1^{\rm SCET}$ by $-\delta Z_1^{\rm SCET}/(4C_F)$. After this replacement, since only the finite part is taken, the logarithm within square brackets can be dropped. Furthermore, we have written
\begin{equation}
H^{\rm SCET}_{2,\rm ren} = F_{2,\rm ren}^{m} + H_2^{\rm bHQET}+\frac{1}{3}\biggl\{\!\biggl(F_{1,\rm ren}^{m} + \frac{\delta Z_1^m}{4C_F}\biggr)\! \biggl[\! \biggl( \frac{\mu^2e^{\gamma_E}}{M^2} \biggr)^{\!\!\varepsilon} \Gamma (\varepsilon) -\frac{1}{\varepsilon} \biggr]\!\biggr\}_{\rm fin}\,,
\end{equation}
grouped both $Z$ factors and noted that $\delta Z_1^m+\delta Z_1^{\rm SCET}=\delta Z_1^{\rm bHQET}\propto 1/\varepsilon$. Since only finite terms are kept and $F_{1,\rm ren}^{m}$ is already finite, we only need to consider the $\mathcal{O}(\varepsilon^0)$ piece of the expression within brackets that multiplies it (that is, $L_M$). Likewise, it suffices keeping only the $\mathcal{O}(\varepsilon)$ piece of the term within square brackets [\,that is, $\varepsilon(L_M^2/2+\pi^2/12)$\,] that multiplies $\delta Z_1^{\rm bHQET}$. Our result is in complete agreement with Eq.~(4.30) of Ref.~\cite{Hoang:2015vua}, which is a highly nontrivial check on many computations carried out in various sections of this article.

\subsection{Jet function}\label{sec:Bjet}
The last set of computations that are discussed in this article correspond to the bHQET hemisphere jet function, which can be used for $2$-jettiness, heavy-jet-mass and \mbox{$C$-jettiness}. The momentum-space jet function depends on the variable $\hat s$, which has dimensions of energy. Most of the comments made for the SCET function apply here as well: distributions show up in momentum space, but turn into regular functions for the cumulative jet function defined as
\begin{equation}
\Sigma^B_n(\hat s_c,\mu) = \int_0^{\hat s_c} {\rm d}\hat s\,B_n(\hat s,\mu)\,.
\end{equation}
The virtual diagrams contain only distributions, while the real-radiation contribution is composed entirely by regular functions. The former can be isolated from the residue of the only pole on the positive real axis, that is, at $h=\varepsilon$, while the latter is the sum of poles for $h\leq 0$ obtained once $\varepsilon$ has been set to zero. Although the jet function $B_n$ has dimensions of the inverse of an energy squared, it is proportional to the mass of the primary quark $M$, hence in this section we will show results for $MB_n$, which has dimensions of the inverse of an energy. In particular, the following relations concerning the renormalization of the bHQET jet function must be observed:
\begin{align}
B_n^{\rm bare}(\hat s) =\, & \int {\rm d}\hat s' Z_B(\hat s-\hat s',\mu) B_n(\hat s',\mu)\,, \\
Z_B(\hat s,\mu) \doteq\, & \delta(\hat s) + \frac{\alpha_s(\mu)}{\pi}C_F\delta Z_1^B + \biggl[\frac{\alpha_s(\mu)}{\pi}\biggr]^2C_FT_F\delta Z_2^B \,,\nonumber\\
\delta Z_i^B =\, & \delta Z_{i,\rm nc}^B \delta(s) + \frac{\delta Z_{i,\rm cusp}^B}{\mu}\biggl[\dfrac{\mu}{\hat s}\biggr]_{+} \nonumber\,,\\
\delta Z_{i,\rm cusp}^B =\, & \sum_{j=1}^i \frac{\delta Z_{i,j,\rm cusp}^B}{\varepsilon^j}\,,\qquad
\delta Z_{i,\rm nc}^B = \, \sum_{j=1}^{i+1} \frac{\delta Z_{i,j,\rm nc}^B}{\varepsilon^j}\,.\nonumber
\end{align}
The evolution of the renormalized bHQET jet function involves a convolution with the corresponding anomalous dimension, which consists on two distributions:
\begin{align}\label{eq:BGamma}
\mu \dfrac{\rm d}{{\rm d}\mu} B_n(\hat s,\mu) \, = & \int {\rm d}\hat s' \gamma_B(\hat s-\hat s') B_n(\hat s',\mu)\,,\\
\gamma_B(\hat s,\mu) \, = & -\! \frac{2\Gamma_{\rm cusp}}{\mu}\biggl[\dfrac{\mu}{\hat s}\biggr]_{+} + \gamma^B_{\rm nc} \delta(\hat s)\,,\nonumber\\
\gamma_{\rm nc}^B \, = &\sum_{n = 1} \gamma^B_n \biggl[\frac{\alpha_s (\mu)}{4 \pi} \biggr]^n .\nonumber
\end{align}
The anomalous dimensions are proportional to the $1/\varepsilon$ terms in the $Z$ factor:
\begin{align}\label{eq:BGammaZ}
\Gamma_0 =\, & -\!4 C_F\delta Z_{1,1,\rm cusp}^B\,, &\Gamma_1^{(n_f)} &= - 32 C_FT_F\delta Z_{2,1,\rm cusp}^B\,,\\
\gamma_0^B =\, & 8 C_F\delta Z_{1,1,\rm nc}^B\,, &\gamma_{1,n_f}^B &= 64 C_FT_F\delta Z_{2,1,\rm nc}^B\,,\nonumber
\end{align}
where we have assumed the expected cancellation between $1/\varepsilon^n$ terms with $n>1$ takes place.
The master piece for our final analysis is the one-loop computation of $M \Sigma_n^B$ with a modified gluon propagator. In this case we identify $\mathcal{Q}=\hat s_c$, and use the result computed in Ref.~\cite{Gracia:2021nut}:
\begin{equation}
m_1 (h,\varepsilon) = \frac{\Gamma (2 + h - \varepsilon)}{2(h - \varepsilon)^2\Gamma (1 - h) \Gamma (2 + 2 h - 2 \varepsilon)} \,,
\end{equation}
where a double pole is located at $h=\varepsilon$. We label quantities related with the differential and cumulative bHQET jet functions with $B$ and $\Sigma_B$ superscripts, respectively.

\subsubsection{Massive vector Boson}
To simplify notation as much as possible, we define the dimensionless variable $\tilde s_g=\hat s/m_g$. The most relevant results obtained after multiplying by the $\Gamma(h)\Gamma(1-h)$ kernel necessary to compute the massive vector boson corrections are collected below:
\begin{align}\label{eq:MgB}
\mathcal{M}_1^{\Sigma_B}(h, 0,\tilde s_g ) \!=\, & \frac{(h+1) \Gamma^2(h) \tilde s_g^{2 h}}{2 h \Gamma (2 h+2)}
\xrightarrow[| h | \gg 1]{} \frac{\sqrt{\pi }}{4h^{5/2}} \biggl(\frac{\tilde s_g}{2}\biggr)^{\!\!2 h},\\
\delta Z_1^B(\hat s, \mu, \varepsilon) =\, &\frac{1}{2}\biggl(\frac{1}{\varepsilon^2} + \frac{1}{\varepsilon}\biggr)\delta(\hat s)
- \frac{1}{\varepsilon}\frac{1}{\mu}\biggl(\frac{\mu}{\hat s}\biggr)_{\!\!+},\nonumber\\
F_{1,\rm ren}^{B}(0, \hat s, \mu) =\, & \biggl(1 - \frac{\pi^2}{8}\biggr)\delta(\hat s) -\frac{1}{\mu}\biggl[\dfrac{\mu}{\hat s}\biggr]_+
+ \frac{2}{\mu}\biggl[\dfrac{\mu}{\hat s}\log\biggl(\frac{\hat s}{\mu}\biggr)\biggr]_+ \,,\nonumber\\
m_g\Delta_0^\infty F_1^B (\tilde s_g)=\, & \biggl(\frac{\pi^2}{12} -1\biggr)\delta(\tilde s_g) + \biggl[\frac{1}{\tilde s_g}\biggr]_+
-2 \biggl[\frac{\log(\tilde s_g)}{\tilde s_g}\biggr]_+,\nonumber
\end{align}
where from the first line we determine that $\tilde s_g=2$ is the boundary between the large- and small-mass expansions, where both will converge due to the overall $h^{-5/2}$ suppression factor. The results in the second and third lines agree with Ref.~\cite{Fleming:2007xt}. Furthermore, the result for the $Z$ factor reproduces the one-loop cusp anomalous dimension and predicts the one-loop non-cusp piece: $\gamma_0^B=4C_F=16/3$.

To obtain the Mellin-Barnes transform for the differential bHQET jet function we apply a derivative with respect to $\hat s$ and find $\mathcal{M}^{B}_1(h,\tilde{s}_g,0) = 2h \mathcal{M}^{\Sigma_B}_1(h,\tilde s_g,0)/\hat s$. Even after setting $\varepsilon=0$ there is infinite number of poles on the real negative axis. The virtual contribution to the bare bHQET jet function is given by the pole at $h=\varepsilon$:
\begin{align}
F_1^{B,\rm virt} =\, & \frac{\Gamma (\varepsilon )}{2} \biggl(\frac{e^{\gamma_E}\mu ^2}{m_g^2}\biggr)^{\!\!\varepsilon }
\biggl\{\biggl[1-H_{\varepsilon -1}- \log \biggl(\frac{\mu^2}{m_g^2}\biggr)\biggr]\delta(\hat s) -\frac{2}{\mu} \biggl(\frac{\mu}{\hat s}\biggr)_{\!\!+}\biggr\}\\
= \,&\delta Z_B^1(\hat s,\mu,\varepsilon) + F^{B,\rm virt}_{1,\rm ren}(\hat s,m_g,\mu) +\mathcal{O}(\varepsilon)\,,\nonumber \\
F^{B,\rm virt}_{1,\rm ren} = \,& \biggl[\log \biggl(\frac{\mu }{m_g}\biggr)-\log ^2\biggl(\frac{\mu }{m_g}\biggr)-\frac{\pi ^2}{24}\biggr]\delta(\hat s)
- 2\log \biggl(\frac{\mu }{m_g}\biggr) \frac{1}{\mu}\biggl(\frac{\mu}{\hat s}\biggr)_{\!\!+}\,.\nonumber
\end{align}

\begin{figure*}[t!]
\subfigure[]
{\includegraphics[width=0.49\textwidth]{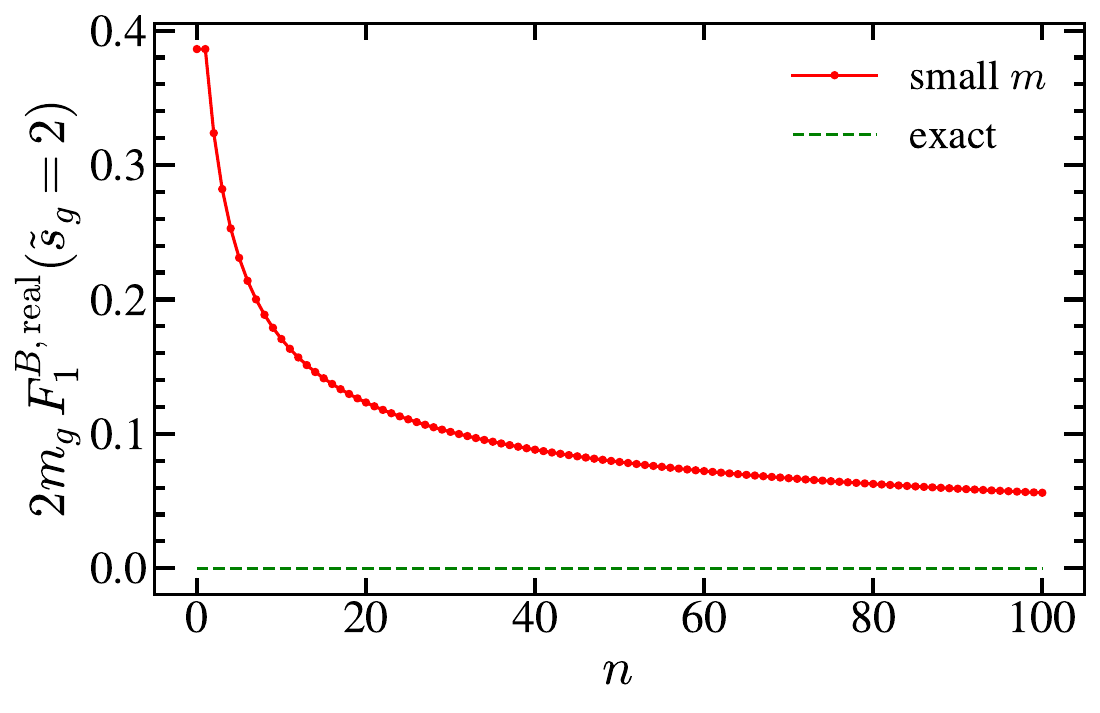}
\label{fig:BJetMg1}}
\subfigure[]{\includegraphics[width=0.49\textwidth]{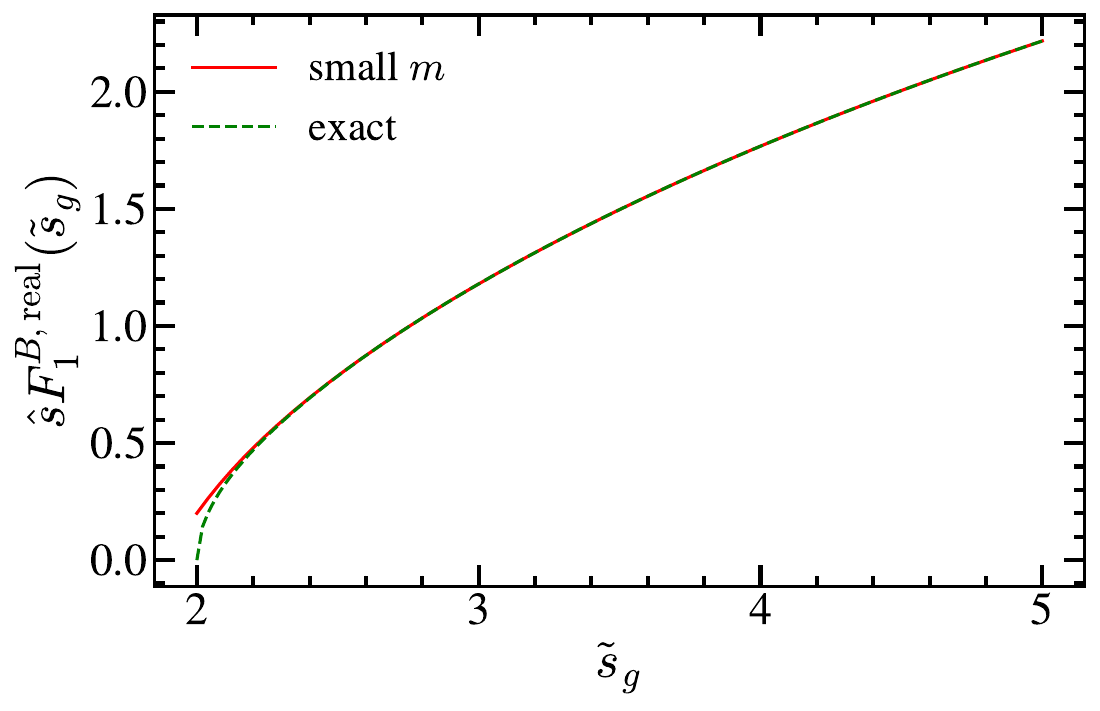}
\label{fig:BJetMg}}
\caption{Real-radiation term of the one-loop massive vector boson bHQET jet function. We show exact results as green dashed lines, and the expansions for small masses as red solid lines. Left panel: small mass expansion of $2m_g F^{B,\rm real}_1$ at threshold $\tilde s_g=2$ as a function of the expansion order $n$. Right panel: Dependence of $\hat s F^{B,\rm real}_1$ with $\tilde s_g$, including $6$ non-zero terms in the expansion.}
\label{fig:MgBJet}
\end{figure*}

To obtain the $\mu$-independent and UV-finite real-radiation contribution we simply sum all poles on the real axis for $h\leq 0$, that is, the double pole at $h=0$ and the simple poles at all negative integer values of $h\leq -2$. The resulting series can be summed up:
\begin{align}
F_1^{B,\rm real} =\, & \frac{1}{\hat s}\biggl[2\log(\tilde s_g) - 1 - 2 \sum_{n=2}\frac{(n-1) (2 n-2)! }{(n!)^2\,\tilde s_g^{2 n}}\biggr]\\
= \,&\frac{1}{\hat s}\Biggl\{2 \log \biggl[\frac{1}{2} \biggl(\sqrt{\tilde s_g^2-4}+ \tilde s_g\biggr)\biggr]-\sqrt{1-\frac{4}{\tilde s_g^2}}\,\Biggr\}\theta(\tilde s_g -2) \,, \nonumber
\end{align}
where in the second line every term is manifestly real for the function's domain and the theta has been obtained from the convergence radius of the series. It is not hard to see that, as expected, the real-radiation contribution vanishes at $\tilde s_g=2$. In Fig.~\ref{fig:MgBJet} we compare the series expansion of the real-radiation contribution for small gluon masses with the exact result. We observe an excellent convergence everywhere except for $\hat s_g = 2$ where convergence is rather slow.

\begin{figure*}[t!]
\subfigure[]
{\includegraphics[width=0.495\textwidth]{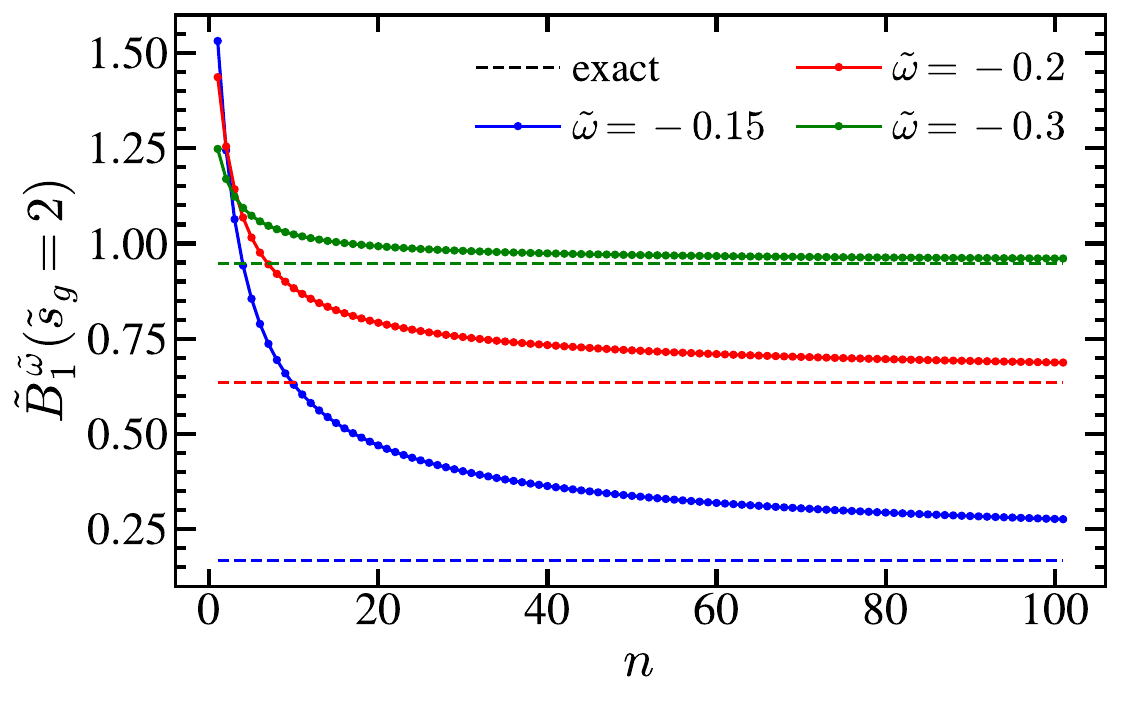}
\label{fig:BJetMgRG1}}
\subfigure[]{\includegraphics[width=0.485\textwidth]{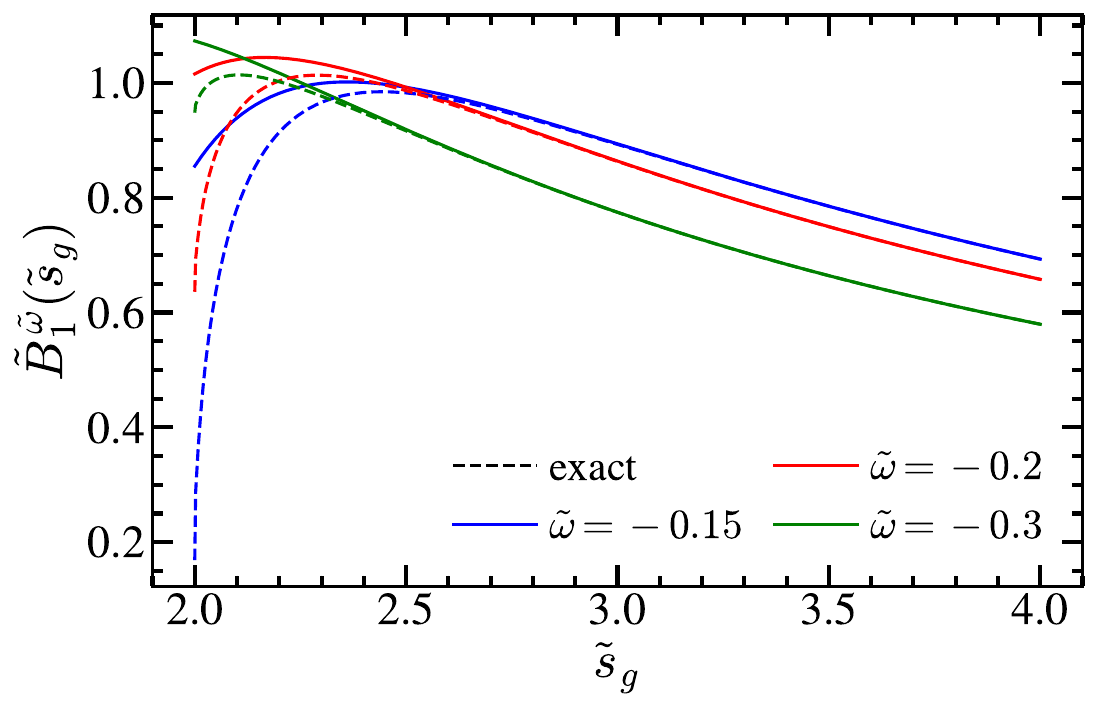}
\label{fig:BJetRGMg}}
\caption{Gluon mass correction to the massless RG-evolved bHQET jet function in its exact form (dashed), and expansions for small masses (solid lines) for three values of $\tilde \omega$: $-0.15$ (blue), $-0.2$ (red), and $-0.3$ (green). Left panel: small mass expansion for $\tilde B_1^{\tilde\omega}$ at threshold $\tilde s_g=2$ as a function of the expansion order $n$. Right panel: Dependence of $\tilde B_1^{\tilde\omega}$ with $\tilde s_g$, including $6$ non-zero terms in the small-mass expansion.}
\label{fig:MgBJetRG}
\end{figure*}

The one-loop correction to the bHQET jet matching coefficient relating the jet functions in the theories with and without massive vector bosons is given by
\begin{equation}
\! B_n^{(n_g)}(\hat s)=\!\!\int_0^{\hat s} \!{\rm d}\hat s' \mathcal{M}^{(n_g\to n_\ell)}_B(\hat s-\hat s') B_n^{(n_\ell)}(\hat s'),\qquad\mathcal{M}^{(n_g\to n_\ell)}_B=\delta(\hat s)+\frac{\alpha_s(\mu)}{\pi} C_FF^{B,\rm virt}_{1,\rm ren}.
\end{equation}
The gluon mass correction to the renormalized one-loop massless bHQET jet function can be written as
\begin{equation}\label{eq:TildeB}
M\Delta_0 B_1(\hat s, m_g) \equiv \frac{1}{\hat s} \tilde B_1(\tilde s_g)\,,
\end{equation}
and the RG-evolved mass correction for the bHQET jet function is expressed in terms of $\tilde B_1$ as follows
\begin{align}\label{eq:BRGEvolved}
\tilde B_1^{\tilde \omega}(\tilde s_g) =\ \int_0^{\tilde s_g} \frac{{\rm d}\hat s'}{\hat s'}
\biggl(1-\frac{\hat s'}{\tilde s_g}\biggr)^{\!-1-\tilde \omega} \tilde B_1(\hat s')
= \int_{c - i \infty}^{c + i \infty} \frac{{\rm d} h}{2 \pi i} \frac{(h+1) \Gamma^2(h) \tilde s_g^{2 h}}{2 h (2 h+1) (-\tilde \omega)_{2 h}}\,,
\end{align}
where $-1/2<c<0$. Using the converse mapping theorem we find
\begin{align}\label{eq:MBjetRGMg}
\tilde B_1^{\tilde \omega}(\tilde s_g) =\, &\tilde L_g -\tilde L_g^2+\psi ^{(1)}(-\tilde \omega)-\frac{\pi ^2}{12}-1 \\
=\, & \frac{\pi (\tilde \omega+1)}{\tilde s_g}
-\sum_{n=1} \frac{ (\tilde \omega+1)_{2 n}\tilde s_g^{-2 n}}{n (2 n-1) (n!)^2} \bigg\{(n-1)(\tilde L_g + H_n)+\frac{2 n^2-4 n+1}{2 n (2 n-1)}\nonumber\\
& +(n-1)\bigl[\psi ^{(0)}(\tilde \omega+1)-\psi ^{(0)}(2 n+\tilde \omega+1)\bigr]\biggr\}\,,\nonumber
\end{align}
with $\tilde L_g = \log(\tilde s_g) - \gamma_E - \psi ^{(0)}(-\tilde \omega)$. The leading-mass correction comes from the simple pole at $h=-1/2$, directly related to the $u=1/2$ renormalon of the bHQET jet function. All poles at negative integer values are double except for that at $h=-1$, which is simple. To obtain the previous result we have used the following relation among digamma functions:
$\psi ^{(0)}(-2 n-\tilde \omega) - \psi ^{(0)}(-\tilde \omega)= \psi ^{(0)}(2 n+\tilde \omega+1)-\psi ^{(0)}(\tilde \omega+1)$.
For a better numerical implementation, it is convenient to write
\begin{equation}
\psi ^{(0)}(\tilde \omega+1)-\psi ^{(0)}(2 n+\tilde \omega+1) = - \sum_{i=1}^{2n}\frac{1}{\tilde \omega + i}\,.
\end{equation}
We have not been able to sum up the series in Eq.~\eqref{eq:MBjetRGMg}, therefore we consider ``exact'' the sum of the first $140$ terms. In Fig.~\ref{fig:MgBJetRG} we study the convergence of the series for the RG-evolved bHQET jet function for three values of the running parameter $\tilde \omega$. We observe that, analogously to our findings for the real-radiation bHQET jet function, convergence is not great at $\tilde s_g=2$, although it becomes better for values of $\tilde \omega$ further away from zero. Likewise, the series converges rapidly for $\tilde s_g>2$, and the more negative $\tilde \omega$ is, the faster it converges.
The Fourier transform of the bHQET jet function is defined in analogy to Eq.~\eqref{eq:jet-fourier} replacing $J\to MB$ and $s\to \hat s$, such that the dimensions of the Fourier variable $y$ is of the inverse of an energy. Using the converse mapping theorem on the corresponding Mellin transform, closing the contour integral towards the negative real axis we find for the Fourier transform of $M\Delta_0 B_1(\hat s, m_g)$ the following expression:
\begin{align}
\Delta_0 \tilde B_1(y_g) =\, & \int_{c - i \infty}^{c + i \infty} \frac{{\rm d} h}{2 \pi i} \frac{(h+1) \Gamma^2(h) (i y_g)^{-2 h}}{2 h (2 h+1)} \\
=\, & -\!i\pi y_g +\sum_{n=1} \frac{(i y_g)^{2 n}}{n (2 n-1) (n!)^2}
\biggl[(n-1) (L_y-H_n)+\frac{4 n-2 n^2-1}{2 n (2
n-1)}\biggr]\,,\nonumber
\end{align}
where $-1/2 < c < 0$, and $y_g$ and $L_y$ were defined after Eq.~\eqref{eq:FourJMg} and no analytic form could be found since the infinite sum cannot be expressed in terms of standard functions.

\subsubsection{Secondary massive bubble}
The last application of the method introduced in this article, the computation of the contribution from secondary massive quark bubbles to the bHQET jet function, will again produce new results and confirm earlier computations. To write expressions as simple as possible, we use the dimensionless variable $\tilde s = \hat s/m$, which is used for both the differential and cumulative versions of the bHQET jet function. A summary of the most relevant pieces for the cumulative and differential bHQET jet functions follows:
\begin{align}\label{eq:MassBubJetbHQET}
\mathcal{M}^{\Sigma_B}_2(h,\tilde{s}_c,0) =\, & \frac{(1+h)^2 \Gamma^4(h)\tilde s_c^{2h}}{2 (2 h+3) \Gamma^2(2 h+2)}
\xrightarrow[| h | \gg 1]{} \frac{\pi }{16 h^4}\left(\frac{\tilde{s}_c}{4}\right)^{2h} \,,\\
\delta Z^B_{2}(\hat{s},\mu,\varepsilon) = \,& \biggl[\frac{1}{8 \varepsilon^3} + \frac{1}{72\varepsilon^2} + \frac{1}{\varepsilon}\left(\frac{\pi^2}{144} - \frac{29}{108}\right) \biggr]\delta(\hat{s})
+\frac{1}{6} \biggl(\frac{5}{3\varepsilon}-\frac{1}{\varepsilon^2}\biggr)
\dfrac{1}{\mu}\biggl[\dfrac{\mu}{\hat{s}}\biggr]_+\,,\nonumber\\
F^B_{2, \rm ren}(0,\hat{s},\mu,\varepsilon) =\, & \biggl(\frac{17\zeta_3}{36} + \frac{59 \pi^2}{432} - \frac{281}{162}\biggr)\delta(\hat{s}) + \biggl(\frac{47}{27}- \frac{\pi^2}{9} \biggr) \dfrac{1}{\mu}\biggl[\dfrac{\mu}{\hat{s}}\biggr]_+ \nonumber\\
& - \frac{16}{9} \dfrac{1}{\mu}\biggl[\dfrac{\mu}{\hat{s}}\log\biggl(\frac{\hat{s}}{\mu}\biggr)\biggr]_+
+ \frac{2}{3} \dfrac{1}{\mu}\biggl[\dfrac{\mu}{\hat{s}}\log^2\biggl(\frac{\hat{s}}{\mu}\biggr)\biggr]_+ \,,\nonumber\\
m \Delta^\infty_0 F^B_2(\tilde s) = \, & -\!\frac{2}{3} \biggl[\dfrac{\log^2(\tilde s)}{\tilde s}\biggr]_++\frac{16}{9}\biggl[\dfrac{\log(\tilde s)}{\tilde s}\biggr]_+
+\biggl(\frac{\pi^2}{9} - \frac{61}{27}\biggr) \biggl[\dfrac{1}{\tilde s}\biggr]_+ \nonumber\\
& + \biggl(\frac{223}{81}-\frac{2 \zeta_3}{3}-\frac{4 \pi ^2}{27} \biggr)\delta(\tilde{s}) \,.\nonumber
\end{align}
\begin{figure*}[t!]
\subfigure[]
{\includegraphics[width=0.503\textwidth]{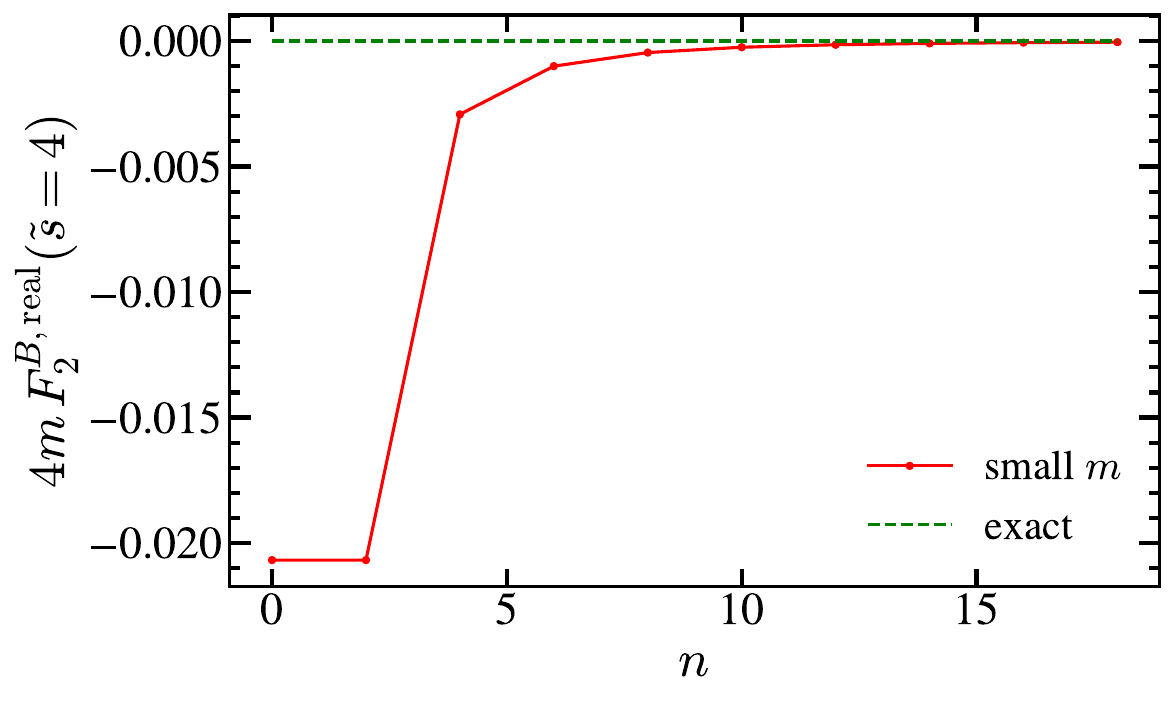}
\label{fig:BJetSec1}}
\subfigure[]{\includegraphics[width=0.4707\textwidth]{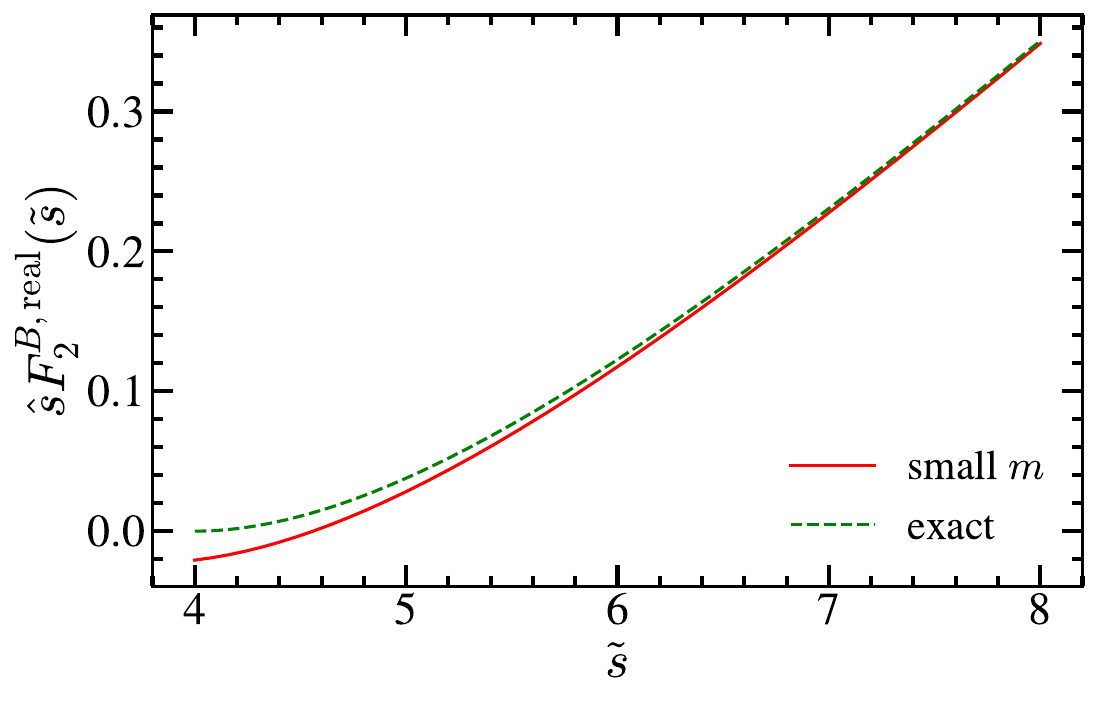}
\label{fig:BJetSec}}
\caption{Real-radiation contribution to the two-loop bHQET jet function due to a secondary massive quark bubble. Exact results are shown as green dashed lines, and the expansions for small masses as red solid lines. Left panel: small mass expansion of $4m F^{B,\rm real}_2$ at threshold $\tilde s=4$ as a function of the expansion order $n$. Right panel: Dependence of $\hat s F^{B,\rm real}_2$ with $\tilde s$, including a single non-zero term in the expansion.}
\label{fig:SecBJet}
\end{figure*}
The limit taken on the first line establishes the boundary between the two expansions at $\tilde{s}=4$. The second and third lines agree with the massless result computed in Ref.~\cite{Jain:2008gb}. From the divergent pieces we confirm the flavor piece of the two-loop cusp anomalous dimension, and obtain the corresponding term for the non-cusp anomalous dimension:
\begin{equation}
\gamma_{1,n_f}^B = C_FT_F\biggl(\frac{4 \pi ^2}{9}-\frac{464}{27}\biggr) = \frac{8 \pi ^2}{27}-\frac{928}{81}\,,
\end{equation}
in agreement with Eq.~(41) of Ref.~\cite{Jain:2008gb}. For $\tilde s < 4$ only virtual diagrams contribute, given by minus the residue of the pole at $h=\varepsilon$. After removing the massless limit (which also removes the $\mu$ dependence and UV singularities), this contribution coincides with $\Delta^\infty_0 F^B_2(\tilde s)$ and can be obtained as the residue of the pole at $h=0$ computed after setting $\varepsilon=0$. The real-radiation correction to the massless limit is obtained as the sum of the infinite poles for $h \leq 0$, that is, a triple pole sitting at $h=0$ plus double poles at all integer negative values of $h<-1$. For that, we use the MB transform of the differential bHQET jet function, and obtain
\begin{align}
\hat{s}\,\Delta_0 F_{2,\rm real}^B =&\biggl\{\frac{61}{27}-\frac{\pi^2}{9}-\frac{16}{9}\log(\tilde{s})+\frac{2}{3}\log^2(\tilde{s})
+32\sum_{n=2}^{\infty} \frac{ (2 n-3) [(2 n-4)!]^2 \tilde s^{-2 n}}{n^3[(n-2)!]^4}\\
& \times\!\biggl[\log(\tilde s) + 2 H_n-2 H_{2 n-2}+\frac{9 n-3-4 n^2}{2 n (n-1) (2 n-3)}\biggr]\biggr\} \theta(\tilde{s}-4)\,,\nonumber
\end{align}
where the Heaviside theta appears from the condition of the convergence of the sum. For $\tilde s = 4$ the series indeed sums to zero, as can be seen in Fig.~\ref{fig:BJetSec1}, and this happens at a fast pace, signaling the excellent convergence of the series. We have not been able to find a closed form for the sum, but at the sight of its fast convergence, we assume that after summing up 160 terms the result can be regarded as ``exact''. In Fig.~\ref{fig:BJetSec} we show that including only the zeroth order term, the approximation to the exact result is great.

\begin{figure*}[t!]
\subfigure[]
{\includegraphics[width=0.49\textwidth]{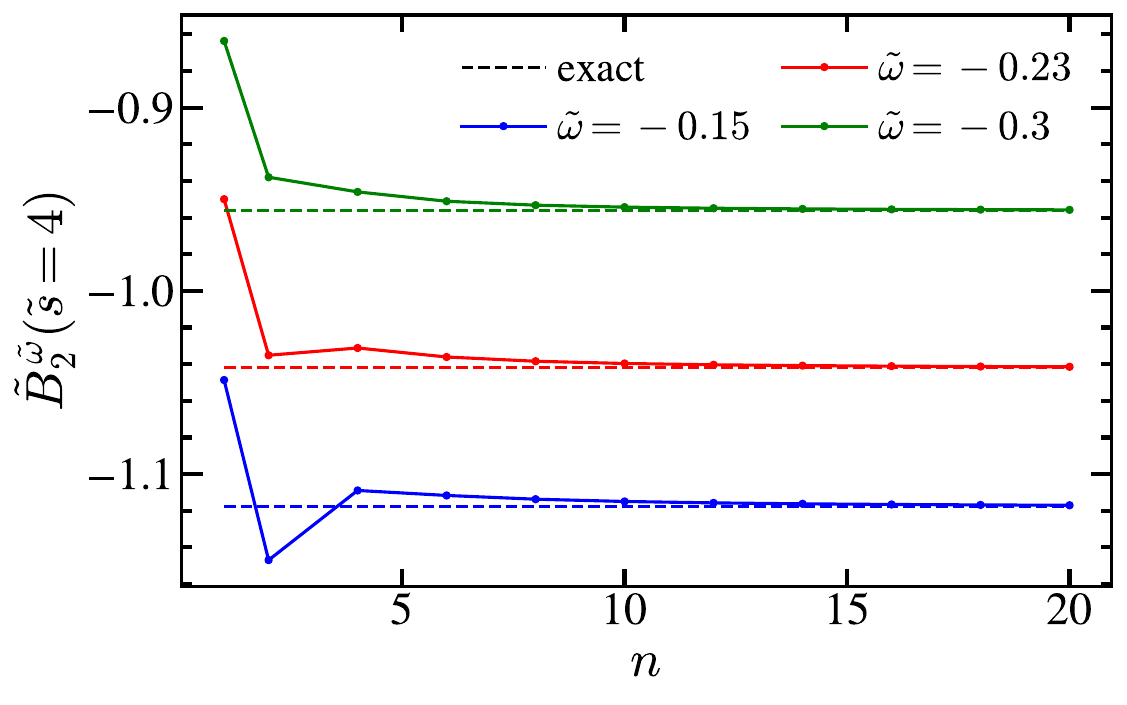}
\label{fig:BJetSecRG1}}
\subfigure[]{\includegraphics[width=0.49\textwidth]{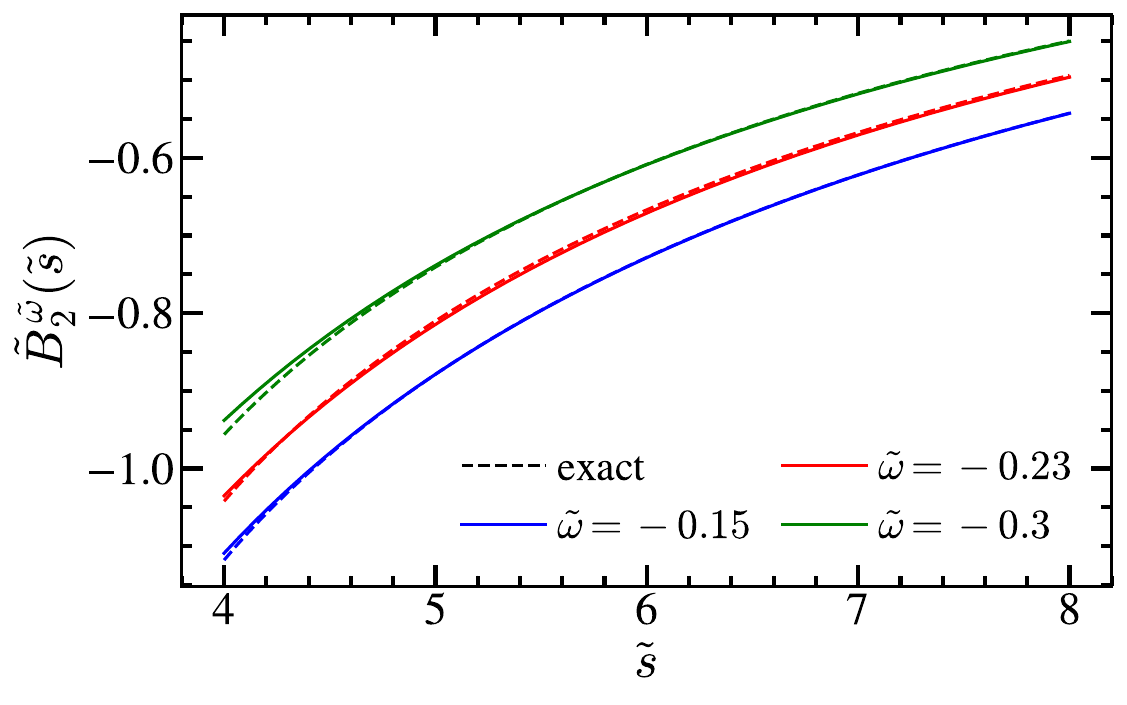}
\label{fig:BJetRGSec}}
\caption{Secondary quark mass correction to the massless RG-evolved bHQET jet function in its exact form (dashed), and expansions for small masses (solid lines) for three values of $\tilde \omega$: $-0.15$ (blue), $-0.23$ (red), and $-0.3$ (green). Left panel: small mass expansion of $\tilde B_2^{\tilde\omega}$ at threshold $\tilde s=4$ as a function of the expansion order $n$. Right panel: Dependence of $\tilde B_2^{\tilde\omega}$ with $\tilde s$, including 3 (2) non-zero terms in the small-mass expansion for $\tilde\omega = -0.15$ (other values).}
\label{fig:SecBJetRG}
\end{figure*}
The RG-evolved massive two-loop bHQET function is defined in complete analogy to Eq.~\eqref{eq:BRGEvolved}, and can be easily cast in the form of a MB integral:
\begin{equation}
\tilde B_2^{\tilde \omega}(\tilde s) =\!\! \int_0^{\tilde s} \frac{{\rm d}\hat s'}{\hat s'}
\biggl(1-\frac{\hat s'}{\tilde s}\biggr)^{\!-1-\tilde \omega} \!\!\tilde B_2(\hat s')
= \!\! \int_{c - i \infty}^{c + i \infty} \! \frac{{\rm d} h}{2 \pi i}
\frac{(h+1)^2 s^{2 h} \Gamma (h)^4}{4 h (2 h+1)^2 (2 h+3) \Gamma (2 h) (-\tilde \omega)_{2 h}}\,,
\end{equation}
with $-1/2<c<0$. We can use the converse mapping theorem to obtain the series for large and small masses. The former, as expected, has only one term, obtained from the quartic pole sitting at $h=0$, while the latter has simple poles at $h=-1/2$ and $-1$, and triple poles at all other negative integer values of $h$:
\begin{align}
\tilde B_2^{\tilde \omega}(\tilde s) = \, &\frac{8}{9} \tilde L^2 -\frac{2}{9} \tilde L^3+\frac{\tilde L}{3} \biggl[2 \psi ^{(1)}(-\tilde \omega)
-\frac{61}{9}\biggr]-\frac{8}{9} \psi ^{(1)}(-\tilde \omega)
+\frac{2}{9} \psi^{(2)}(-\tilde \omega)+\frac{223}{81}-\frac{2\zeta_3}{9} \nonumber\\
= \, & -\!\frac{\pi ^2 (\tilde \omega+1)}{2 \tilde s} + \sum_{n=1}\frac{(2 n)! (\tilde \omega+1)_{2 n}\tilde s^{-2 n}}{n(2 n - 1)^2 (2 n-3) (n!)^4} \biggl\{(n-1)^2\bigl\{\bigl[\tilde L+K_n(\tilde \omega)\bigr]^2+H_n^{(2)}\nonumber\\
&-\!H_{2 n}^{(2)}+\psi ^{(1)}(2 n+\tilde \omega+1)-\psi ^{(1)}(\tilde \omega+1)-\psi ^{(1)}(-\tilde \omega)\bigr\}\nonumber\\
& + \frac{(n-1) (8 n^3-22 n^2+19 n-3)}{n (2 n-3) (2 n-1)}[\tilde L + K_n(\tilde \omega)] \nonumber\\
& +\frac{48 n^6-264 n^5+596 n^4-670 n^3+372 n^2-90 n+9}{2 n^2 (2 n-3)^2 (2 n-1)^2}\biggr\}\,,\nonumber\\
K_n(\tilde \omega) =\, & 2 H_n-H_{2 n}-\sum_{i=1}^{2n}\frac{1}{\tilde\omega+i}\,,
\end{align}
with $\tilde L = \log(\tilde s) - \gamma_E - \psi ^{(0)}(-\tilde \omega)$ and where, for numerical performance, it is convenient to use the following identity:
\begin{equation}
\psi ^{(1)}(2 n+\tilde \omega+1) -\psi ^{(1)}(\tilde \omega+1) = -\sum_{i=1}^{2n}\frac{1}{(\tilde\omega+i)^2}\,.
\end{equation}
Since the infinite sum cannot be carried out, no analytic, all-order result for the RG-evolved function is obtained. Given the excellent convergence of the series, for our numerical studies, we consider summing the first $160$ terms, for all practical purposes, as the ``exact solution''. The excellent convergence of the series is studied in Fig.~\ref{fig:SecBJetRG}, where one can see that at the point where convergence should be slower, that is for $\tilde s = 4$, for all values of $\tilde \omega$ we have explored, the sum quickly converges to the expected result (given by the large-mass expression). At any other value of $\tilde s$, very good accuracy is already attained including 2 or 3 terms in the expansion.

The very last result we present in this article is the small-mass expansion of the Fourier transform of $M\Delta_0 B_2(\hat s, m)$
\begin{align}
\Delta_0 \tilde B_2(x_m) =\, & \int_{c - i \infty}^{c + i \infty} \frac{{\rm d} h}{2 \pi i}\frac{(h+1)^3 (i x_m)^{-2 h} \Gamma (h)^4}{(2 h+1) \Gamma (2 h+4)}\\
=\, & \frac{1}{2} i \pi ^2 x_m + 2\sum_{n=1}\frac{ (2 n-2)!(i x_m)^{2 n}}{(2n-3)(2 n-1) (n!)^4}
\biggl\{ (n-1)^2\bigl[\tilde L_x^2 + H_n^{(2)}-H_{2 n-2}^{(2)}\bigr]\nonumber\\
&-\frac{2(n-1) \tilde L_x}{(2 n-3) (2 n-1)}+\frac{12 n^2-24 n+13}{2 (2 n-3)^2 (2 n-1)^2}\biggr\}\,,\nonumber
\end{align}
with $L_x = \log(i x_m) + \gamma_E$, $-1/2 < c < 0$, and $\tilde L_x = L_x +H_{2 n-2}-2 H_n$.

\section{Conclusions}\label{sec:conclusions}
In this article we have presented a novel technique to compute quantum corrections due to either massive vector bosons or secondary massive quark bubbles in the form of series expansions for small and large (vector boson or secondary quark) masses, in both cases virtual. The method relies on the Mellin Barnes transform, and yields series to arbitrarily high orders whose convergence radii exactly matches up. Each term in the expansion corresponds to a pole along the positive (large-mass expansion) or negative (small mass series) real axis, and depending on the pole's multiplicity, logarithms of the mass might appear. In fact, at the mass value which acts as a boundary between the two expansions, both series converge and should agree: this has been extensively used as a sanity check of our results. In many cases, it is possible to sum up the series to all orders, finding closed analytic expressions that, in some cases, to the best of our knowledge are impossible to obtain with a direct computation. The advantage of the method is that the MB transform is applied at a very early stage of the computation: before computing the loop integral associated to the momentum running along the gluon leg (either massive by itself, or containing the secondary mass bubble as a sub-diagram). Therefore, the difficulty of the calculation is reduced to that of massless one-loop diagrams with a modified gluon propagator. Furthermore, these exactly coincide with the computations one has to carry out in the large-$\beta_0$ expansion at leading order, hence many known results can be recycled and physical connections between renormalons and mass corrections can be easily made: for example, a $u=1/2$ renormalon implies that the expansion for small masses will involve a linear term. We have observed that for the case of massive vector bosons, such renormalons seem to come in hand with oscillatory behavior and slow convergence of the small-mass series, although this IR sensitivity does no affect the convergence of the series associated to the secondary quark mass correction. Even though the computations must be carried out in $d=4-2\varepsilon$ dimensions in order to recover the $m\to 0$ limit, once the massless result is subtracted one can set $\varepsilon\to 0$ and use the converse mapping theorem to obtain both expansions.

We have applied this new method to a plethora of examples, mostly confined to the context of effective field theories for jets (SCET and bHQET), but as a warm-up exercise also to the massive-quark vacuum polarization function and the relation between the pole and $\MSb$ masses. In all cases, we have recovered known results for the massless limit, $Z$ factors and anomalous dimensions. We have also reproduced known results for the Wilson coefficient and jet function in SCET. For the case of the SCET and bHQET jet functions, it is very simple to isolate the contributions from virtual and real radiation of massive particles: the large mass expansion has a single term (the pole located at $\varepsilon$) which encompasses the entire virtual contribution. Furthermore, we have also found expansions for the RG-evolved version of the various jet functions, which seems to be the only possible approach to obtain analytic results. Except for the cases laid out in the previous paragraph, we have found excellent convergence in all cases under study, such that using the expansions is more practical than evaluating the exact result (if it is known), and becomes especially relevant for numeric implementations. We have coded all our results in \texttt{Python}~\cite{Rossum:1995:PRM:869369}, using the \texttt{numpy}~\cite{oliphant2006guide} and \texttt{scipy}~\cite{Virtanen:2019joe} modules, to carry out all our numerical studies, which have been cross-checked against an independent \texttt{Mathematica} code~\cite{mathematica}. All plots have been generated using the \texttt{matplotlib} module~\cite{Hunter:2007}.

The downside of the method is that it can only be applied to quantities which have no infrared or collinear divergences at one-loop with massless gluons, but for the examples we have worked out in this article, consistency conditions have been applied to circumvent this difficulty, and the individual pieces entering the matching conditions between QCD and SCET, or SCET and bHQET, have been obtained. In case such conditions do not apply for a particular method, one can use an IR regulator such as off-shellness to obtain the result. After the MB transform is applied, the regulator can be safely set to zero. We have not explored this possibility in detail in this article, but plan to do so in upcoming works.

The results we have obtained are highly relevant for the theoretical description of jet cross sections involving massless or massive secondary quarks, and can be easily adapted to other examples such as $B$-meson decays, Higgs cross sections, DIS, etc. The expansions we have found will play an important role in future updates of the Monte Carlo top quark mass calibration initiated in Refs.~\cite{Butenschoen:2016lpz,Dehnadi:2023msm}, and follow the efforts started in Refs.~\cite{Lepenik:2019jjk,Bris:2020uyb}, where fixed-order and resummed results where found for massive primary quarks. We have set the basis for a variable flavor number scheme for $e^+e^-\to t\bar t + X$ cross section in the peak, which must be taken into account when the mass of the bottom quark is not neglected ---\,this is necessary when aiming at high precision. A thorough numerical analysis of these effects, along with the computations of the necessary pieces using the dispersive framework, will be presented in the near future. The ideas introduced in this article can be extended in a number of ways: one can consider two massive bubbles inserted into a massless gluon line, either with equal or different masses, or the insertion of a massive bubble followed by an arbitrary number of massless quark bubbles. One could also try to adapt the method for the case in which the secondary quarks are produced on-shell, which would be relevant for instance to compute an expansion for the soft function. Finally, one could have both primary and secondary quarks massive, and work out double expansions. The MB procedure can be modified to write the convolution of shape functions with partonic cross sections as an expansion of (inverse) powers of $\Lambda_{\rm QCD}$ in the (peak) tail of the distribution. Since we have worked out all the necessary pieces to compute the N$^3$LL resummed singular cross section for the secondary radiation of massive quarks, the next natural step would be providing the two-loop fixed-order contribution to jet cross sections. These shall be presented in future publications.

\begin{acknowledgments}
This work was supported in part by the Spanish MECD grants Nos.\ FPA2016-78645-P, PID2019-105439GB-C22, and PID2022-141910NB-I00, the Spanish MCI grant No.\ PID2022-141910NB-I00, the EU STRONG-2020 project under Program No.\ H2020-INFRAIA-2018-1, Grant Agreement No.\ 824093, the COST Action No.\ CA16201 PARTICLEFACE, and the IFT Centro de Excelencia Severo Ochoa Program under Grant SEV-2012-0249. A.\,B.\ is supported by an FPI scholarship funded by the Spanish MICINN under grant no. BES-2017-081399. We are grateful to the Erwin-Schr\"odinger International Institute for Mathematics and Physics for partial support during the Programme ``Quantum Field Theory at the Frontiers of the Strong Interactions'', July 31 - September 1, 2023. A.\,B.\ thanks the University of Salamanca for hospitality while parts of this work were completed.
\end{acknowledgments}

\bibliography{thrust3}
\bibliographystyle{JHEP}

\end{document}